\documentclass[twocolumn]{aastex701}

\shortauthors{K. Biazzo et al.}
\received{Apr 15, 2026}
\revised{Jun 19, 2026}
\accepted{Jul 16, 2026}

\newcommand{\gtsim}{\protect\raisebox{-0.5ex}{$\:\stackrel{\textstyle >}
        {\sim}\:$}}

\newcommand{\teff}{$T_{\rm eff}$}

\newcommand{\Msun}{{$M_{\odot}$}}

\usepackage{aas_macros}
\usepackage{amssymb} 

\begin{document}
 
\title{LBT-MODS spectroscopy of young stellar objects in the distant metal-poor star forming region Sh2-284: Stellar and accretion properties\footnote{Based on observations collected at the Large Binocular Telescope under the LBT programmes IT-2021B-013 and IT-2023B-026 (PI: K. Biazzo).} 
\footnote{This paper is dedicated to the memory of Prof. Nino Panagia, who passed away in January 2025}}

\author[orcid=0000-0002-1892-2180]{K. Biazzo}
\affiliation{INAF - Osservatorio Astronomico di Roma, Via Frascati 33, I-00040 Monte Porzio Catone (RM), Italy}
\email[show]{katia.biazzo@inaf.it}  

\author[orcid=0000-0001-8657-095X]{J. M. Alcal\'a}
\affiliation{INAF - Osservatorio Astronomico di Capodimonte, Salita Moiariello 16, I-80131 Napoli, Italy}
\email[show]{juan.alcala@inaf.it}  

\author[orcid=0000-0003-2910-6565]{F. Cusano}
\affiliation{INAF-Osservatorio di Astrofisica e Scienza dello Spazio, Via Piero Gobetti, 93/3, I-40129 Bologna, Italy}
\email[]{felice.cusano@inaf.it}  

\author[orcid=0000-0002-3010-2310]{M. G. Guarcello}
\affiliation{INAF – Osservatorio Astronomico di Palermo, Piazza del Parlamento 1, 90134 Palermo, Italy}
\email[]{mario.guarcello@inaf.it}  

\author[orcid=0000-0002-7409-8114]{D. Paris} 
\affiliation{INAF - Osservatorio Astronomico di Roma, Via Frascati 33, I-00040 Monte Porzio Catone (RM), Italy}
\email[]{diego.paris@inaf.it}  

\author[orcid=0000-0002-6894-1267]{E. Marini} 
\affiliation{INAF - Osservatorio Astronomico di Roma, Via Frascati 33, I-00040 Monte Porzio Catone (RM), Italy}
\email[]{ester.marini@inaf.it}  

\author[orcid=0000-0002-7409-8114]{R. Carini} 
\affiliation{INAF - Osservatorio Astronomico di Roma, Via Frascati 33, I-00040 Monte Porzio Catone (RM), Italy}
\email[]{roberta.carini@inaf.it}  

\author[orcid=0000-0002-7035-8513]{T. Giannini} 
\affiliation{INAF - Osservatorio Astronomico di Roma, Via Frascati 33, I-00040 Monte Porzio Catone (RM), Italy}
\email[]{teresa.giannini@inaf.it}  

\author[orcid=0000-0002-9190-0113]{B. Nisini} 
\affiliation{INAF - Osservatorio Astronomico di Roma, Via Frascati 33, I-00040 Monte Porzio Catone (RM), Italy}
\email[]{}  

\author[orcid=0000-0002-0666-3847]{S. Antoniucci} 
\affiliation{INAF - Osservatorio Astronomico di Roma, Via Frascati 33, I-00040 Monte Porzio Catone (RM), Italy}
\email[]{simone.antoniucci}  

\author[orcid=0000-0001-7906-3829]{G. De Marchi} 
\affiliation{European Space Research and Technology Centre, Keplerlaan 1, 2200 AG Noordwijk, The Netherlands}
\email[]{Guido.De.Marchi@esa.int}  

\author[orcid=0000-0002-1860-2304]{M. G. Navarro Ovando}
\affiliation{INAF - Osservatorio Astronomico di Roma, Via Frascati 33, I-00040 Monte Porzio Catone (RM), Italy}
\email[]{maria.navarro@inaf.it}

\begin{abstract}

We present a spectroscopic survey of young stellar objects (YSOs) in Sh2-284, a distant ($\sim$4.5\,kpc), low-metallicity ($Z\,\sim\,1/3\,Z_\odot$) star-forming region (SFR) toward the Galactic anticenter. Candidate YSOs were selected using mid-infrared Spitzer/IRAC data with optical and near-infrared photometry. Follow-up spectroscopy was conducted with MODS at the Large Binocular Telescope. We characterize the stellar and accretion properties of the disk-bearing population in a metal-poor environment, probing stellar masses from $\sim$2.95\,$M_\odot$ to $\sim$0.35\,$M_\odot$. This work constitutes the first wide-field ($\sim\,45'\times\,45'$) spectroscopic investigation of YSOs in Sh2-284, providing a comprehensive view of star formation in this Galactic environment. We provide tentative first estimates of iron abundance for three low-mass targets and lithium abundance for a limited number of objects, offering an initial characterization of the chemical properties of these stars. We adopt a multi-diagnostic approach based on nine tracers, exploiting the spectral coverage of the instrument, using H$\alpha$, H$\beta$, H$\gamma$, Ca{\sc ii} infrared triplet, Pa$\eta$, Pa$\zeta$, and Pa$\epsilon$ emission lines. We derive key accretion properties and compare them with those measured in nearby, solar-metallicity SFRs to explore potential metallicity-driven differences in accretion behavior and disk evolution. We tentatively find indications of a flattening in the flux-flux relations of the metal-poor YSOs compared to solar-metallicity samples, a behavior that is recovered across diagnostics. Our observations indicate that the selected disk-bearing YSOs in this metal-poor environment exhibit resilient accretion activity, spanning from $\sim\,2.3\,\times\,10^{-10}\,M_\odot$\,yr$^{-1}$ and $\sim\,1.0\,\times\,10^{-6}\,M_\odot$\,yr$^{-1}$, and a median rate of $\sim\,2.2\,\times 10^{-8}\,M_\odot$\,yr$^{-1}$. Overall, this demonstrates that gas accretion can be efficiently sustained in sub-solar metallicity environments. 

\end{abstract}


\section{Introduction} 
\label{sec:intro}

During the final stages of their formation, low-mass ($<$2\,\Msun) stars are still surrounded by a dense and mostly gaseous circumstellar disk, even after having shed most of the dust and gas envelope from which they were born \citep{Hartmannetal2016}. These circumstellar disks are not only the remnants of the star formation process, but also the cradles where planetary systems emerge to take shape. Infrared observations of star-forming regions (SFRs) with solar metallicity in the solar vicinity show that optically thick circumstellar disks disappear in $\sim$1-3\,Myr in almost half of young low-mass stars and are almost absent around members of $\gtrsim$15\,Myr old associations (e.g., \citealt{Fedeleetal2010, CurrieSicilia2011, Polnitzkyetal2025}), the latter age approximately corresponding to the shutdown of mass accretion (e.g., \citealt{Calvetetal2005, Sicilia-Aguilaretal2005, Manaraetal2023}). The disappearance of gas and dust places stringent constraints on the timescale of planet formation. In fact, since the formation and evolution of stars and planets are strongly connected, learning about the properties of young stars and the evolution of their circumstellar accretion disks is key to understanding the formation of planets (\citealt{MorbidelliRaymond2016}). 

In the magnetospheric accretion scenario, a central low-mass star grows in mass over time through the accretion of material from a circumstellar disk of dust and gas funneled by the stellar magnetic field (e.g., \citealt{Koenigl1991, Hartmannetal2016}). In this context, metallicity may play an important role in the evolution and following dispersal of the accretion disks, but its influence is still far from being clear.

In one hand, observational studies of young stellar populations ($\sim$1\,Myr) in SFRs of the outer Galaxy and at low metallicity ($Z \sim 0.1-0.3\,Z_\odot$) indicate that metal-poor stars dissolve their  disks significantly faster than solar-Z YSOs (\citealt{Yasuietal2009, Yasuietal2016}). This would suggest that a low-metallicity environment may induce higher mass accretion rates, shorter disk lifetimes and/or a very efficient external photoevaporation of the circumstellar matter due to UV and/or X-ray radiation from OB stars, which contributes to a rapid disk erosion (\citealt{ErcolanoClarke2010}, \citealt{Gehrigetal2023}). As a consequence, stars forming in a low-metallicity environment should experience disk dispersal on timescales shorter than 1\,Myr, much shorter than in a solar-metallicity environment. 

On the other hand, studies performed using HST photometry in young open clusters of the Magellanic Clouds ($Z \sim 0.1-0.3\,Z_\odot$), 
demonstrate that mass accretion rates decrease more slowly with time than what is observed in solar-metallicity SFRs (\citealt{DeMarchietal2010, DeMarchietal2013, Biazzoetal2019, Carinietal2022}). More recently, \cite{DeMarchietal2024}  performed JWST observations of solar mass stars in NGC\,346, a metal-poor ($Z = 1/8 Z_\odot$) SFR of the Small Magellanic Cloud, and found indications that YSOs at ages older than $\sim$10 Myr are still accreting gas with typical rates of $\sim 10^{-8} M_\odot$ yr$^{-1}$. The spectra also revealed near-infrared excess and molecular hydrogen excitation lines, consistent with the presence of disks around these stars. These findings suggested that in a low-metallicity environment circumstellar disks can live longer than previously thought.

These two types of results are apparently in contradiction, because the former foresees shorter timescales for the disk dispersal, and the latter longer timescales for the decrease of mass accretion rate (and, in principle, longer disk lifetime). However, these studies are not directly comparable because they scrutinize different properties. From the one side, the dust content of circumstellar disks is used as a proxy for the total mass of the disks and their lifetimes, while from the other side the infall of the more abundant inner gas onto the stars is used as a proxy of the mass accretion rate. Very recently, \cite{Yasuietal2026} reported disk fractions, derived from diagnostics at wavelengths $\geq 3\,\mu$m, in a distant, metal-poor Galactic SFR that are consistent with those observed in the solar neighborhood. This result contrasts with their earlier findings based on near-infrared ($\lesssim 3\,\mu$m) diagnostics. They therefore concluded that the innermost disk material, which dominates the emission at shorter wavelengths, may have already dissipated or become undetectable even in sources that still exhibit clear disk excess at longer wavelengths.

Another possibility has been explored that low metallicity might enhance the effectiveness of photoevaporation in removing gas and small dust grains from circumstellar disks, therefore impacting disk erosion (\citealt{Nakatanietal2018}). On the other hand, recent observations hint at the effect of stellar density, intense local UV fields, and close stellar encounters as additional factors regulating the intensity of the accretion process (\citealt{TsiliaDeMarchi2023, VlasblomDeMarchi2023, Guarcelloetal2023}). Therefore, a clear, unified picture is still missing.

Complete censuses of metal-poor YSOs in low-metallicity regions, like the one studied in this work, are essential to perform a detailed investigation of mass accretion rate and to compare the results with similar spectroscopic studies led in solar metallicity Galactic YSOs. This will allow us to investigate the star formation history in metallicity environments different from those of the solar vicinity and to shed light on possible inconsistencies claimed in the literature.

\subsection{Sh2-284: a distant metal-poor star forming region in the direction of the Galactic anticenter}
\label{sec:sh2-284_intro}

Located at the Galactic anticenter ($l \simeq 212.0^{\circ}$, $b \simeq -1.2^{\circ}$) and at a distance of $\sim$4.5\,kpc from the Sun (e.g., \citealt{Negueruelaetal2015}), Sharpless\,2-284 (hereafter Sh2-284; \citealt{Sharpless1959}), is an H{\sc ii} region excited by the ultraviolet radiation of massive OB stars residing in the young ($\sim$3-6 Myr) open cluster Dolidze\,25 (\citealt{TurbideMoffat1993, Negueruelaetal2015}). \cite{Lennonetal1990} found for individual elements of OB-type stars abundances ranging from $-0.5$ to $-0.9$\,dex, implying an overal metallicity of roughly  one-sixth of the solar value. More recently, \cite{Negueruelaetal2015} measured abundances of $\sim -0.3$\,dex and $\sim -0.5$\,dex below solar for silicon and oxygen, respectively, placing the cluster at a metallicity of $Z\sim1/3-1/2\,Z_\odot$. Consequently, the cluster's metallicity is comparable to that of the Large Magellanic Cloud (LMC; \citealt{Coluccietal2012}) or intermediate between the LMC and the Small Magellanic Cloud (\citealt{Leeetal2005}). Given that Dolidze\,25 is associated with Sh2-284, this implies that the latter is also characterized by low metallicity, making it one of the most metal-poor star-forming regions currently known in the Milky Way. Sh2-284 therefore represents a unique testbed for studying the evolution of accreting circumstellar disks at low metallicity, under the combined effects of reduced metal content and intense ionization radiation from massive OB stars.

\cite{Pugaetal2009} identified class I and class II YSO candidates through IR Spitzer/IRAC colors. They also found indications of triggered and sequential star formation of the region by expansion of its individual H{\sc ii} regions. \cite{Cusanoetal2011} presented the results of a photometric and spectroscopic follow-up of 23 YSOs observed with VIMOS@VLT, claiming that a significant fraction of YSOs may have preserved their disks/envelopes. \cite{KalariVink2015} used the spectra acquired by \cite{Cusanoetal2011} to perform spectroscopic estimates of the mass accretion rates ($\dot M_{\rm acc}$) of 24 YSOs in Sh2-284, finding no evidence for a systematic difference in the mean value of $\dot M_{\rm acc}$ with metallicity. A comprehensive work based on {\it Chandra} observations combined with existing optical/infrared catalogues was led by \cite{Guarcelloetal2021}, who selected the disk-bearing population in a circular region with a diameter of 1\,$^\circ$ centered on Dolidze\,25. The authors claim that the disk evolution in the cluster is dictated more by the low metallicity than by external photoevaporation or dynamical encounters. Recently, ALMA and JWST observations of the most active star-forming sites in Sh2-284 performed by \cite{Chengetal2025} and \cite{Jadhavetal2025} revealed CO and H$_2$ outflows, indicating ongoing active star formation, while the same research group found a dependence of the initial mass function on $Z$ (\citealt{Andersenetal2025}). Very recently, \cite{Ashrafetal2026} observed the central $2\arcmin \times 1\arcmin$ core area of Dolidze\,25 with MUSE@VLT deriving mass accretion rates from H$\alpha$ emission for 55 sources in the $0.1-0.7$\,$M_\odot$ mass range. They concluded that the accretion rates in Dolidze\,25 are broadly comparable to those in solar-metallicity regions, finding no clear evidence of a strong metallicity dependence within the cluster core. 

In this work, we investigate the accretion properties of YSOs across a much wider spatial field ($\sim 45'\times 45'$), allowing us to extend the core analysis to the outer cluster environment and possibly examine sequential star formation. Moreover, our analysis, which is based on spectroscopic observations performed at the Large Binocular Telescope (LBT), is complementary to those of the latter authors in terms of stellar mass, investigating YSOs with masses of $\sim 0.35-2.95$\,$M_\odot$.

To our knowledge, this work constitutes the first spectroscopic investigation of the YSO population across a wide field on the Dolidze\,25 region, extending the accretion studies beyond the cluster core and providing a more comprehensive view of star formation in this low-metallicity Galactic environment. However, characterizing star formation in such a distant and metal-poor cluster introduces observational and analysis challenges. To actively address these concerns, this study presents the first multi-diagnostic analysis of this region, adopting an approach based on nine tracers and fully exploiting the capabilities of the instrument over a broad spectral range, from H$\beta$ to Pa$\epsilon$. This comprehensive strategy allows us to cross-check our diagnostics and mitigate systematic uncertainties. Our analysis is focused on the connection between accretion and stellar properties, and on how these relations compare with those observed in nearby Galactic star-forming regions, as well as in more distant Galactic regions and metal-poor environments in the Magellanic Clouds recently probed by JWST spectroscopy.

\section{Stellar sample, observations, and data reduction}
\label{sec:sample_selection}
\subsection{Sample selection}

We considered a $\sim 45'\times 45'$ region centered on Dolidze\,25 and used the catalogue of \citet{Guarcelloetal2021}, which includes 1097 sources. Given the large physical scale spanned by our survey ($\sim 60$ pc at a distance of $\sim$4.5\,kpc), Sh2-284 should not be viewed as a single ``monolithic'' entity but rather as a structured star-forming complex containing multiple sub-clusters, as suggested by previous spatial analyses (e.g., \citealt{Patraetal2024}). Combining YSOs across such an extended region could smooth over localized variations, and we therefore consider both the global properties of the sample and the characteristics of individual sources in our analysis.

From the sample by \citet{Guarcelloetal2021}, we selected targets satisfying the following criteria:
\begin{itemize}
\item[-] disk-bearing objects (666) identified by the authors (DISK = 1 and XMEMBER = 0 flags in their catalogue);
\item[-] sources (152 over 666) with $16 < R_{\rm PANSTARRS} < 20$ mag, a range chosen to ensure a signal-to-noise ratio ($S/N$) greater than $\sim 30$ in 1\,hr exposures, which is sufficient to detect emission lines in young, low-mass stars, while keeping photometric uncertainties low and minimizing spurious detections;
\item[-] sources (104 over 152) falling within the class I and class II region in the [3.6]$–$[4.5] versus [5.8]$–$[8.0] color–color diagram, as defined by \citet{Allenetal2004} using Spitzer data;
\item[-] objects (91 over 104) exhibiting H$\alpha$ excess based on photometry from the INT Photometric H$\alpha$ Survey (IPHAS).\footnote{We used IPHAS data for the entire sample, except for two sources not detected in IPHAS H$\alpha$, for which we adopted VPHAS$+$ (VST Photometric H$\alpha$ Survey) photometry.}
\end{itemize}

The adopted selection criteria naturally favor optically bright and actively accreting disk-bearing objects, in line with previous spectroscopic surveys of individual YSOs in nearby star-forming regions (see, e.g., \citealt{Alcalaetal2017, Gangietal2022}, and references therein), thus ensuring us a meaningful comparison of our analysis with these previous surveys. Interestingly, these requirements do not introduce a bias against embedded or highly obscured sources; indeed, the mean visual extinction of our spectroscopic sample ($\langle A_V \rangle = 2.5$ mag) is consistent with that of the entire parent catalog ($\langle A_V \rangle = 2.8$ mag) of \cite{Guarcelloetal2021}. Instead, our selection strategy is primarily tailored to maximize the recovery of the actively accreting population across the widest possible mass range, ensuring the high signal-to-noise ratios necessary to robustly characterize their individual accretion properties. As said above, such a targeted spectroscopic follow-up approach is similar to the one in previous cornerstone surveys of accreting young populations across a variety of star-forming environments (e.g., \citealt{Alcalaetal2017, Manaraetal2015, Gangietal2022, DeMarchietal2024, Rogersetal2024}).

\begin{figure*}[!t]
\begin{center}
\includegraphics[width=18cm]{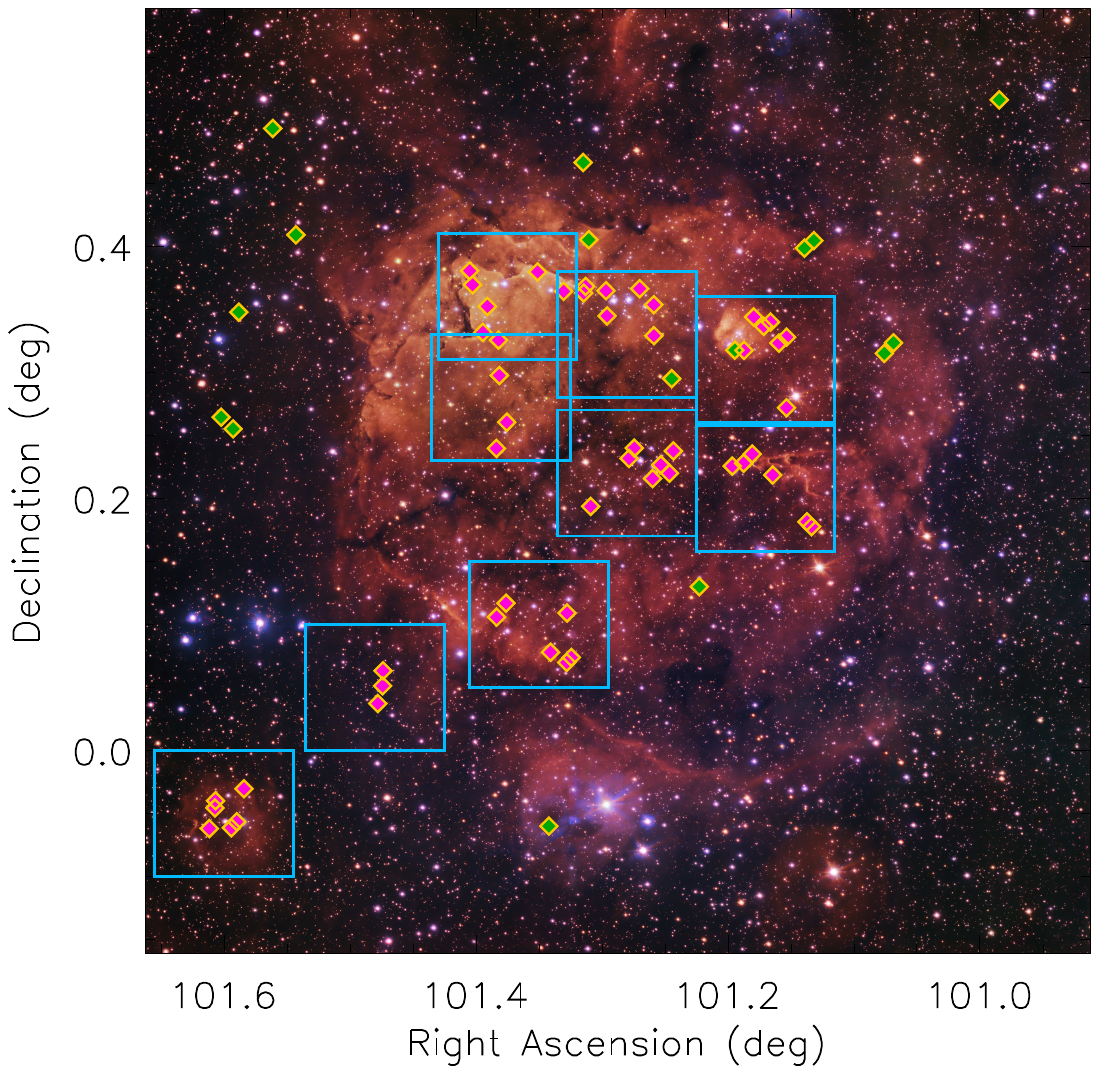}
\vspace{-5.9cm}
\caption{VPHAS+ image of the Sh2-284 SFR. VPHAS+ uses VST/OmegaCAM to survey the southern Galactic plane in $ugri$H$\alpha$ bands. Blue squares represent the nine MODS fields observed in multi-object mode (see upper part of Table\,\ref{tab:log}). YSO candidates observed in multi-object mode are represented with filled magenta diamonds, while those observed in single-slit mode are represented with filled green diamonds. The covered area is $\sim 45'\times 45'$. North is up and east to the left. Credits: ESO/VPHAS+ team.} 
\label{fig:spatial_distrib} 
\end{center}
\end{figure*}

Among the 91 YSOs originally selected from the \cite{Guarcelloetal2021} catalogue using the above criteria, we were able to observe 68 (see Sect.\,\ref{sec:observations}). Although bad weather conditions prevented the observation of the entire list, this final sample of 68 targets remains highly representative and provides a robust statistical baseline to reliably characterize the accreting population of Sh2-284. Of these 68 targets, 46 were also included in the sample of Class\,I or Class\,II sources observed with {\it Spitzer} and analyzed by \cite{Pugaetal2009}. Figure\,\ref{fig:spatial_distrib} shows the location of the 68 selected YSO candidates superimposed on the VPHAS+ image, while Fig.\,\ref{fig:ccd_all} shows the position of our sample in the 2MASS $J$ versus $J-K$, Spitzer [3.6]$-$[4.5] versus [5.8]$-$[8.0], and IPHAS $r-H\alpha$ versus $r-i$ diagrams.

\subsection{Observations}
\label{sec:observations}

\begin{figure*}[!t]
\begin{center}
\includegraphics[width=1\linewidth]{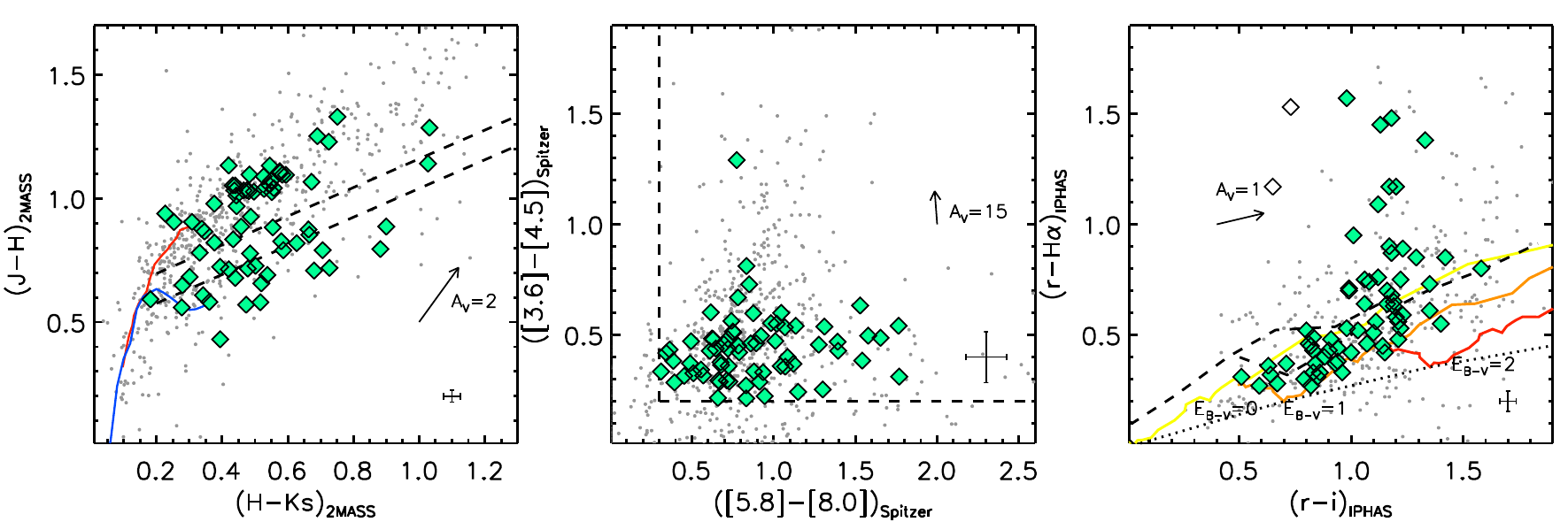}
\caption{Color-color diagrams of the Sh2-284 targets. Small grey dots are sources from the catalogue by \cite{Guarcelloetal2023}, while green diamonds represent the 68 YSO candidates selected and then observed by us. {\it Left:} 2MASS $J-H$ versus $J-K_{\rm S}$ color-color diagram. The dwarf (lower branch; \citealt{BessellBrett1988}) and the giant (upper branch; \citealt{KenyonHartmann1995}) sequences are marked by solid lines. The CTTs locus (\citealt{Meyeretal1997}) is delimited by the dashed lines. The arrow indicates the reddening vector for $A_{\rm V}=2$\,mag. The mean 2MASS photometric errors are overplotted on the lower-right corner of the panel. {\it Middle:}  Spitzer [3.6]$-$[4.5] versus [5.8]$-$[8.0] color-color diagram. The locus of stars with disks is delimited with dashed lines (\citealt{Allenetal2004}). The arrow represents the reddening vector for $A_{\rm V}=15$\,mag. {\it Right:} IPHAS $r-H\alpha$ vs $r-i$ color-color diagram. Solid (yellow, orange, red) lines are the ZAMS at increasing extinction values and dotted line marks the locus of A-type stars according to \cite{Drewetal2005}, while the dashed line is the ZAMS defined by \cite{Barentsenetal2011} assuming $EW_{H\alpha}$=10\,\AA\, and $E_{B-V}=1$. Open diamonds identify the two targets selected through VPHAS+ colors.}
\label{fig:ccd_all}
\end{center}
\end{figure*}

Observations were conducted between 2021 November 22 and 2024 November 30 using the Multi-Object Double Spectrograph (MODS; \citealt{Poggeetal2010}) mounted on the 8.4\,m Large Binocular Telescope (LBT; Mount Graham, Arizona, USA). Data were acquired in dual grating mode, employing two channels optimized for the blue ($\sim 320-590$\,nm) and red ($\sim 540-1000$\,nm) portions of the spectrum. Depending on the target position, we performed either multi-object (FoV: 6$\arcmin\times$6$\arcmin$) or segmented long-slit (total length: 60$\arcmin$) spectroscopy. In both configurations, a 0.$\arcsec 6$ slit was used, yielding a nominal spectral resolution of $R_{\rm BLUE} \sim 1850$ in the blue channel and $R_{\rm RED} \sim 2300$ in the red channel. The resulting $S/N$ ratio for the brightest sources ranges from $\sim$40 at $\sim$4600\,\AA\,(blue channel) to $\sim$50 at $\sim$6200\,\AA\,(red channel). A log of the observations is provided in Table\,\ref{tab:log}, where the targets (ID) observed in multi-object (Field) and long-slit (LS) modes are listed in the upper and the lower part of the table, respectively. The positions of the sources are illustrated in Fig.\,\ref{fig:ccd_all}.

All targets were also imaged in the $u$ and $g$ filters (blue channel) and $r$, $i$, and $z$ filters (red channel) with the aim to perform spectrophotometry of the YSOs and optimize the flux calibration (see Sects.\,\ref{sec:photometric_data}, \ref{sec:spectroscopic_data}).

\begin{table}  
\caption{Summary of the observations.}
\label{tab:log} 
\tiny
\vspace{-0.5cm}
\begin{center} 
\setlength{\tabcolsep}{1.4pt}
\begin{tabular}{lcccrccl} 
\hline 
\hline 
\multicolumn{8}{c}{{Multi-Object mode}}\\
\hline
Field & ID & $N$ &    RA      &     DEC     &  Date       &    UT   & $t_{\rm exp}$ \\
         &  &  &   (deg)    &   (deg)     &    (yy-mm-dd) & (hh:mm:ss)   & (s)  \\ 
\hline
F1	 & 10,20,21,45,51,66,76 & 7 & 101.3733 & 0.3595    &   2022-02-06 & 02:29:25.1 & 3600 \\   
F10	 & 18,22,52& 3 & 101.4742 & 0.0509    &   2022-02-05 & 06:24:48.1 & 3600 \\   
F3	 & 3,5,25,53,60,80,83& 7 & 101.2752 & 0.2202    &   2021-11-03 & 10:06:31.3 & 3600 \\   
F4	 & 4,16,29,38,65,70& 6 & 101.3562 & 0.0910    &   2021-11-03 & 11:23:42.2 & 3600 \\   
F5$^\ast$& 7,19,20,32,36,57,85,86& 8 & 101.2919 & 0.3493    &   2021-12-21 & 09:43:51.7 & 4800 \\  
F6 &9,14,46,48,64,75,81&  7 & 101.1718 & 0.3020    &   2021-11-05 & 09:39:46.7 & 4800 \\   
F7 &15,26,28,42,54,63 & 6 & 101.1610 & 0.2111    &   2022-02-05 & 05:33:15.3 & 3600 \\   
F8$^\bullet$& 11,58,67,69,72,78 & 6 & 101.6068 &$-$0.0404  &  2021-12-21 & 07:57:59.1 & 6000 \\ 
F9 &21,40,44,87 & 4 & 101.3796 & 0.2814    &  2021-11-06 & 10:13:44.6 & 3600 \\   
\hline
\multicolumn{8}{c}{{Long-Slit mode}}\\
\hline
LS& ID & $N$ &    RA      &     DEC     &  Date       &    UT   & $t_{\rm exp}$ \\
&          &   &   (deg)    &   (deg)     &    (yy-mm-dd) & (hh:mm:ss)   & (s)  \\ 
\hline
12 & 12 & 1 & 101.2226 & 0.1297    &   2022-02-06 & 05:37:39.0 & 3600 \\   
47 & 47,61 & 2 & 101.1871 & 0.3173 &   2022-02-06 & 06:07:12.9 & 4800 \\		   
6 & 6 & 1 & 100.9845 & 0.5162    &   2022-02-06 & 03:31:29.8 & 3600 \\   
8 & 8,13 & 2 & 101.0758 & 0.3149 &   2022-02-06 & 04:23:01.9 & 3600 \\   
17	& 17 & 1 & 101.0758 & 0.3149    &   2023-10-19 & 11:40:39.5 & 3600 \\   
24	& 24 & 1 & 101.5612 & 0.4935    &   2023-10-17 & 10:34:40.1 & 3000$^a$ \\      
27	& 27 & 1 & 101.5428 & 0.4090    &   2023-10-17 & 11:20:57.9 & 3600 \\   
33	& 33 & 1 & 101.3148 & 0.4665    &   2023-10-19 & 10:00:35.6 & 3600 \\   
34	& 34 & 1 & 101.3420 &$-$0.0603  &   2023-10-17 & 12:07:05.5 & 3600 \\   
37	& 37 & 1 & 101.5882 & 0.3477    &   2023-10-19 & 10:50:29.6 & 3600 \\   
23 & 23,62  & 2 & 101.6022 & 0.2643 &   2024-01-10 & 05:38:18.0 & 4800 \\   
30 & 30,31 & 2 & 101.1317 & 0.4045 &   2024-01-10 & 06:44:34.6 & 6000 \\   
71	& 71 & 1 & 101.2445 & 0.2949    &  2024-02-16 & 04:38:25.7 & 3600$^b$ \\   
77	& 77 & 1 & 101.3103 & 0.4052    &  2024-02-16 & 03:36:40.8 & 4800 \\   
\hline
\end{tabular} 
\end{center} 
Notes: ID20 was observed twice (both in F5 and F1). ID21 was observed twice (both in F9 and F1). ID47 in slit mode was also observed in multi-object mode (named ID48 in F6). ID17 and ID8 observed in slit mode correspond to the same target.\\
$^\ast$ No observations in the blue channel because of bad weather conditions. $^\bullet$ No useful data in the blue channel with the exception of ID11 because of bad weather conditions throughout the observing night. $^a$ The blue channel observations were done with $t_{\rm exp}$=3600\,s. $^b$ The blue channel observations were done with $t_{\rm exp}$=7200\,s.
\normalsize 
\end{table}

\subsection{Data reduction}

\subsubsection{Photometry}
\label{sec:photometric_data}
The imaging data were reduced at the LBT Imaging Data Center\footnote{https://lsc.oa-roma.inaf.it} of the {\it Osservatorio Astronomico di Roma}, utilizing the pipeline described in \cite{Fontanaetal2014}.


Each raw science image was de-biased by subtracting a median stack bias image (masterbias) and flat-fielded using a normalized median stack flat image (masterflat). The de-biased science images were then divided by the masterflat in order to homogenize the response of each image pixel. During this pre-reduction, dedicated algorithms were also applied to mask artifacts like saturated regions, cosmic rays, and hot/cold pixels. Background maps were then generated with SExtractor (\citealt{BertinArnouts1996}) on pre-reduced science images by masking the flux of the sources, their outskirts and artifacts. These maps were used to subtract additive background structures from science images.


Astrometric solution was computed using SCAMP (\citealt{Bertin2006}) and applied to the calibrated background-subtracted science images to correct for geometrical distortions, by correcting for relative positional offsets between exposures and by refining the absolute global calibration, anchoring all the images to Gaia-DR3 (\citealt{GaiaDR32023}), taken as a reference. Flux scales were computed, allowing for the final co-addition of images rescaled to the unit exposure time of 1\,s.



Absolute noise maps were generated from the raw images, assuming the Poisson statistics. These maps were propagated taking into account the scaling applied to each pixel during the processing (flat fielding, normalization of exposure time, etc.) and used as during the co-addition stage. Finally, the calibrated science images were resampled to a common grid and co-added using a weighted average with SWarp (\citealt{Bertin2010}). The individual weights $w_i(x,y)$ were derived from the previously computed noise maps. The final weight and rms maps for the co-added image were calculated as $W(x,y)=\sum_{i=1}^n{w_i(x,y)}$ and $rms(x,y)= {1}/{\sqrt{\sum_{i=1}^n{w_i(x,y)}}}$, respectively.

\subsubsection{Spectroscopy}
\label{sec:spectroscopic_data}

The spectroscopic data were reduced using the Spectroscopic Interactive Pipeline and Graphical Interface (SIPGI; \citealt{Gargiuloetal2022}), a dedicated tool developed for processing MODS and LUCI spectra. Initially, we applied a bad pixel mask, created from imaging flats, to each science frame and corrected for cosmic ray hits. Each frame was then independently processed for bias subtraction and flat-fielding, using a master flat built from spectroscopic flat exposures. Wavelength calibration was performed by applying the inverse dispersion solution, obtained from arc lamp spectra and stored in the master calibration file. This procedure also accounted for optical distortions across the spectral field. The typical wavelength calibration precision was $\sim$0.067\,\AA\,in the red channel and $\sim$0.052\,\AA\,in the blue channel. 

Once the calibration steps were completed, we extracted the two-dimensional spectra and subtracted the sky background. Relative flux calibration was carried out using the response function derived from observations of spectro-photometric standard stars acquired shortly after our science target. The resulting spectra, wavelength- and flux- calibrated, and sky-subtracted, were finally combined, and one-dimensional extractions were performed for each source. 

To further refine the flux calibration, we performed aperture photometry on field stars using the MODS $g$- and $r$-band imaging described in Sect.\,\ref{sec:photometric_data}. The photometric measurements, extracted with SExtractor (\citealt{BertinArnouts1996}), were calibrated against the ATLAS ALL-Sky Stellar Reference Catalog (ATLAS-REFCAT2; \citealt{Tonryetal2018}) using stars in common, thus providing an additional photometric anchor to the spectral flux scale. The reduced spectra, and the comparison with the photometric fluxes, are shown in Figs.\,\ref{fig:spectra1234}, \ref{fig:spectra5678}, \ref{fig:spectra910}.

\section{Data analysis}
\label{sec:data_analysis}
\subsection{Stellar properties}
\label{sec:stellar_parameters}

\subsubsection{Proper motion}
\label{sec:proper_motions}

\begin{figure}[h!]
\begin{center}
\includegraphics[width=8.5cm, trim=0 0 0 0]{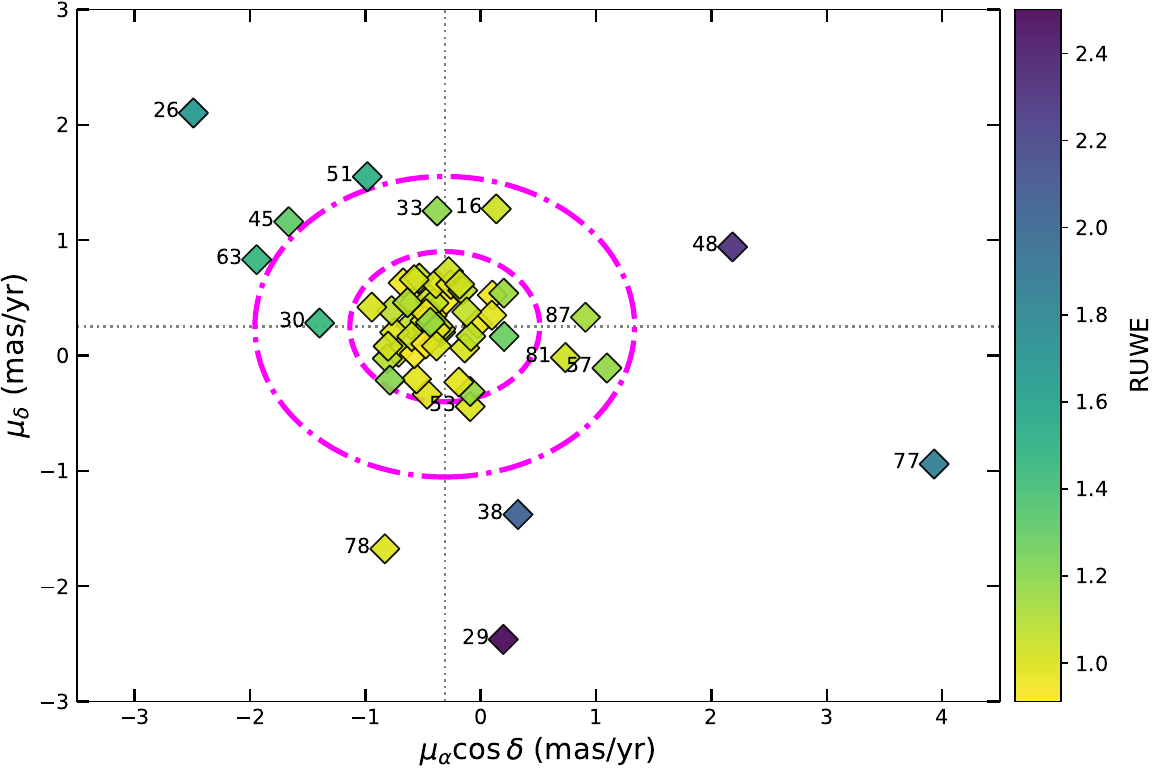}
\vspace{-0.2cm}
\caption{Proper motions vector point diagram for the observed sample. Points are color-coded according to their Gaia RUWE value showed in the colored right bar. The dotted cross-hair represents the position of the center ($<\mu_\alpha cos\delta>=-0.312\pm0.823$\,mas/yr, $<\mu_\delta>=0.249\pm0.651$\,mas/yr) where most of the targets are clustered. The dashed and dot-dashed magenta concentric ellipses denote the empirical boundaries at 1$\sigma$ and 2$\sigma$, respectively.}
\label{fig:proper_motion} 
\end{center}
\end{figure}

Figure\,\ref{fig:proper_motion} shows the vector point diagram of the Gaia DR3 proper motions of our sample ($\mu_{\alpha}\cos\delta$, $\mu_{\delta}$). We also used the renormalized unit weight error (RUWE), encoded by the color scale on the right panel, as an indicator of the astrometric solution  quality. 

The mean values of the proper motions are at $<\mu_\alpha cos\delta>=-0.312\pm0.823$\,mas/yr, $<\mu_\delta>=0.249\pm0.651$\,mas/yr, which can be tentatively considered as the cluster center. Overall, the proper-motion distribution appears to support the presence of a kinematically coherent population, with a limited number of outliers that deserve individual consideration. The majority of the targets are indeed tightly clustered around the center position and are within 1$\sigma$ from the center (see the dashed magenta ellipse), indicating broadly similar kinematics. Seven targets are within 1$\sigma$ and 2$\sigma$ (dot-dashed magenta ellipse), while nine are outside the 2$\sigma$ region. Among the nine targets, only one (namely, ID78) does not show evidence of accretion (see Sect.\,\ref{sec:accr_stmass}). Furthermore, its proper motion falls outside the cluster locus. Given its low RUWE value ($\sim$1.00), which rules out astrometric noise or unresolved binarity and confirms the reliability of the kinematic measurement, both the lack of accretion and the discrepant proper motion point toward the possibility of non-membership.

Most (60) of our previously selected targets have a RUWE $< 1.4$, which typically indicates a good single-star astrometric fit. Eight targets (ID26, 29, 30, 38, 48, 51, 63, 77) exhibit RUWE values higher than 1.4, with ID29, 38, 48 having RUWE$> 2.0$ and being located far from the main kinematic locus. These high RUWE values, associated with sources that lie away from the main proper-motion membership locus, may indicate unresolved multiplicity, astrometric perturbations, or other complexities affecting the Gaia solutions. Among these eight targets, only one (namely, ID30) does not show evidence of accretion (see Sect.\,\ref{sec:accr_stmass}).

We note, however, that the use of Gaia DR3 proper motions introduces an implicit selection effect due to the large distance of Sh2-284. This is because, at this distance, Gaia astrometry is limited to relatively bright, low-extinction sources. Consequently, our sample is primarily representative of the less embedded, intrinsically brighter YSO population. While this limitation prevents a fully unbiased census of the entire cluster, this sample perfectly matches the requirements of our optical and near-infrared spectroscopy, ensuring a self-consistent and robust characterization of the actively accreting, optically/NIR visible disk systems in the region.

\subsubsection{Stellar effective temperature}
\label{sec:teff}

We derived the effective temperature $T_{\rm eff}$ from two different spectroscopic methods, which include the use of the Na\,\textsc{i} doublet at $\sim$5890-5896\,\AA\,and the H$\beta$ line when observed in absorption, and a photometric method based on colour-$T_{\rm eff}$ relations. 

\begin{figure} 
\begin{center}
\includegraphics[width=0.45\textwidth, trim=0 0 0 0]{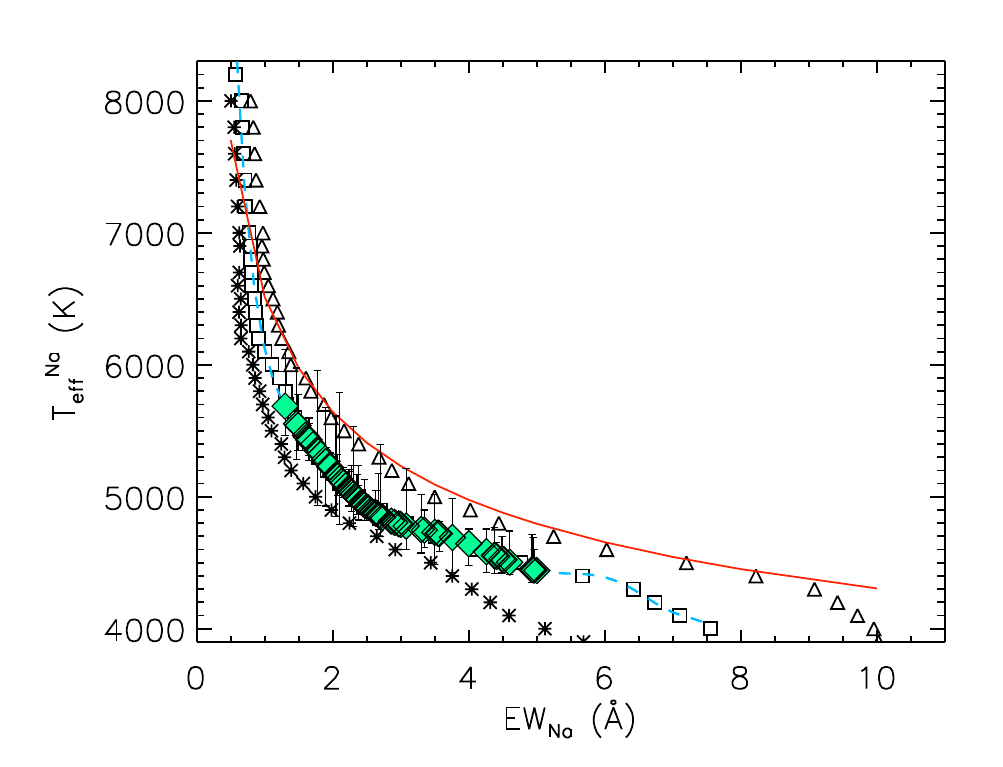}
\caption{Measured $T_{\rm eff}^{\rm Na}$ versus $EW_{\rm Na}$ of our targets (green diamonds). The triangles, squares, and asterisks refer to values obtained using BT-Settl synthetic spectra at [Fe/H]=0.0, $-0.5$, $-1.0$\,dex, respectively. Dashed line marks the least-squares fit to the values obtained at [Fe/H]=$-0.5$ and used for the $T_{\rm eff}^{\rm Na}$. Continuous line represents the relation by \cite{Tripicchioetal1997} obtained using atmospheric models at solar metallicity.}
\label{fig:EW_Na}
\end{center}
\end{figure}

In the first case, we used the resonance absorption lines of Na\,\textsc{i}, which serves as sensitive temperature indicators for the upper photosphere and low chromosphere, particularly for late-type stars (spectral types K and M; \citealt{Tripicchioetal1997}). This diagnostic method was successfully applied to 59 out of the 68 targets in our sample. We therefore considered the BT-Settl model grid of theoretical spectra (\citealt{Allardetal2012}), which we then degraded to the MODS resolution. Thirty-three targets in our sample had gravities derived from Gaia-DR3 (\citealt{GaiaDR32023}), with the peak of the distribution at $\log g = 4.7 \pm 0.2$, thus we adopted synthetic spectra with $\log g = 4.7$. Furthermore, given that the cluster metallicity is consistent with $[\text{Fe}/\text{H}] = -0.5$\,dex (see \citealt{Cusanoetal2011} and Sect.\,\ref{sec:sh2-284_intro}), we decided to adopt synthetic spectra with this metallicity value. Assuming a greater or smaller value for $\log g$ or [Fe/H] by, e.g., $\pm$0.2\,dex could introduce a shift in the derived $T_{\rm eff}$ of approximately $\mp$60\,K and $\mp$120\,K, respectively. These temperature variations propagate into the derived mass accretion rate as a minor source of uncertainty that remains well within our nominal errors (see Table\,\ref{tab:stellar_parameters}). From these synthetic spectra, and following the procedure by \cite{Tripicchioetal1997}, we measured the sodium equivalent widths (EWs) for each $T_{\rm eff}^{\rm Na}$ and thus obtained $T_{\rm eff}$-$EW_{\rm Na}$ calibration relations adopting a least-squares fitting procedure (see Fig.\,\ref{fig:EW_Na}). We then measured equivalent widths of the Na\,\textsc{ii} doublet lines of our targets, which are blended at the MODS resolution and then computed the $T_{\rm eff}$ from our calibration (Fig.\,\ref{fig:EW_Na}).  

Conversely, the absorption wings of Balmer lines are optimized for earlier-type stars (spectral types G and early K), provided that the profile core is not completely filled-in or contaminated by strong accretion emission. In particular, for the case of YSOs with H$\beta$ in absorption (19/68 targets), the wings of the line profile are sensitive to temperature, depending only weakly on metallicity and gravity (\citealt{Barklemetal2002}). In particular, the spectral region extending from 3 to 5 \AA\, around the H$\beta$ line center provides a good diagnostic of the effective temperature (\citealt{Barklem2008}). We therefore adopted this region as a temperature indicator for stars showing the line in absorption or partially filled-in. For each star, we compared the H$\beta$ line profile outside the core in our real spectrum against the BT-Settl synthetic spectra considered above. To reduce subjectiveness in the analysis, we quantified the agreement between the observed and synthetic profiles by minimizing the root-mean-square (rms) of their residuals. The procedure and the associated uncertainties are described in 
detail in \citet[see also \citealt{Pasquinietal2008}]{Barklemetal2002}.

We also derived a photometric \teff\ using several colour combinations based on Gaia DR3 $G$, $BP$, and $RP$ magnitudes, together with 2MASS $J$, $H$, and $K_{\rm s}$ photometry. We adopted the same mean metallicity and surface gravity as above, and corrected the photometry for extinction using the $A_V$ derived by \cite{Guarcelloetal2021} and the extinction law adopted by the same authors (i.e. \citealt{Cardellietal1989, ODonnell1994}). The photometric effective temperatures $T_{\rm eff}^{\rm phot}$ were computed with the {\sc colte}\footnote{https://github.com/casaluca/colte}
 code, which implements the colour-\teff\ calibrations described in \cite{Casagrandeetal2021}. The final photometric effective temperature was obtained as the weighted mean of the individual \teff\ estimates derived from the different colour indices. We could apply this method to 63 targets.

To quantitatively evaluate the consistency among these separate 
temperature indicators, we performed a statistical cross-comparison across the 
sample. The comparison between $T_{\rm eff}^{\rm Na}$ and $T_{\rm eff}^{\rm phot}$ ($N=56$) reveals a median systematic offset of $\Delta T_{\rm med} \sim +60$\,K with a scatter (Median Absolute Deviation, MAD) of $\sim$300\,K, confirming 
a good overall alignment of our baseline temperature scale. Conversely, for the 
19 targets where the H$\beta$ absorption profiles were accessible, $T_{\rm eff}^{\rm H\beta}$ is found to be systematically warmer than both $T_{\rm eff}^{\rm Na}$ and and $T_{\rm colte}$ ($\Delta T_{\rm med} \sim -800$\,K; \text{MAD} = 700\,K). This pronounced systematic overestimation is physically consistent with the presence of heated stellar regions or localized hotspots, thus shifting the derived temperature solutions toward higher values.

Since H$\beta$ and Na\,\textsc{i} absorption lines probe different regions of the stellar photosphere (\citealt{Gray2005}), and noting that H$\beta$ is present in absorption or partially filled-in in the core across 19 targets (all accretor candidates with the exception of ID16 and ID37; see Sect.\,\ref{sec:Maccr}), we adopted as \teff\ the average of the corresponding estimates, when available (see Table\,\ref{tab:stellar_parameters}). Although the $S/N$ ratio and spectral resolution of the data are not optimal for a precise determination of individual stellar parameters, this averaging approach does not introduce significant systematic flaws into our derivation of the accretion properties. Indeed, the uncertainty in \teff\ due to the averaging procedure propagates into a variation of only $\sim$0.1\,dex in the final mass accretion rate ($\log \dot{M}_{\rm acc}$), which is well within our nominal statistical error bounds (see Sect.\,\ref{sec:Maccr}).

\begin{deluxetable}{llrcccrcccrr}
\label{tab:stellar_parameters}
\tabletypesize{\tiny}
\tablecaption{Coordinates, stellar parameters, and accretion properties for the target sample.}
\tablewidth{0pt}
\setlength{\tabcolsep}{2pt}
\tablehead{
\colhead{ID} & \colhead{RA} & \colhead{DEC} & \colhead{Gaia Name} & \colhead{2MASS Name} & \colhead{$T_{\rm eff}$} & \colhead{$L_{\star}$} & \colhead{$M_{\star}$} & \colhead{$\log R_{\star}$} & \colhead{$\log(\text{Age})$} & \colhead{$L_{\rm acc}$} & \colhead{$\log \dot M_{\rm acc}$} \\
\colhead{} & \colhead{(deg)} & \colhead{(deg)} & \colhead{} & \colhead{} & \colhead{(K)} & \colhead{($L_{\odot}$)} & \colhead{($M_{\odot}$)} & \colhead{($R_{\odot}$)} & \colhead{(yr)} & \colhead{($L_{\odot}$)} & \colhead{($M_{\odot}$/yr)}
}
\startdata
3 & 101.2784 & 0.2315 & 33125505670478609920 & 06450681+0013535 & 5690$\pm$650 (N,H,p) & 28.95$\pm$3.57 & 2.95$\pm$0.75 & 0.75 & 5.77 & 0.16 $\pm$ 0.22 & $-$6.97 $\pm$ 0.32 \\
4 & 101.3243 & 0.0736 & 33125498862949890048 & 06451782+0004249 & 5310$\pm$150 (N,p) & 17.59$\pm$2.56 & 2.52$\pm$0.64 & 0.72 & 5.38 & 0.06 $\pm$ 0.22 & $-$7.05 $\pm$ 0.26 \\
5 & 101.2432 & 0.2375 & 33125506456454197120 & 06445837+0014151 & 5235$\pm$150 (N,p) & 12.95$\pm$1.84 & 2.12$\pm$0.54 & 0.64 & 5.59 & $-$0.39 $\pm$ 0.22 & $-$7.48 $\pm$ 0.27 \\
6 & 100.9845 & 0.5162 & 33125572878126840192 & 06435629+0030581 & 5720$\pm$740 (N,H,p) & 7.67$\pm$1.01 & 1.91$\pm$0.48 & 0.52 & 6.20 & $-$0.52 $\pm$ 0.23 & $-$7.76 $\pm$ 0.35 \\
7 & 101.2587 & 0.3290 & 33125524735835124224 & 06450208+0019443 & 5130$\pm$250 (N,p) & 12.19$\pm$1.70 & 1.77$\pm$0.45 & 0.63 & 5.51 & $-$0.35 $\pm$ 0.22 & $-$7.35 $\pm$ 0.28 \\
8 & 101.0757 & 0.3149 & 33125522334951775232 & 06441818+0018537 & 6125$\pm$830 (N,H,p) & 10.48$\pm$1.52 & 1.79$\pm$0.45 & 0.47 & 6.52 & $-$0.20 $\pm$ 0.22 & $-$7.39 $\pm$ 0.35 \\
9 & 101.1534 & 0.2718 & 33125518349222110080 & 06443682+0016186 & 5390$\pm$920 (N,H,p) & 9.18$\pm$0.95 & 2.21$\pm$0.60 & 0.51 & 5.95 & $-$0.85 $\pm$ 0.23 & $-$8.05 $\pm$ 0.39 \\
10 & 101.3510 & 0.3796 & 33125514977671490944 & 06452424+0022448 & 5110$\pm$95 (N,H,p) & 16.27$\pm$2.31 & 1.83$\pm$0.46 & 0.72 & 5.34 & $-$0.37 $\pm$ 0.22 & $-$7.32 $\pm$ 0.27 \\
11 & 101.6066 & $-$0.0402 & 33113473542937253376 & 06462558$-$0002247 & 5400$\pm$400 (N,H,p) & 18.14$\pm$2.56 & 2.70$\pm$0.68 & 0.70 & 5.61 & $-$0.57 $\pm$ 0.23 & $-$7.71 $\pm$ 0.30 \\
12 & 101.2226 & 0.1297 & 33125503123559486720 & 06445343+0007469 & 6030$\pm$980 (N,H,p) & 11.16$\pm$1.52 & 1.90$\pm$0.48 & 0.50 & 6.34 & $-$0.87 $\pm$ 0.23 & $-$8.06 $\pm$ 0.38 \\
13 & 101.0685 & 0.3233 & 33125522433732647040 & 06441644+0019239 & 5850$\pm$490 (N,H,p) & 6.03$\pm$0.35 & 1.65$\pm$0.42 & 0.38 & 6.60 & $-$0.58 $\pm$ 0.23 & $-$7.82 $\pm$ 0.31 \\
14 & 101.1594 & 0.3230 & 33125519684953577088 & 06443827+0019229 & 6075$\pm$860 (N,H,p) & 8.26$\pm$1.17 & 1.69$\pm$0.43 & 0.39 & 6.66 & $-$0.32 $\pm$ 0.22 & $-$7.53 $\pm$ 0.36 \\
15 & 101.1806 & 0.2351 & 33125506220234475008 & 06444335+0014062 & 5185$\pm$50 (N,H,p) & 6.91$\pm$0.97 & 1.71$\pm$0.43 & 0.52 & 5.76 & $-$0.33 $\pm$ 0.22 & $-$7.45 $\pm$ 0.26 \\
16 & 101.3407 & 0.0777 & 33125500340418641792 & 06452177+0004398 & 5320$\pm$70 (N,H,p) & 3.74$\pm$0.35 & 1.61$\pm$0.41 & 0.35 & 6.30 & ... & ... \\
18 & 101.4737 & 0.0630 & 33113489928234237824 & 06455369+0003468 & 5204$\pm$150 (N,H,p) & 9.38$\pm$1.34 & 1.85$\pm$0.47 & 0.58 & 5.67 & $-$0.33 $\pm$ 0.22 & $-$7.42 $\pm$ 0.27 \\
19 & 101.2959 & 0.3447 & 33125525079430509312 & 06451100+0020407 & 5810$\pm$380 (N,p) & 12.28$\pm$1.25 & 2.24$\pm$0.57 & 0.52 & 6.36 & $-$0.74 $\pm$ 0.23 & $-$7.96 $\pm$ 0.30 \\
20 & 101.3303 & 0.3642 & 33125513504498867712 & 06451927+0021510 & 5880$\pm$480 (N,H,p) & 8.90$\pm$0.86 & 1.88$\pm$0.48 & 0.50 & 6.40 & $-$1.29 $\pm$ 0.24 & $-$8.50 $\pm$ 0.32 \\
21 & 101.3820 & 0.3254 & 33125512400687025664 & 06453168+0019316 & 5050$\pm$75 (N,H,p) & 4.98$\pm$0.63 & 1.40$\pm$0.35 & 0.46 & 5.84 & $-$1.76 $\pm$ 0.26 & $-$8.85 $\pm$ 0.29 \\
22 & 101.4778 & 0.0370 & 33113486874515756928 & 06455467+0002132 & 5120$\pm$40 (N,H,p) & 5.67$\pm$0.80 & 1.55$\pm$0.39 & 0.48 & 5.89 & $-$1.28 $\pm$ 0.24 & $-$8.39 $\pm$ 0.28 \\
23 & 101.6022 & 0.2643 & 33113500133077230720 & 06462452+0015514 & 4630$\pm$170 (N,p) & 9.41$\pm$1.16 & 0.88$\pm$0.22 & 0.67 & 5.08 & $-$0.91 $\pm$ 0.24 & $-$7.58 $\pm$ 0.29 \\
24 & 101.5612 & 0.4935 & 33125541022349339520 & 06461468+0029364 & 5260$\pm$150 (p) & 3.92$\pm$0.51 & 1.61$\pm$0.41 & 0.34 & 6.45 & ... & ... \\
25 & 101.3089 & 0.1935 & 33125502367643279744 & 06451412+0011364 & 5180$\pm$190 (N,H,p) & 8.64$\pm$0.84 & 1.78$\pm$0.45 & 0.56 & 5.73 & $-$0.64 $\pm$ 0.32 & $-$7.73 $\pm$ 0.37 \\
26 & 101.1372 & 0.1813 & 33125504467889008512 & ... & 4560$\pm$420 (N,p) & 15.57$\pm$1.01 & 0.85$\pm$0.22 & 0.81 & 4.48 & 0.54 $\pm$ 0.37 & $-$5.98 $\pm$ 0.46 \\
27 & 101.5428 & 0.4090 & 33125537899913487360 & 06461026+0024323 & 5030$\pm$240 (N,p) & 6.27$\pm$0.84 & 1.40$\pm$0.35 & 0.52 & 5.69 & $-$1.41 $\pm$ 0.25 & $-$8.44 $\pm$ 0.31 \\
28 & 101.1875 & 0.2277 & 33125506215936019072 & ... & 4710$\pm$50 (N,p) & 5.26$\pm$0.57 & 0.90$\pm$0.23 & 0.54 & 5.43 & $-$0.14 $\pm$ 0.29 & $-$6.96 $\pm$ 0.32 \\
29 & 101.3761 & 0.1167 & 33125500512217351936 & 06453025+0006598 & 4720$\pm$70 (N,p) & 5.04$\pm$0.54 & 0.93$\pm$0.24 & 0.52 & 5.43 & $-$0.88 $\pm$ 0.23 & $-$7.73 $\pm$ 0.27 \\
30 & 101.1317 & 0.4045 & 33125529271320710400 & 06443160+0024160 & 5220$\pm$170 (N,p) & 2.58$\pm$0.08 & 1.44$\pm$0.36 & 0.30 & 6.43 & ... & ... \\
31 & 101.1391 & 0.3985 & 33125529202603169408 & 06443339+0023545 & 4060$\pm$380 (N,p) & 3.76$\pm$0.50 & 0.42$\pm$0.11 & 0.60 & 4.60 & $-$0.73 $\pm$ 0.23 & $-$7.16 $\pm$ 0.33 \\
32 & 101.3144 & 0.3628 & 33125525216869490816 & 06451545+0021462 & 4475$\pm$280 (p) & 5.08$\pm$0.45 & 0.68$\pm$0.17 & 0.60 & 5.00 & $-$0.47 $\pm$ 0.23 & $-$7.13 $\pm$ 0.30 \\
33 & 101.3148 & 0.4665 & 33125528274887496448 & 06451556+0027592 & 3860$\pm$100 (p) & 9.45$\pm$0.39 & 0.35$\pm$0.09 & 0.70 & 4.48 & $-$0.54 $\pm$ 0.23 & $-$6.64 $\pm$ 0.27 \\
34 & 101.3420 & $-$0.0603 & 331134737076374680192 & 06452208$-$0003370 & 3870$\pm$90 (p) & 2.56$\pm$0.18 & 0.35$\pm$0.09 & 0.52 & 4.85 & 0.12 $\pm$ 0.37 & $-$6.28 $\pm$ 0.41 \\
36 & 101.3129 & 0.3681 & 33125525221169655552 & 06451509+0022050 & 5160$\pm$230 (p) & 17.47$\pm$2.25 & 2.05$\pm$0.52 & 0.72 & 5.38 & $-$0.82 $\pm$ 0.23 & $-$7.81 $\pm$ 0.29 \\
37 & 101.5882 & 0.3477 & 33125533806804059776 & 06462117+0020515 & 5310$\pm$215 (N,H,p) & 3.48$\pm$0.47 & 1.58$\pm$0.40 & 0.35 & 6.40 & ... & ... \\
38 & 101.3279 & 0.0696 & 33125498867250807552 & 06451868+0004104 & 4990$\pm$50 (N,p) & 4.92$\pm$0.33 & 1.31$\pm$0.33 & 0.47 & 5.78 & $-$1.42 $\pm$ 0.25 & $-$8.46 $\pm$ 0.27 \\
40 & 101.3813 & 0.2974 & 33125511988370132352 & 06453152+0017505 & 4780$\pm$285 (p) & 4.74$\pm$0.62 & 0.99$\pm$0.25 & 0.50 & 5.57 & ... & ... \\
42 & 101.1964 & 0.2255 & 33125506147218155520 & ... & 4840$\pm$120 (N,p) & 7.90$\pm$1.07 & 1.10$\pm$0.28 & 0.62 & 5.30 & $-$0.38 $\pm$ 0.30 & $-$7.22 $\pm$ 0.34 \\
44 & 101.3837 & 0.2395 & 33125508449317013888 & 06453208+0014220 & 5810$\pm$345 (N,H,p) & 2.47$\pm$0.33 & 1.23$\pm$0.31 & 0.20 & 6.90 & $-$2.34 $\pm$ 0.27 & $-$9.64 $\pm$ 0.34 \\
45 & 101.3945 & 0.3316 & 33125513878155782272 & 06453468+0019537 & 4220$\pm$320 (N,p) & 3.34$\pm$0.26 & 0.51$\pm$0.13 & 0.55 & 5.18 & $-$1.70 $\pm$ 0.25 & $-$8.27 $\pm$ 0.34 \\
46 & 101.1715 & 0.3364 & 33125519788032808192 & 06444116+0020108 & 4220$\pm$300 (N,p) & 2.90$\pm$0.23 & 0.51$\pm$0.13 & 0.50 & 5.23 & $-$1.23 $\pm$ 0.24 & $-$7.83 $\pm$ 0.32 \\
48 & 101.1871 & 0.3173 & 33125518963401038336 & 06444489+0019022 & 5085$\pm$220 (N,H,p) & 2.52$\pm$0.32 & 1.37$\pm$0.35 & 0.30 & 6.36 & $-$1.36 $\pm$ 0.25 & $-$8.59 $\pm$ 0.30 \\
51 & 101.4049 & 0.3805 & 33125514874588269056 & 06453718+0022497 & 4750$\pm$110 (N,p) & 4.96$\pm$0.63 & 0.96$\pm$0.24 & 0.51 & 5.54 & $-$1.07 $\pm$ 0.24 & $-$7.93 $\pm$ 0.28 \\
52 & 101.4740 & 0.0510 & 33113489825154973696 & 06455375+0003038 & 5215$\pm$50 (N,p) & 3.61$\pm$0.48 & 1.57$\pm$0.40 & 0.34 & 6.34 & $-$0.51 $\pm$ 0.22 & $-$7.74 $\pm$ 0.25 \\
53 & 101.2597 & 0.2155 & 33125505258161769984 & 06450233+0012558 & 4725$\pm$150 (N,p) & 3.65$\pm$0.47 & 0.92$\pm$0.23 & 0.44 & 5.60 & $-$0.75 $\pm$ 0.32 & $-$7.66 $\pm$ 0.37 \\
54 & 101.1335 & 0.1774 & 33125504463589316736 & 06443204+0010385 & 5580$\pm$100 (N,p) & 9.50$\pm$0.96 & 2.23$\pm$0.56 & 0.50 & 6.30 & 0.66 $\pm$ 0.36 & $-$6.57 $\pm$ 0.40 \\
57 & 101.2700 & 0.3663 & 33125525354310068224 & 06450479+0021585 & 4560$\pm$55 (N,p) & 2.14$\pm$0.04 & 0.78$\pm$0.20 & 0.37 & 5.80 & $-$0.92 $\pm$ 0.24 & $-$7.84 $\pm$ 0.26 \\
58 & 101.5840 & $-$0.0306 & 33113473710437319296 & 06462017$-$0001502 & 4320$\pm$420 (N,p) & 2.99$\pm$0.30 & 0.57$\pm$0.14 & 0.50 & 5.41 & $-$1.39 $\pm$ 0.25 & $-$8.06 $\pm$ 0.35 \\
60 & 101.2531 & 0.2265 & 33125506387734708736 & 06450075+0013356 & 4430$\pm$210 (N,p) & 2.81$\pm$0.00 & 0.65$\pm$0.17 & 0.46 & 5.45 & $-$0.92 $\pm$ 0.23 & $-$7.68 $\pm$ 0.28 \\
61 & 101.1945 & 0.3175 & 33125518963401135872 & 06444666+0019029 & 4960$\pm$300 (N,p) & 4.19$\pm$0.47 & 1.25$\pm$0.32 & 0.45 & 5.79 & $-$0.07 $\pm$ 0.22 & $-$7.12 $\pm$ 0.29 \\
62 & 101.5926 & 0.2550 & 33113500029999219968 & 06462223+0015180 & 5310$\pm$240 (N,p) & 3.20$\pm$0.34 & 1.54$\pm$0.39 & 0.35 & 6.40 & $-$0.45 $\pm$ 0.29 & $-$7.72 $\pm$ 0.35 \\
63 & 101.1644 & 0.2184 & 33125504704107519104 & 06443944+0013062 & 5075$\pm$385 (N,p) & 6.10$\pm$0.73 & 1.48$\pm$0.37 & 0.51 & 5.71 & 0.02 $\pm$ 0.22 & $-$7.05 $\pm$ 0.30 \\
64 & 101.1661 & 0.3398 & 33125519994191242624 & 06443986+0020230 & 4650$\pm$200 (N,p) & 5.03$\pm$0.54 & 0.85$\pm$0.22 & 0.55 & 5.34 & $-$0.13 $\pm$ 0.21 & $-$6.92 $\pm$ 0.27 \\
65 & 101.3835 & 0.1056 & 33113490894605152768 & ... & 4305$\pm$220 (N,p) & 6.18$\pm$0.23 & 0.56$\pm$0.14 & 0.63 & 4.78 & $-$0.08 $\pm$ 0.34 & $-$6.58 $\pm$ 0.40 \\
66 & 101.4027 & 0.3694 & 33125514771510635392 & 06453664+0022095 & 4540$\pm$95 (N,p) & 3.18$\pm$0.28 & 0.75$\pm$0.20 & 0.45 & 5.58 & $-$1.21 $\pm$ 0.24 & $-$8.02 $\pm$ 0.28 \\
67 & 101.6116 & $-$0.0623 & 33113473366839791744 & 06462677$-$0003441 & 4820$\pm$265 (N,p) & 5.13$\pm$0.61 & 1.05$\pm$0.27 & 0.51 & 5.53 & $-$0.77 $\pm$ 0.23 & $-$7.68 $\pm$ 0.30 \\
69 & 101.5898 & $-$0.0568 & 33113473405493573632 & 06462155$-$0003246 & 4420$\pm$355 (N,p) & 2.62$\pm$0.15 & 0.65$\pm$0.17 & 0.45 & 5.48 & $-$1.05 $\pm$ 0.24 & $-$7.83 $\pm$ 0.32 \\
70 & 101.3276 & 0.1088 & 33125500653957195264 & 06451862+0006317 & 5150$\pm$295 (N,p) & 4.80$\pm$0.56 & 1.58$\pm$0.40 & 0.43 & 6.00 & $-$0.14 $\pm$ 0.22 & $-$7.30 $\pm$ 0.28 \\
71 & 101.2445 & 0.2949 & 33125518654161386112 & 06445867+0017415 & 6755$\pm$145 (p) & 21.37$\pm$1.13 & 1.98$\pm$0.50 & 0.54 & 6.46 & ... & ... \\
72 & 101.5941 & $-$0.0616 & 33113473401199677696 & 06462257$-$0003418 & 4870$\pm$260 (N,p) & 2.54$\pm$0.21 & 1.13$\pm$0.29 & 0.35 & 6.00 & $-$0.40 $\pm$ 0.31 & $-$7.51 $\pm$ 0.37 \\
75 & 101.1795 & 0.3438 & 33125519822394595712 & 06444305+0020375 & 4448$\pm$180 (N) & 1.52$\pm$0.18 & 0.72$\pm$0.18 & 0.30 & 5.78 & $-$0.57 $\pm$ 0.28 & $-$7.51 $\pm$ 0.34 \\
76 & 101.3909 & 0.3522 & 33125514668429797504 & 06453380+0021078 & 4550$\pm$170 (N) & 1.82$\pm$0.22 & 0.79$\pm$0.20 & 0.33 & 5.89 & $-$0.42 $\pm$ 0.42 & $-$7.39 $\pm$ 0.47 \\
77 & 101.3103 & 0.4051 & 33125527003579931776 & 06451448+0024185 & 5080$\pm$260 (N,p) & 4.98$\pm$0.61 & 1.45$\pm$0.37 & 0.47 & 5.84 & 0.43 $\pm$ 0.20 & $-$6.67 $\pm$ 0.26 \\
78 & 101.6071 & $-$0.0459 & 33113473435558614016 & 06462569$-$0002451 & 6035$\pm$65 (p) & 2.83$\pm$0.07 & 1.20$\pm$0.30 & 0.19 & 6.97 & ... & ... \\
80 & 101.2742 & 0.2398 & 33125505670475460480 & 06450581+0014231 & 5095$\pm$340 (N,p) & 5.21$\pm$0.62 & 1.49$\pm$0.38 & 0.46 & 5.85 & 0.05 $\pm$ 0.21 & $-$7.05 $\pm$ 0.28 \\
81 & 101.1533 & 0.3279 & 33125519689248742912 & ... & 4950$\pm$220 (N,p) & 3.16$\pm$0.38 & 1.24$\pm$0.31 & 0.39 & 5.96 & $-$0.52 $\pm$ 0.28 & $-$7.63 $\pm$ 0.34 \\
83 & 101.2464 & 0.2198 & 33125505258161774848 & 06445913+0013115 & 4395$\pm$185 (N,p) & 1.56$\pm$0.16 & 0.68$\pm$0.17 & 0.40 & 5.78 & $-$1.08 $\pm$ 0.24 & $-$7.98 $\pm$ 0.29 \\
85 & 101.2588 & 0.3536 & 33125524976353328256 & 06450212+0021127 & 4740$\pm$575 (p) & 1.46$\pm$0.18 & 1.02$\pm$0.26 & 0.27 & 6.20 & $-$1.13 $\pm$ 0.24 & $-$8.28 $\pm$ 0.36 \\
86 & 101.2967 & 0.3647 & 33125525182510973568 & ... & 4755$\pm$595 (N) & 4.40$\pm$0.38 & 0.96$\pm$0.24 & 0.49 & 5.60 & ... & ... \\
87 & 101.3754 & 0.2603 & 33125511507337903744 & 06453011+0015369 & 5055$\pm$50 (N,p) & 2.39$\pm$0.27 & 1.34$\pm$0.34 & 0.30 & 6.30 & $-$0.85 $\pm$ 0.22 & $-$8.08 $\pm$ 0.26 \\
\enddata
\end{deluxetable}

\subsubsection{Elemental abundances}
\label{sec:abundances}
 {\it \underline{Iron.}} We attempted to estimate the iron abundance ([Fe/H]) for the three targets showing purely photospheric spectra (see Sect.~\ref{sec:emission_lines} and Table\,\ref{tab:line_fluxes_id_ordered}), namely ID16, ID30, and ID37. Since these specific sources exhibit purely photospheric profiles devoid of active accretion signatures, any contribution from continuum veiling, which may affect the depth of the photospheric absorption line profiles, can be safely neglected (see \citealt{Biazzoetal2011}). For these objects, the signal-to-noise ratio of the spectra is $\sim30–50$ at $\sim600$\,nm, which is not optimal for a robust abundance determination, particularly in the very metal-poor regime and for spectra acquired at relatively moderate spectral resolution. We therefore employed a spectral synthesis approach using the {\sc iSpec} tool (\citealt{blancocuaresma2014}, \citealt{blancocuaresma2019}), fixing the stellar parameters ($T_{\rm eff}$ and $\log g$) to the values derived above. We adopted the sixth version of the Gaia-ESO Survey atomic line list (\citealt{Heiteretal2021}), the Spectroscopy Make Easy radiative transfer code (\citealt{ValentiPiskunov1996}; version 4.23), and the MARCS grid of model atmospheres (\citealt{Gustafssonetal2008}). We then employed the non-linear least-squares Levenberg-Marquardt fitting algorithm (\citealt{Markwardt2009}) to iteratively minimise the $\chi^2$ value between the synthetic and observed spectra. Considering the 470–700\,nm spectral range, we derived the following values: [Fe/H]$_{\rm ID16} = -0.6 \pm 0.4$\,dex, [Fe/H]$_{\rm ID30} = -0.7 \pm 0.5$\,dex, and [Fe/H]$_{\rm ID37} = -0.7 \pm 0.5$\,dex, where the formal error budget accounts for both internal sources of uncertainty (e.g., continuum placement, stellar parameters) and external systematic factors (see \citealt{DOrazietal2011} for details). We explicitly note that this chemical analysis is intended as a tentative exploration rather than a definitive characterization, as our data are not primarily optimized for high-precision abundance work. Our estimates should be therefore regarded as strictly indicative and must be interpreted with caution. However, to the best of our knowledge, this work represents the first attempt to constrain the iron abundance in solar-type stars of this metal-poor cluster. Although affected by significant uncertainties, our measurements are broadly consistent with recent determinations reported in the literature for two massive main sequence stars in the Dolidze\,25 cluster (\citealt{Ashrafetal2026}). 

\begin{figure*}[!t]
\begin{center}
\includegraphics[width=0.95\linewidth]{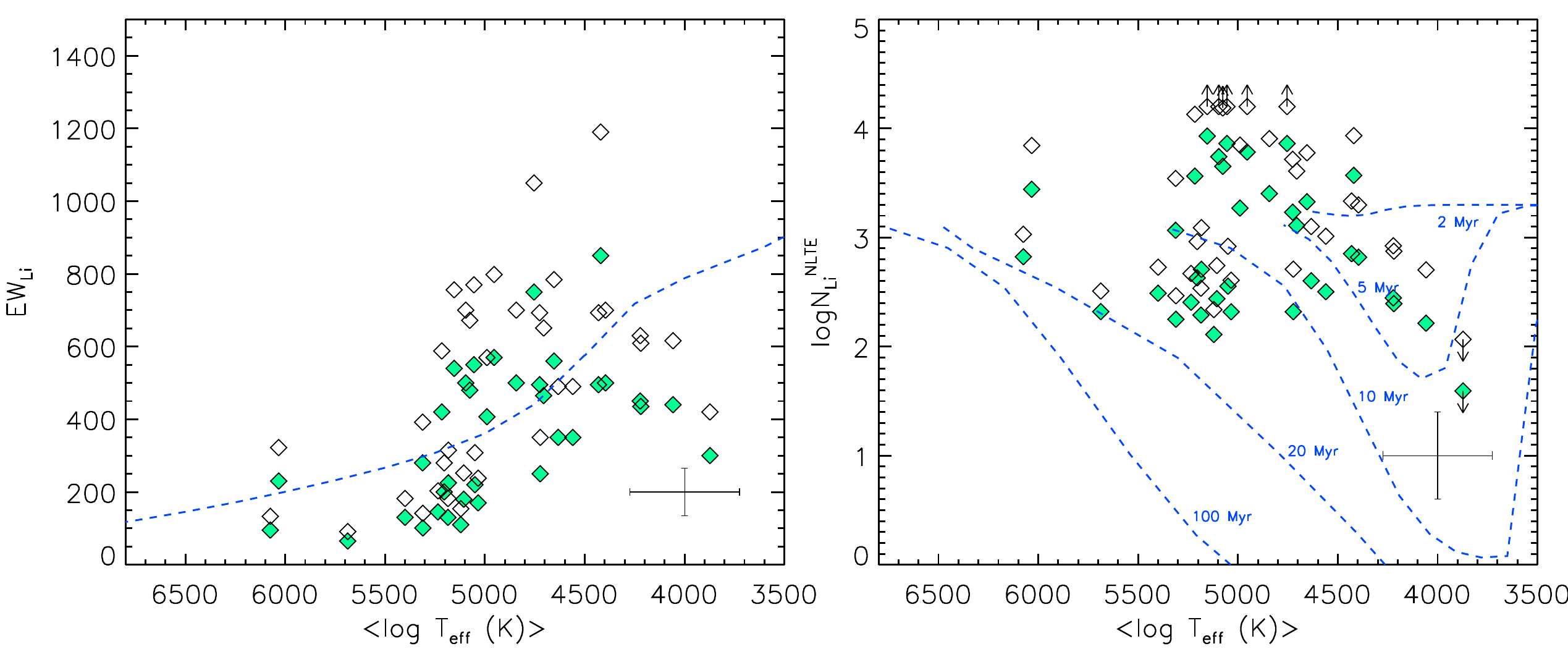}
\vspace{0cm}
\caption{Lithium equivalent width ({\it left panel}) and Li abundance ({\it right panel}) as a function of the mean effective temperature. Filled symbols denote measured equivalent widths (EWs), while open symbols indicate EWs corrected for spectral veiling. In the left panel, the dashed line shows the curve of growth (COG) for FGKM stars with lithium abundance of 3.3\,dex from \cite{Franciosinietal2022}. In the right panel, lithium isochrones from \cite{Baraffeetal2015}, spanning ages from 2 to 100 Myr, are overplotted as dashed lines. Arrows mark lower limits. Mean uncertainties are reported in the lower-right corner of each panel.
}
\label{fig:lithium}
\end{center}
\end{figure*}

{\it \underline{Lithium.}} We also attempted to estimate the lithium abundance from the Li {\sc i} line at $\sim 6707.8$\,\AA. Lithium is a fragile element, destroyed at relatively low temperatures, and is therefore widely used as an age indicator of low-mass stars (e.g., \citealt{Bildstenetal1997}, \citealt{Jeffriesetal2023}). We detected the lithium line in 36 targets. In the remaining objects, no lithium feature could be reliably identified. This is likely the result of a combination of observational and physical effects, including the relatively low signal-to-noise ratio of some spectra ($\lesssim$25--30), the moderate spectral resolution, the low metallicity, and possible continuum veiling associated with ongoing accretion. The latter compromises line identification by adding an excess continuum emission from accretion shocks that partially fills-in photospheric absorption features, thereby reducing their observed EWs and preventing reliable measurements. Consequently, we provide approximate veiling estimates around $\sim 6700$\,\AA, following the procedure described in \cite{Biazzoetal2011} and using both the calcium line Ca\,\textsc{i} at $\sim 6717.7$\,\AA, which is nearby the lithium line. We could estimate the veiling for the targets using the relation $r_{\rm \lambda}=\frac{EW_{\rm synt}}{EW_{\rm obs}}-1$, where $EW_{\rm obs}$ and $EW_{\rm synt}$ are the Ca equivalent widths measured in the observed and synthetic spectra (see Sect.\,\ref{sec:teff}). The synthetic templates natively account for the blending with the adjacent Fe\,\textsc{i} line at 6707.4\,\AA, meaning no additional correction was required. The final veiling estimates, together with the non detections, exhibit no correlation with either the stellar parameters derived above or the accretion properties discussed in Sect.~\ref{sec:Maccr}; however, this lack of trend may simply stem from our sample size and uncertainties in the measurements. The sample-averaged mean veiling factor is $\sim$0.4--0.5, which is comparable to the typical values found in young stars in nearby SFRs (\citealt{Carinietal2026}), although we caution that this estimate remains uncertain. The left panel of Fig.~\ref{fig:lithium} displays the $EW_{\rm Li}$ versus $T_{\rm eff}$ diagram before (filled symbols) and after (empty symbols) applying the veiling correction. The targets lie close to, or slightly above, the curves of growth corresponding to meteoritic lithium abundances. We also utilized the {\sc eagles} code \citep{Jeffriesetal2023}\footnote{\url{https://github.com/robdjeff/eagles}, which is an empirical model of age-dependent photospheric lithium depletion calibrated using a large, homogeneously analyzed sample of stars. This tool provides age estimates and posterior probability distributions specifically for pre-main sequence stars in a temperature range of $3000 < T_{\rm eff}\,({\rm K}) < 6500$, and is primarily optimized for solar-metallicity targets. By inputting our measured EWs and $T_{\rm eff}$ values into the code, we tentatively derived indicative upper limits of the cluster age of $\sim$6\,Myr and $\sim$4\,Myr before and after the veiling correction, respectively. This shift underscores the results of \cite{Carinietal2026} and reinforces the critical dependence of lithium-based age estimates on proper veiling treatments.} The right panel of the same figure shows the corresponding lithium abundance corrected for NLTE effects by \cite{Lindetal2009} and derived using the stellar parameters from previous sections. Overall, the inferred abundances are compatible with very young evolutionary stages, although significant uncertainties remain. 

\subsubsection{Distance, mass and luminosity}
\label{sec:mass_lum}

Based on the various determination of the effective temperature, we derived the stellar luminosity $L_{\star}$ using the $J$-band photometry\footnote{We note that while the $J$ band is a standard baseline choice to mitigate circumstellar continuum excess, highly accreting YSOs can still exhibit a residual NIR excess that might systematically overestimate $L_{\star}$. To evaluate the impact of this effect, we performed a consistency check by independently calculating the stellar luminosities using the optical Gaia DR3 broadband $G$ photometry. The two independent methods yield remarkably consistent results, agreeing in average within $0.01~\log L_{\odot}$ across the analyzed sample.}, under the assumption that the emission in this band is only minimally affected by accretion hotspots and circumstellar disk emission (see, e.g., \citealt{Gianninietal2026}). The bolometric magnitude was computed using the formula $M_{\rm bol}=J - A_J + 5 - 5\log d + BC_J$, where $J$ is the 2MASS $J$ magnitude, $A_J$ is the extinction in the $J$ band derived by \cite{Guarcelloetal2021}, $d$ is the cluster distance in parsec, and $BC_J$ is the bolometric correction in the $J$ band. For $BC_J$, we adopted the values tabulated by \cite{PecautMamajek2013}. 

\begin{figure} 
\begin{center}
\includegraphics[width=0.95\linewidth, trim=0 0 0 0, clip]{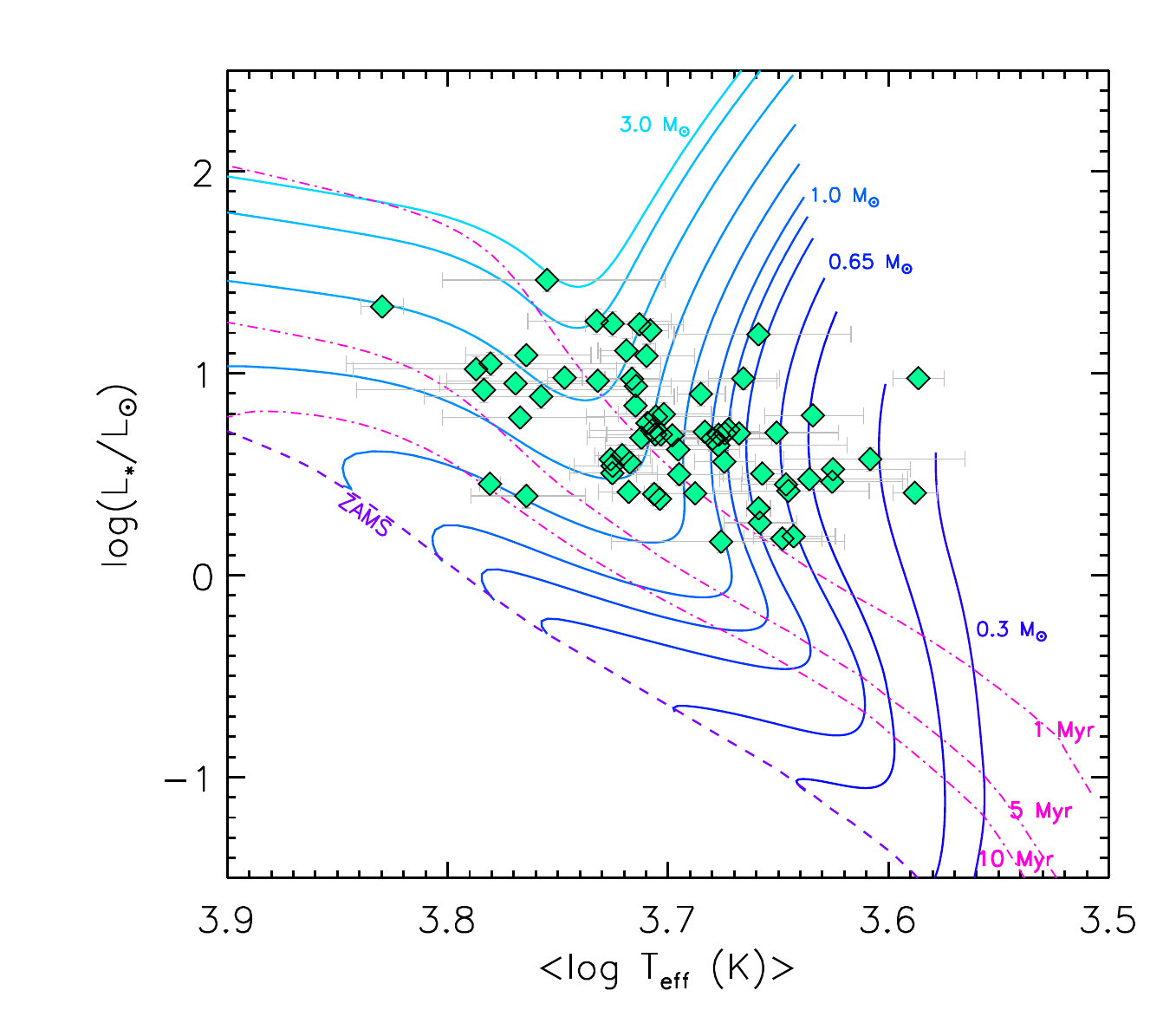}
\vspace{0cm}
\caption{Location of the YSOs candidates in the HR diagram. Superimposed are the MIST evolutionary tracks (colored solid lines) and theoretical isochrones (dot-dashed magenta lines) of \cite{Choietal2016} for metallicity $Z \sim 0.005$. The masses in $M_\odot$ and ages in Myr are indicated next to each track and isochrone, respectively. The position of the ZAMS is marked with a dashed violet line.}
\label{fig:HRdiagram}
\end{center}
\end{figure}

As for the distance, we adopted a value of $4.7 \pm 0.5$\,kpc, calculated as the weighted average of the Gaia DR3 parallaxes for our target sample, after following the prescriptions by \cite{Bailer-Jonesetal2021} and the parallax zero-point offset from \cite{Lindegrenetal2021}. At the  distance of the SFR the expected physical depth of the cluster is much smaller than the individual parallax uncertainties thus, we adopted the average distance $d \sim 4.7$\,kpc. Indeed, given that the physical diameter of the Sh2-284 complex is estimated at $\sim 60$ pc, any line-of-sight physical depth variations ($\pm 30$ pc) introduce an intrinsic distance uncertainty of less than 0.6\%, which is negligible compared to the $\sim 10\%$ statistical uncertainty of the collective Gaia parallax ($\pm 0.5$ kpc). Therefore, adopting a common distance prevents the introduction of artificial scatter in the resulting stellar luminosities and masses \citep[see][]{Lurietal2018}. Moreover, the adopted value is consistent with recent estimates reported in the literature \citep[e.g.,][and references therein]{Guarcelloetal2021, Ashrafetal2026}. The uncertainty in stellar luminosity was then derived by propagating the errors in effective temperature, distance, photometry, and extinction (see Table\,\ref{tab:stellar_parameters}). Fig.\,\ref{fig:HRdiagram} shows the position of our targets in the HR diagram.

The stellar mass and age were estimated adopting the theoretical MESA Isochrones and Stellar Tracks (MIST, v1.2; \citealt{Choietal2016}) for $Z \sim 0.005$, [Fe/H]=$-0.5$\,dex and available down to $M_\star = 0.1\,M_\odot$. We applied the method described \cite{Alcalaetal2014}, which is based on the interpolation of the PMS evolutionary models. 
As for the \teff, we adopted the mean values obtained in Sect.\,\ref{sec:teff}. The stellar radius was calculated from the effective temperature and stellar luminosity. Mean uncertainty in $M_\star$ is around 25\% (see Table\,\ref{tab:stellar_parameters}), while typical error in $R_\star$ is around 20-25\%.

Using the method by \cite{Alcalaetal2014}, the age distribution of our targets on the HR diagram derived from the PMS evolutionary tracks shows a mean value of $<\log({\it Age})>$=5.80\,yr and a standard deviation of 0.55\,yr in $\log({\it Age})$ (see Fig.\,\ref{fig:HRdiagram}). The typical error in age is around 20-25\% (see \citealt{Alcalaetal2014} for further details on the procedure). The large apparent dispersion is a combination of both an intrinsic age spread, potentially driven by sequential or triggered star formation across the complex, and observational error propagation. Specifically, the mean uncertainties in $T_{\rm eff}$ and luminosity propagate into errors of $\sim 0.2$ dex in $\log({\it Age})$. Thus, while a real age spread cannot be ruled out, a fraction of the observed dispersion could be observational in origin, and the population may be more coeval than the distribution suggests. Our age estimate is consistent with the median age of $\log({\it Age}) \sim$ 6.2\,yr (standard deviation 0.3 yr) found by \cite{Guarcelloetal2021} from the color–magnitude diagrams of members with masses between 0.7 and 2.5\,$M_\odot$. Comparable age estimates have also been reported in the literature: \cite{Negueruelaetal2015} found ages below 3 Myr from photometry of massive stars, while \cite{Cusanoetal2011} and \cite{KalariVink2015} estimated ages of 1–2\,Myr for cluster members.

We emphasize that pre-main sequence (PMS) evolutionary tracks in the low-mass and low-metallicity regimes are highly sensitive to the underlying input physics (e.g., convection efficiency, atmospheric boundary conditions). To evaluate the impact of these model-dependent systematic uncertainties, we independently derived stellar masses and ages using the isochrones and tracks by PARSEC (Padova-Trieste Stellar Evolution Code; \citealt{Nguyenetal2022}). We found excellent agreement between the two grids, with discrepancies tightly bounded within 0.03\,dex in $\log M_\star$ and 0.1\,dex in $\log({\it Age})$, thus confirming the robustness of our derived parameters against the choice of evolutionary models.

\subsection{Line fluxes and luminosities}
\label{sec:emission_lines}

We detected several emission lines that are useful tracers of accretion (\citealt{Alcalaetal2017}, and references therein), including hydrogen lines from both the Balmer and Paschen series, as well as calcium lines. In a small number of cases, we also detected forbidden emission lines, including [O{\sc i}] and [S{\sc ii}]. The study of these features, which are diagnostics of jets and winds (see, e.g., \citealt{Nisinietal2018}), lies beyond the scope of this paper thus, we do not analyze them here.

\begin{figure*}[!t]
\begin{center}
\includegraphics[width=18cm, trim=0 350 0 150,clip]{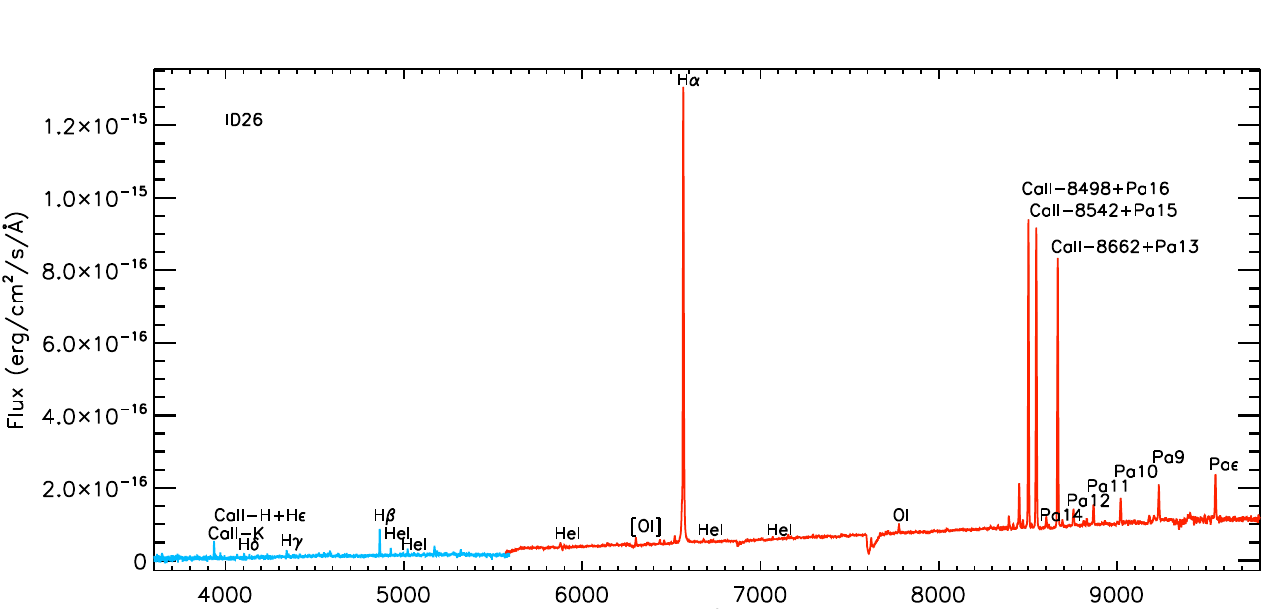}
\caption{Example of a MODS spectrum of a YSO candidate obtained with the blue ($\sim$320-590 nm) and red ($\sim$540 nm-1 mm) gratings. Emission of accretion diagnostics, such as H$\alpha$, H$\beta$, H$\gamma$, Ca{\sc ii} H$\&$K lines and infrared triplet, and the Paschen series is evident, as marked in the plot.}
\label{fig:spectrum}
\end{center}
\end{figure*}

The detection/non-detection of the lines depends on the instrumental sensitivity, signal-to-noise of the spectra, and on the accretion rate of the individual target. In Figure\,\ref{fig:spectrum}, we show the example of a MODS flux-calibrated spectrum obtained using both blue and red gratings (for the spectra of the whole sample, see Figs.\,\ref{fig:spectra1234}, \ref{fig:spectra5678}, \ref{fig:spectra910}). Emission lines, such as hydrogen Balmer and Paschen lines or Ca\,{\sc ii} lines, which are typical tracers of accreting YSOs, are clearly detected. 

H$\alpha$ line was detected in emission in 61 young stellar objects, H$\beta$ in 16 targets, H$\gamma$ in 7 YSOs, calcium lines in 13 YSOs, and Paschen lines in four. When present in the spectra, we measured the equivalent widths and the integrated fluxes of the H$\alpha$, H$\beta$, H$\gamma$, Ca{\sc ii}-$\lambda\lambda$8498, 8542, 8662\,\AA, Pa$\eta$ ($\equiv$ Pa\,10) at $\sim \lambda$9015\,\AA, Pa$\zeta$ ($\equiv$ Pa\,9) at $\sim$9229\,\AA, and Pa$\epsilon$ ($\equiv$ Pa\,8) at $\sim$9546\,\AA. In 16 objects, the H$\alpha$ line is detected as an emission reversal superposed on the broad photospheric absorption lines (namely, ID3, 6, 8, 9, 11, 13, 14, 22, 27) or as an inverse P-Cygni profile (ID10, 38) or a filled-in absorption (ID19, 20, 21, 36, 44). Therefore, since in the latter cases the flux measurement is influenced by photospheric absorption, we considered the spectral subtraction method to remove the photospheric flux and emphasize the emission of the line core (see, e.g., \citealt{FrascaCatalano1994, Biazzoetal2009, Frascaetal2017}). This correction is negligible for objects with very strong emission lines, but it is needed in cases where the photospheric profile is only filled-in with emission or the latter is weak. In particular, we considered as template for each source a BT-Settl model with the same stellar parameters as the target and degraded to the MODS resolution, as done in Sect.\,\ref{sec:stellar_parameters}. Any potential mismatch in the template selection due to the uncertainties in the adopted $T_{\rm eff}$ (and the corresponding spectral type) has a negligible impact on our final results. Indeed, the variation of the underlying photospheric profile within the temperature error bars translates into a marginal discrepancy in the subtracted flux, which remains well within our nominal error bounds. This procedure allowed us to isolate the pure emission that often only fills-in the line cores and allowed us to derive the net H$\alpha$ EWs and fluxes. 

 The flux at the Earth in permitted lines from the observed or photospheric-subtracted spectra was computed by directly integrating the flux-calibrated spectra using the {\it splot} package under {\sc IRAF}. The estimates from photospheric-subtracted spectra for the 5/16 targets with a filled-in H$\alpha$ profile (namely, ID19, 20, 21, 36, 44) resulted to be upper limits, due to the strong contribution of the photospheric flux in absorption. Since the excess emission of these targets is close to the chromospheric level, we consider these objects as weak accretors. The corresponding 1$\sigma$ uncertainties in flux were calculated using the {\sc IRAF} package and considering both the effective readout noise per pixel and the photon noise present in the spectral region encompassing the emission lines \footnote{The effective readout noise per pixel was measured as the rms deviation of the continuum on either side of the H$\alpha$ line. The average rms deviation was then set as the parameter $\sigma_{0}$ in the {\it splot} package. The photon noise was estimated as {\it invgain}$\times${\it I}, where {\it invgain} is the reciprocal of the MODS gain (2.5 {\it e$^-$}/ADU expressed in physical units) and $I$ is the pixel value. The error on the profile fit was then computed by a Monte Carlo simulation, whose iteration number ({\it nerrsample}) was set to 100.} (see \citealt{Gianninietal2024} for details). 
 
 The luminosity of the different emission lines was computed as $L_{\rm line} = 4 \pi d^2 \times f_{\rm line}$, where $d$ is the mean cluster distance and $f_{\rm line}$ is the extinction-corrected flux of the lines. The extinction-corrected fluxes for all diagnostics used in this work, together with their errors, are reported in Table\,\ref{tab:line_fluxes_id_ordered}.

Based on the detection of the H$\alpha$ line in emission or in net emission after the photospheric subtraction, we were not able to derive any line emission in 5/68 (IDs: 24, 40, 71, 78, 80) targets because of very pure quality of the spectrum, while 3/68 (IDs: 16, 30, 37) resulted to have pure photospheric spectra without net emission, even after subtraction of the synthetic photospheric spectrum. Excluding the five targets with spectra of bad quality, we identify about $\sim 95$\% of the targets (60 out of 63) as `{\it bona fide} YSOs', supporting the reliability of our photometric criteria for selecting accreting candidates. 

\begin{figure} 
\centering
\includegraphics[width=0.45\textwidth]{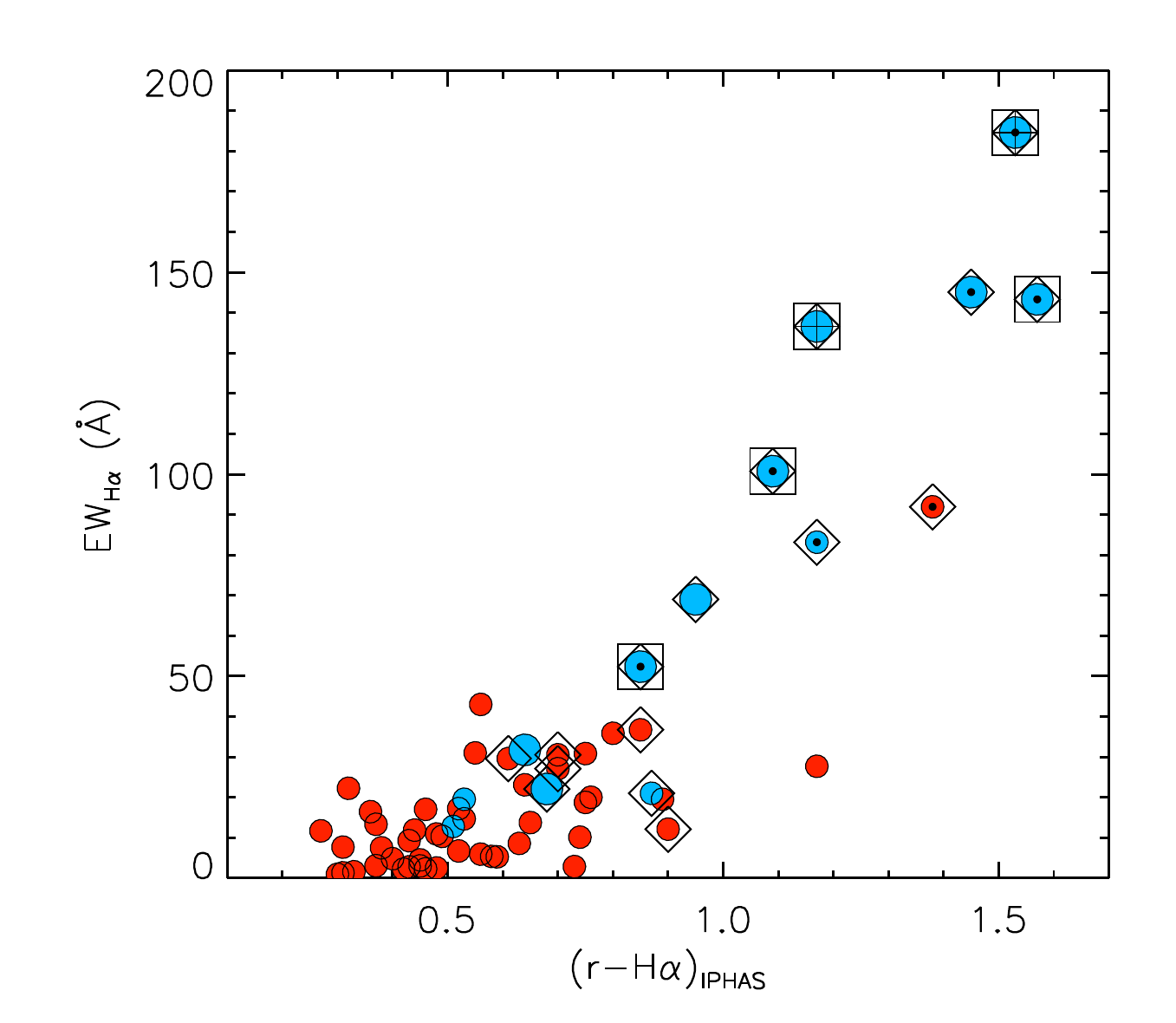} 
\includegraphics[width=0.45\textwidth]{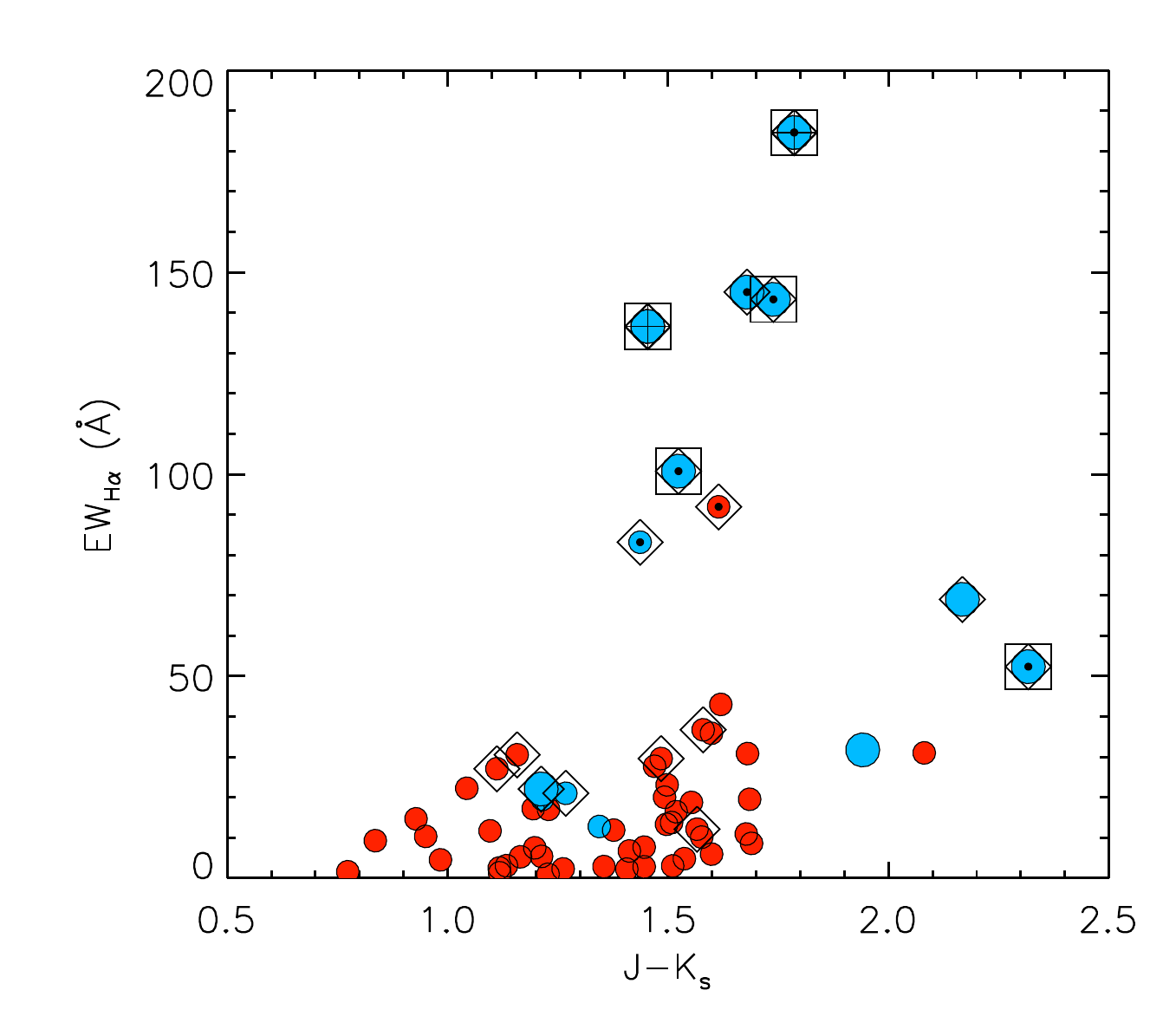} 
\includegraphics[width=0.45\textwidth]{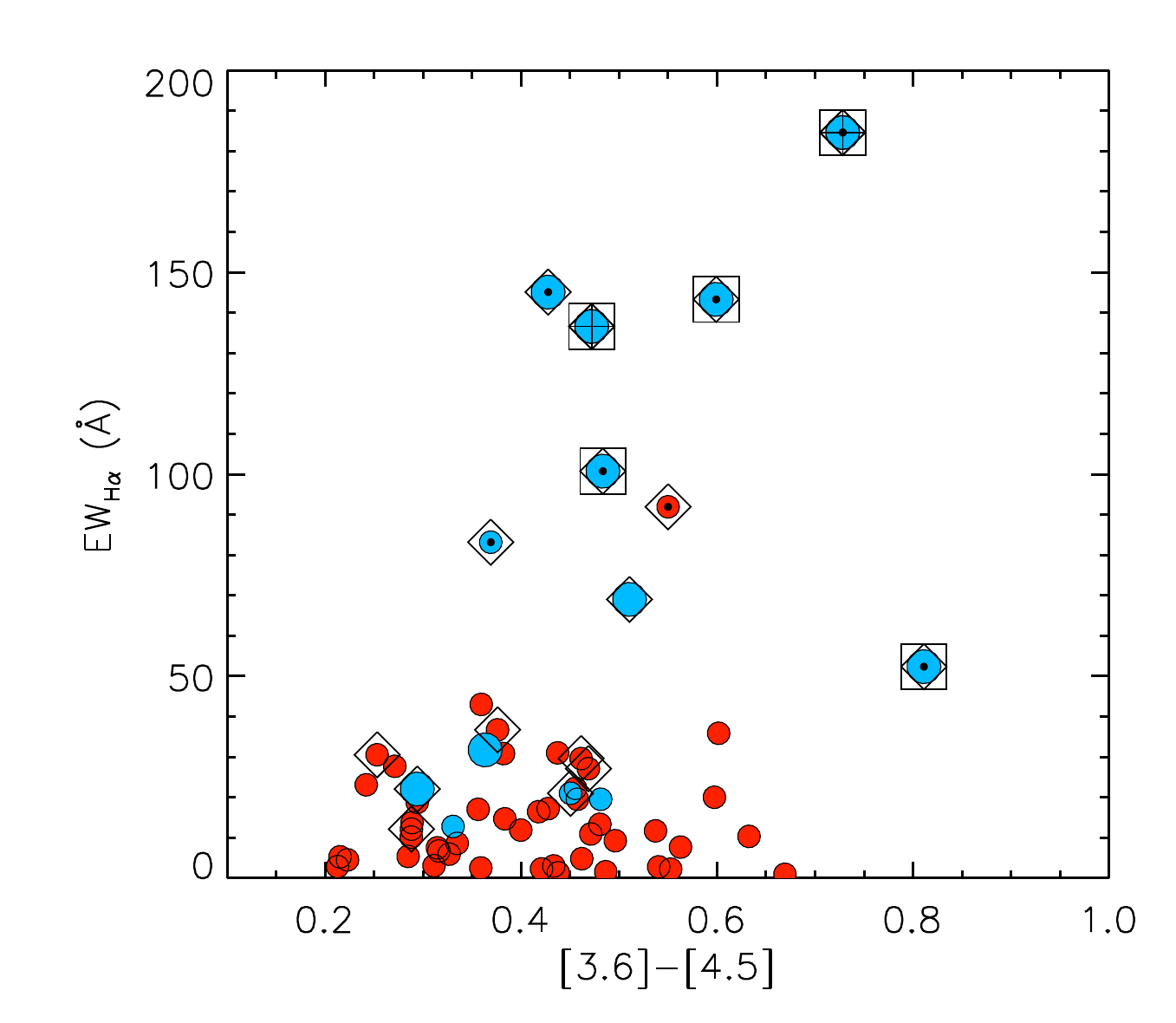} 
\caption{Equivalent width of the H$\alpha$ line versus IPHAS $r-H\alpha$ ({\it up}), 2MASS $J-K$ ({\it middle}), and Spitzer [3.6]$-$[4.5] ({\it below}) colors. Blue dots mark the sources with detected emission of the Ca{\sc ii} infrared triplet lines. Bigger dots represent targets with EW of the Ca{\sc ii} line at $\lambda\sim8662$\,\AA\, greater than 3\,\AA. Diamonds, small dots, and squares are targets with H$\beta$, H$\gamma$, and Paschen lines in emission, respectively. Crosses in the left panel identify the two targets selected through VPHAS+ colors (see Sect.\,\ref{sec:sample_selection}).}
\label{fig:ew_colors} 
\end{figure}

Fig.\,\ref{fig:ew_colors} shows the equivalent width of the H$\alpha$ line as a function of the optical and near-infrared (NIR) colors used for the sample selection ($r-H\alpha$, $J-K$, [3.6]$-$[4.5], from top to bottom). Blue dots mark the sources for which we detected emission in the Ca{\sc ii} infrared triplet (IRT) lines, while larger dots indicate those with Ca{\sc ii} $\lambda\sim8662$\,\AA\, equivalent width greater than 3\,\AA\,(similar results are obtained for the other two lines of the triplet). Diamonds, small dots, and squares mark the targets showing H$\beta$, H$\gamma$, and Paschen lines in emission, respectively. Sources with stronger H$\alpha$ emission also tend to show calcium triplet emission and redder near-infrared colors. Most of these objects exhibit higher Ca{\sc ii} EWs and the presence of H$\beta$, H$\gamma$, and Paschen lines in emission. This correlation between the strength of the accretion diagnostics (H$\alpha$ EW) and infrared colors (tracing disk emission) suggests that objects with detectable accretion are those surrounded by optically thick inner disks (see, e.g., \citealt{Meyeretal1997}, \citealt{Haischetal2001}). 

\begin{figure*}[!t]
\centering
\includegraphics[width=8.5cm, trim=0 0 0 0, clip]{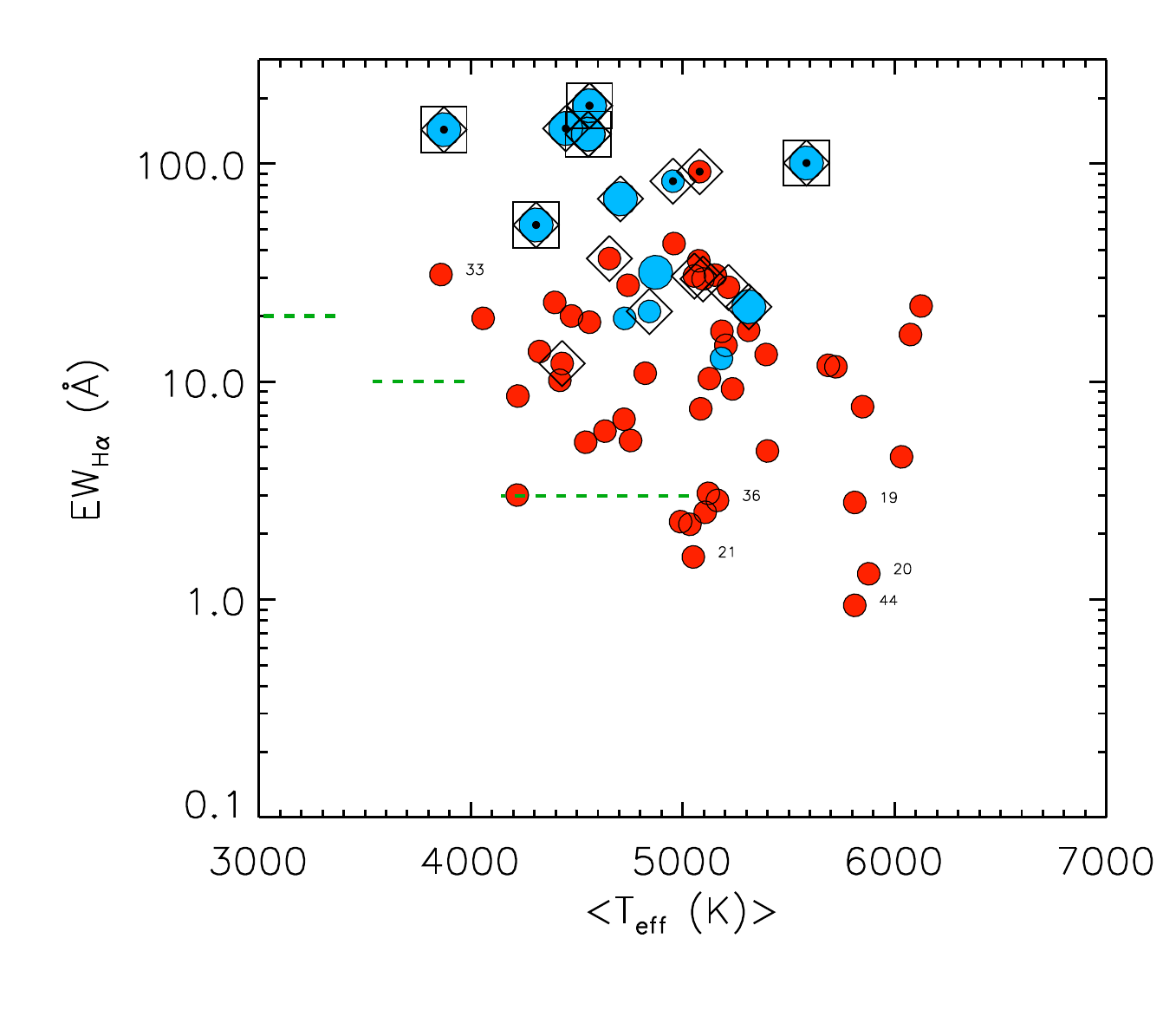} \includegraphics[width=8.5cm, trim=0 0 0 0, clip]{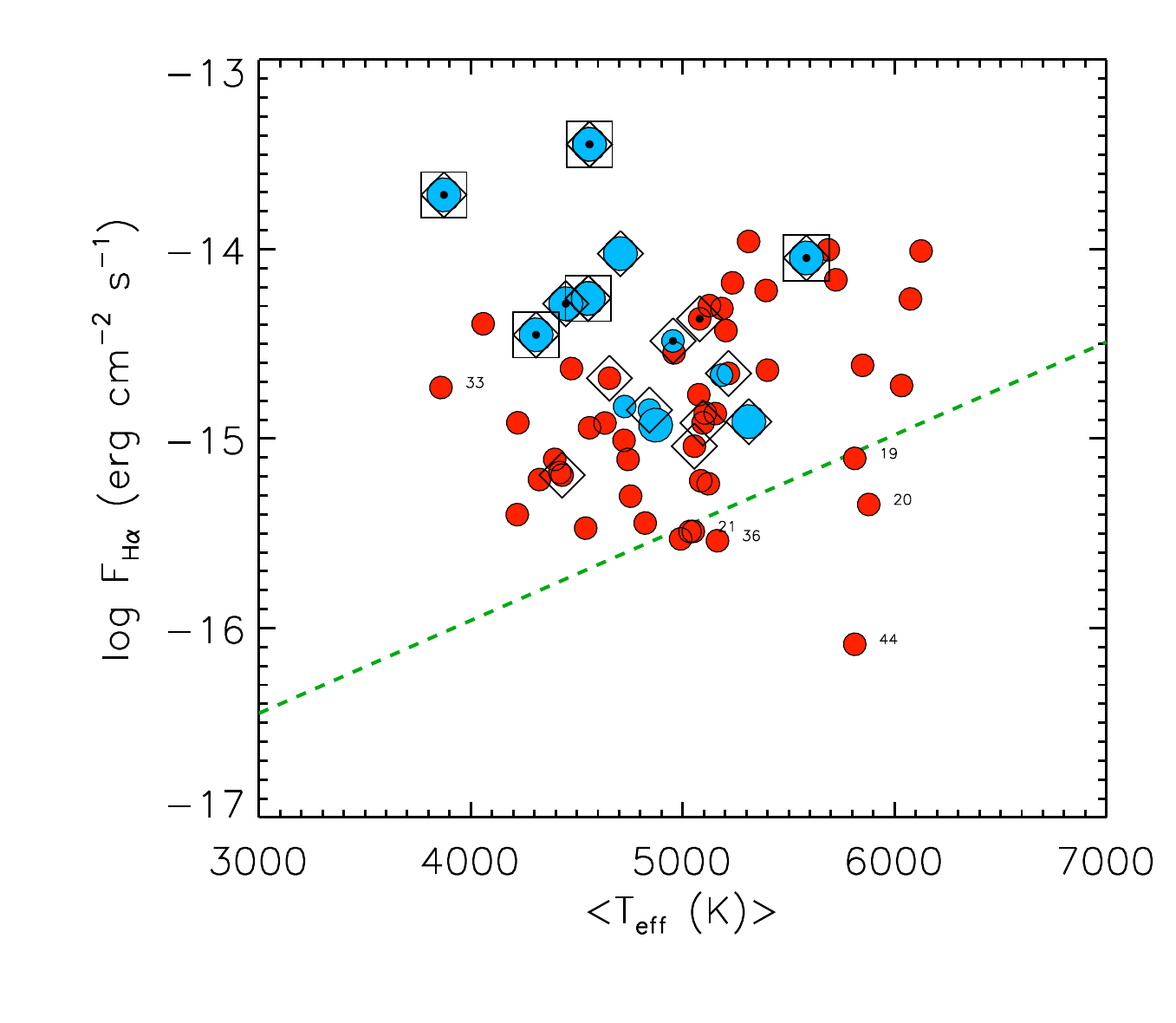}
\caption{H$\alpha$ equivalent width ({\it left panel}) and flux ({\it right panel}) versus \teff. Symbols as in Fig.\,\ref{fig:ew_colors}. Dashed lines in the left panel refers to the threshold above which the objects can be considered accretors, as defined by \cite{WhiteBasri2003}). Dashed line in the right panel is the boundary between chromospheric emission and accretion as derived by \cite{Frascaetal2015} for solar-metallicity Class III sources.}
\label{fig:EW_Flux_Teff} 
\end{figure*}

In Fig.\,\ref{fig:EW_Flux_Teff}, we show the EWs and stellar fluxes of the $H\alpha$ line as a function of the mean $T_{\rm eff}$. All targets we classified as weak accretors lie below the empirical thresholds established by \cite{WhiteBasri2003} and \cite{Frascaetal2015} for separating Class II and Class III objects at solar metallicity. Two targets (ID27 and ID38), which exhibit an $H\alpha$ emission reversal superimposed on the photospheric absorption and an inverse P-Cygni profile, are located in close proximity to these boundaries. All targets displaying H$\beta$ emission (indicated by diamonds in Fig.\,\ref{fig:EW_Flux_Teff}) possess $EW_{\rm H\alpha}$ values exceeding $\sim$10-20\,\AA\ and $H\alpha$ fluxes at least 0.5\,dex above the Class III saturation limit. Furthermore, sources showing H$\gamma$ in emission (small dots) are characterized by $EW_{\rm H\alpha} > 50$\,\AA\ and H$\alpha$ fluxes at least 1.0\,dex above the threshold. This progressive detection of higher Balmer lines with increasing H$\alpha$ flux highlights a clear correlation between the presence of higher-order Balmer lines and the intensity of the H$\alpha$ accretion signal and is consistent with higher excitation conditions and densities in the accretion columns of this low-$Z$ region.


\begin{sidewaystable*}
\setlength{\tabcolsep}{4pt}
\centering
\tiny
\caption{Extinction-corrected fluxes of H$\gamma$, H$\beta$, H$\alpha$, Ca\,{\sc ii} IRT, and P$\eta$, Pa$\zeta$, and Pa$\epsilon$ (in erg\,cm$^{-2}$\,s$^{-1}$) ordered by ID.}
\label{tab:line_fluxes_id_ordered}
\begin{tabular}{lccccccccccc}
\hline\hline
ID & F(H$\gamma$) & F(H$\beta$) & F(H$\alpha$) & F(Ca8498) & F(Ca8542) & F(Ca8662) & F(Pa$\eta$) & F(Pa$\zeta$) & F(Pa$\epsilon$) & Notes \\
\hline
3  & ... & ... & (9.93$\pm$0.05)$\times10^{-15}$ & ... & ... & ... & ... & ... & ... & (a, e) \\
4  & ... & ... & (1.10$\pm$0.06)$\times10^{-14}$ & ... & ... & ... & ... & ... & ... & ... \\
5  & ... & ... & (6.65$\pm$0.04)$\times10^{-15}$ & ... & ... & ... & ... & ... & ... & (e) \\
6  & ... & ... & (6.90$\pm$0.04)$\times10^{-15}$ & ... & ... & ... & ... & ... & ... & (a) \\
7  & ... & ... & (5.04$\pm$0.03)$\times10^{-15}$ & ... & ... & ... & ... & ... & ... & (e) \\
8  & ... & ... & (9.79$\pm$0.05)$\times10^{-15}$ & ... & ... & ... & ... & ... & ... & (a) \\
9  & ... & ... & (6.06$\pm$0.04)$\times10^{-15}$ & ... & ... & ... & ... & ... & ... & (a, e) \\
10 & ... & ... & (1.37$\pm$0.01)$\times10^{-15}$ & ... & ... & ... & ... & ... & ... & (a, e) \\
11 & ... & ... & (2.30$\pm$0.02)$\times10^{-15}$ & ... & ... & ... & ... & ... & ... & (a) \\
12 & ... & ... & (1.91$\pm$0.01)$\times10^{-15}$ & ... & ... & ... & ... & ... & ... & ... \\
13 & ... & ... & (2.44$\pm$0.02)$\times10^{-15}$ & ... & ... & ... & ... & ... & ... & (a) \\
14 & ... & ... & (5.47$\pm$0.03)$\times10^{-15}$ & ... & ... & ... & ... & ... & ... & (a, e) \\
15 & ... & ... & (4.87$\pm$0.03)$\times10^{-15}$ & ... & ... & ... & ... & ... & ... & ... \\
16 & ... & ... & ... & ... & ... & ... & ... & ... & ... & (b) \\
18 & ... & ... & (3.72$\pm$0.02)$\times10^{-15}$ & ... & ... & ... & ... & ... & ... & ... \\
19 & ... & ... & (7.87$\pm$0.01)$\times10^{-16}$ & ... & ... & ... & ... & ... & ... & (a, d) \\
20 & ... & ... & (4.50$\pm$0.01)$\times10^{-16}$ & ... & ... & ... & ... & ... & ... & (a, d) \\
21 & ... & ... & (3.24$\pm$0.01)$\times10^{-16}$ & ... & ... & ... & ... & ... & ... & (a, d) \\
22 & ... & ... & (5.78$\pm$0.01)$\times10^{-16}$ & ... & ... & ... & ... & ... & ... & (a) \\
23 & ... & ... & (1.21$\pm$0.01)$\times10^{-15}$ & ... & ... & ... & ... & ... & ... & ... \\
24 & ... & ... & ... & ... & ... & ... & ... & ... & ... & (c) \\
25 & ... & ... & (2.17$\pm$0.02)$\times10^{-15}$ & (2.78$\pm$0.02)$\times10^{-16}$ & (1.79$\pm$0.01)$\times10^{-16}$ & (2.33$\pm$0.02)$\times10^{-16}$ & ... & ... & ... & ... \\
26 & (1.96$\pm$0.10)$\times10^{-15}$ & (2.87$\pm$0.15)$\times10^{-15}$ & (3.58$\pm$0.02)$\times10^{-14}$ & (1.93$\pm$0.01)$\times10^{-14}$ & (2.09$\pm$0.02)$\times10^{-14}$ & (1.85$\pm$0.01)$\times10^{-14}$ & (1.89$\pm$0.06)$\times10^{-15}$ & (2.57$\pm$0.08)$\times10^{-15}$ & (3.17$\pm$0.10)$\times10^{-15}$ & (e, g) \\
27 & ... & ... & (3.24$\pm$0.02)$\times10^{-16}$ & ... & ... & ... & ... & ... & ... & (a) \\
28 & ... & (3.47$\pm$0.02)$\times10^{-16}$ & (9.48$\pm$0.05)$\times10^{-15}$ & (1.49$\pm$0.01)$\times10^{-15}$ & (1.11$\pm$0.01)$\times10^{-15}$ & (9.00$\pm$0.01)$\times10^{-16}$ & ... & ... & ... & ... \\
29 & ... & ... & (9.80$\pm$0.01)$\times10^{-16}$ & ... & ... & ... & ... & ... & ... & ... \\
30 & ... & ... & ... & ... & ... & ... & ... & ... & ... & (b) \\
31 & ... & ... & (4.04$\pm$0.03)$\times10^{-15}$ & ... & ... & ... & ... & ... & ... & ... \\
32 & ... & ... & (2.35$\pm$0.02)$\times10^{-15}$ & ... & ... & ... & ... & ... & ... & ... \\
33 & ... & ... & (1.87$\pm$0.01)$\times10^{-15}$ & ... & ... & ... & ... & ... & ... & (g) \\
34 & (6.30$\pm$0.04)$\times10^{-16}$ & (1.91$\pm$0.10)$\times10^{-15}$ & (1.94$\pm$0.01)$\times10^{-14}$ & (8.44$\pm$0.05)$\times10^{-15}$ & (8.99$\pm$0.05)$\times10^{-15}$ & (8.10$\pm$0.05)$\times10^{-15}$ & (1.38$\pm$0.05)$\times10^{-15}$ & (1.88$\pm$0.06)$\times10^{-15}$ & (2.16$\pm$0.07)$\times10^{-15}$ & (g) \\
36 & ... & ... & (2.90$\pm$0.02)$\times10^{-16}$ & ... & ... & ... & ... & ... & ... & (a, d) \\
37 & ... & ... & ... & ... & ... & ... & ... & ... & ... & (b) \\
38 & ... & ... & (2.96$\pm$0.02)$\times10^{-16}$ & ... & ... & ... & ... & ... & ... & (a) \\
40 & ... & ... & ... & ... & ... & ... & ... & ... & ... & (c) \\
42 & ... & (7.22$\pm$0.04)$\times10^{-16}$ & (1.42$\pm$0.01)$\times10^{-15}$ & (1.11$\pm$0.01)$\times10^{-16}$ & (1.63$\pm$0.01)$\times10^{-16}$ & (1.89$\pm$0.01)$\times10^{-16}$ & ... & ... & ... & (e) \\
44 & ... & ... & (8.22$\pm$0.01)$\times10^{-17}$ & ... & ... & ... & ... & ... & ... & (a, d) \\
45 & ... & ... & (3.97$\pm$0.02)$\times10^{-16}$ & ... & ... & ... & ... & ... & ... & ... \\
46 & ... & ... & (1.22$\pm$0.01)$\times10^{-15}$ & ... & ... & ... & ... & ... & ... & ... \\
48 & ... & ... & (6.00$\pm$0.01)$\times10^{-16}$ & ... & ... & ... & ... & ... & ... & ... \\
51 & ... & ... & (4.97$\pm$0.03)$\times10^{-16}$ & ... & ... & ... & ... & ... & ... & ... \\
52 & ... & (9.16$\pm$0.05)$\times10^{-17}$ & (2.21$\pm$0.02)$\times10^{-15}$ & ... & ... & ... & ... & ... & ... & ... \\
53 & ... & ... & (1.48$\pm$0.01)$\times10^{-15}$ & (2.05$\pm$0.01)$\times10^{-16}$ & (1.67$\pm$0.01)$\times10^{-16}$ & (2.64$\pm$0.02)$\times10^{-16}$ & ... & ... & ... & (f) \\
54 & (1.32$\pm$0.01)$\times10^{-16}$ & (3.05$\pm$0.02)$\times10^{-16}$ & (8.99$\pm$0.05)$\times10^{-15}$ & (4.45$\pm$0.03)$\times10^{-15}$ & (4.73$\pm$0.03)$\times10^{-15}$ & (4.28$\pm$0.03)$\times10^{-15}$ & (4.58$\pm$0.02)$\times10^{-16}$ & (8.15$\pm$0.03)$\times10^{-16}$ & (8.97$\pm$0.03)$\times10^{-16}$ & ... \\
57 & ... & ... & (1.15$\pm$0.01)$\times10^{-15}$ & ... & ... & ... & ... & ... & ... & ... \\
58 & ... & ... & (6.08$\pm$0.01)$\times10^{-16}$ & ... & ... & ... & ... & ... & ... & ... \\
60 & ... & (8.16$\pm$0.01)$\times10^{-17}$ & (6.42$\pm$0.01)$\times10^{-16}$ & ... & ... & ... & ... & ... & ... & (e, f) \\
61 & ... & ... & (2.84$\pm$0.02)$\times10^{-15}$ & ... & ... & ... & ... & ... & ... & ... \\
62 & ... & (1.04$\pm$0.01)$\times10^{-16}$ & (1.23$\pm$0.01)$\times10^{-15}$ & (5.28$\pm$0.01)$\times10^{-16}$ & (4.94$\pm$0.01)$\times10^{-16}$ & (4.84$\pm$0.03)$\times10^{-16}$ & ... & ... & ... & ... \\
63 & ... & ... & (1.71$\pm$0.01)$\times10^{-15}$ & ... & ... & ... & ... & ... & ... & ... \\
64 & ... & (8.08$\pm$0.01)$\times10^{-17}$ & (2.09$\pm$0.02)$\times10^{-15}$ & ... & ... & ... & ... & ... & ... & ... \\
65 & (6.05$\pm$0.01)$\times10^{-17}$ & (2.03$\pm$0.02)$\times10^{-16}$ & (3.54$\pm$0.02)$\times10^{-15}$ & (2.43$\pm$0.02)$\times10^{-15}$ & (2.73$\pm$0.02)$\times10^{-15}$ & (2.38$\pm$0.02)$\times10^{-15}$ & (2.65$\pm$0.01)$\times10^{-16}$ & (2.78$\pm$0.01)$\times10^{-16}$ & ... & (g) \\
66 & ... & ... & (3.37$\pm$0.01)$\times10^{-16}$ & ... & ... & ... & ... & ... & ... & ... \\
67 & ... & ... & (3.59$\pm$0.01)$\times10^{-16}$ & ... & ... & ... & ... & ... & ... & ... \\
69 & ... & ... & (6.63$\pm$0.01)$\times10^{-16}$ & ... & ... & ... & ... & ... & ... & ... \\
70 & ... & ... & (1.36$\pm$0.01)$\times10^{-15}$ & ... & ... & ... & ... & ... & ... & ... \\
71 & ... & ... & ... & ... & ... & ... & ... & ... & ... & (c) \\
72 & ... & ... & (1.18$\pm$0.01)$\times10^{-15}$ & (3.83$\pm$0.01)$\times10^{-16}$ & (4.96$\pm$0.01)$\times10^{-16}$ & (3.87$\pm$0.01)$\times10^{-16}$ & ... & ... & ... & ... \\
75 & (1.20$\pm$0.01)$\times10^{-16}$ & (3.51$\pm$0.02)$\times10^{-16}$ & (5.16$\pm$0.03)$\times10^{-15}$ & (2.91$\pm$0.02)$\times10^{-16}$ & (3.38$\pm$0.02)$\times10^{-16}$ & (2.55$\pm$0.01)$\times10^{-16}$ & ... & ... & ... & ... \\
76 & ... & (5.86$\pm$0.03)$\times10^{-16}$ & (5.50$\pm$0.03)$\times10^{-15}$ & (3.32$\pm$0.02)$\times10^{-16}$ & (4.94$\pm$0.03)$\times10^{-16}$ & (4.69$\pm$0.03)$\times10^{-16}$ & (1.22$\pm$0.01)$\times10^{-16}$ & (2.53$\pm$0.01)$\times10^{-16}$ & (1.67$\pm$0.01)$\times10^{-16}$ & ... \\
77 & (1.15$\pm$0.01)$\times10^{-16}$ & (3.05$\pm$0.02)$\times10^{-16}$ & (4.30$\pm$0.03)$\times10^{-15}$ & ... & ... & ... & ... & ... & ... & ... \\
78 & ... & ... & ... & ... & ... & ... & ... & ... & ... & (c) \\
80 & ... & (1.09$\pm$0.01)$\times10^{-16}$ & (1.22$\pm$0.01)$\times10^{-15}$ & ... & ... & ... & ... & ... & ... & ... \\
81 & (4.97$\pm$0.01)$\times10^{-17}$ & (2.39$\pm$0.02)$\times10^{-16}$ & (3.28$\pm$0.02)$\times10^{-15}$ & (5.45$\pm$0.01)$\times10^{-17}$ & (9.45$\pm$0.01)$\times10^{-17}$ & (8.77$\pm$0.01)$\times10^{-17}$ & ... & ... & ... & ... \\
83 & ... & ... & (7.77$\pm$0.01)$\times10^{-16}$ & ... & ... & ... & ... & ... & ... & (f) \\
85 & ... & ... & (7.77$\pm$0.01)$\times10^{-16}$ & ... & ... & ... & ... & ... & ... & ... \\
86 & ... & ... & ... & ... & ... & ... & ... & ... & ... & (c) \\
87 & ... & (3.29$\pm$0.01)$\times10^{-17}$ & (9.13$\pm$0.01)$\times10^{-16}$ & ... & ... & ... & ... & ... & ... & ... \\
\hline
\end{tabular}
\begin{flushleft}
\scriptsize (a) H$\alpha$ flux was obtained after the spectral subtraction of the synthetic stellar photosphere (see Sect.\,\ref{sec:emission_lines}). (b) Photospheric spectrum. (c) Bad spectrum. (d) Weak accretor. (e) Target in common with \cite{KalariVink2015}. (f) Target in common with \cite{Ashrafetal2026}. (g) Emission of many other permitted (e.g., Ca\,H\&K, H$\delta$, He\,{\sc i}, Fe\,{\sc ii}) and forbidden (e.g., [O{\sc i}]) lines.
\end{flushleft}
\end{sidewaystable*}

\subsubsection{$EW_{\rm H\alpha}$: comparison with previous works}
\label{sec:ewcomparison} 
Even though accreting YSOs are known to exhibit significant variability (see Sect.~\ref{sec:accr_var}), it is still instructive to compare our measurements with previous spectroscopic studies that targeted a subset of the same sources.

Eight objects in our sample (IDs: 3, 7, 9, 10, 14, 26, 42, 60) were previously observed with the VIsual and Multi-Object Spectrograph (VIMOS) by \citet{Cusanoetal2011}, who measured H$\alpha$ equivalent widths from spectra obtained in February 2006. The mean difference between our $EW_{\rm H\alpha}$ measurements and theirs is $3.7 \pm 20.1$\,\AA. The largest deviation ($\sim$45\,\AA) is observed for ID26, the strongest accretor in our sample. Such a significant discrepancy could be consistent with an accretion burst or strong stochastic variability. While these episodic phenomena are traditionally studied in nearby solar-metallicity environments \citep[e.g.,][]{Costiganetal2014, Venutietal2021}, recent variability surveys of extragalactic metal-poor YSOs \citep[e.g., see][]{Zivkovetal2020} confirm that eruptive accretion processes can be highly efficient even in reduced-metallicity environments. Moreover, high-amplitude variations in emission lines have been well-documented in spectroscopic monitoring of young stars \citep[e.g.,][]{Gianninietal2019}. If ID26 is excluded from the comparison, the mean difference and standard deviation drop to $2.3\pm11.9$\,\AA, indicating a greater level of consistency for the rest of the sample, but still compatible with long-term accretion variability (\citealt{Costiganetal2012}).

Subsequently, in December 2013, \citet{KalariVink2015} obtained spectra of four additional targets in common with our sample (IDs: 3, 5, 7, 9) using the Robert Stobie Spectrograph (RSS) on the Southern African Large Telescope (SALT). Their $EW_{\rm H\alpha}$ values closely match ours, with a mean difference of $1.0\pm$5.7\,\AA, which is within the range expected from accretion variability in classical T~Tauri stars, irrispective to the considered time scale (\citealt{Costiganetal2012}).

Finally, three sources (IDs: 53, 60, 83) were also observed with the Multi Unit Spectroscopic Explorer (MUSE) at the VLT by \citet{Ashrafetal2023} in January 2017. For these targets, the $EW_{\rm H\alpha}$ measurements are again broadly consistent with our results, showing a mean difference of $1.4\pm11.7$\,\AA, but still compatible with short- and mid- term accretion variability.

Overall, the comparison with previous spectroscopic studies shows that our $EW_{\rm H\alpha}$ measurements are consistent with earlier determinations once accretion-driven variability is taken into account.

\subsection{Accretion luminosity}
\label{sec:accr_lum}
We computed the accretion luminosity considering the magnetospheric accretion scenario, and therefore from the measurement of the reradiated energy from the circumstellar gas ionized and heated in funnel flows (\citealt{Hartmannetal2016}). The luminosities of the emission lines generated in this process can be used as diagnostics to derive the accretion luminosity. In our case, we measured the line luminosities of several hydrogen and calcium lines, as described in Sect.\,\ref{sec:emission_lines}. For each line, we considered the empirical linear relationships between the observed line luminosity $L_{\rm line}$ and the accretion luminosity $L_{\rm acc}$ derived by \cite{Alcalaetal2017} by modeling the continuum excess emission and by the simultaneous observations of accretion indicators of T\,Tauri stars in the Lupus association. In particular, we considered the relationships obtained by the latter authors for the $L_{\rm H\alpha}$, $L_{\rm H\beta}$, $L_{\rm Ca\,{\sc II}-8498}$, $L_{\rm Ca\,{\sc II}-8542}$, $L_{\rm Pa\zeta}$, $L_{\rm Pa\zeta}$, and $L_{\rm Pa\epsilon}$ luminosities (see their Table\,B.1). The different line diagnostics we considered yielded consistent values of $L_{\rm acc}$, with mean differences of $0.01\pm0.12$\,dex in $\log L_{\rm acc}$. The overall consistency of the results $L_{\rm acc}$ supports the validity of the adopted extinction correction and justifies the use of all these diagnostics to derive an average accretion luminosity for each YSO. In this way, the error on the averaged accretion luminosity ($<L_{\rm acc}>$) derived from several diagnostics, measured simultaneously, is minimized 
(e.g., \citealt{Rigliacoetal2012, Alcalaetal2014, Biazzoetal2014}). In particular, uncertainty in $L_{\rm acc}$ is due to the error in line luminosity through the $L_{\rm line}$-$L_{\rm acc}$ relationships, which includes the uncertainties in line flux, distance, and extintion (see Table\,\ref{tab:stellar_parameters}).

We remark that this excellent agreement does not constitute an independent validation of our absolute accretion scale, as all line diagnostics are anchored to the empirical calibrations of \cite{Alcalaetal2017}. Instead, this alignment demonstrates a high level of internal consistency within our adopted analytical framework and verifies the accuracy of our extinction corrections across the broad spectral range. When combining multiple line diagnostics, averaging multiple line values minimizes the random measurement errors associated with individual line profiles. However, it does not reduce possible correlated systematic uncertainties, such as distance scale offsets, absolute photometric errors, or potential metallicity-driven deviations from the solar-metallicity-derived calibrations (see Sect.\,\ref{sec:mass_lum}).

\subsection{Mass accretion rate}
\label{sec:Maccr}
Once $<L_{\rm acc}>$ is known, the mass accretion rate $\dot{M}_{\rm acc}$ can be derived from the free-fall equation that links the luminosity released in the impact 
of the accretion flow with the rate of mass accretion according to the following relationship (\citealt{Hartmannetal2016}):

\begin{equation} 
\label{eq:Macc} 
\dot{M}_{\rm acc} = \left(1 - \frac{R_{\star}}{R_{\rm in}} \right)^{-1} \frac{L_{\rm acc} R_{\star}}{G M_{\star}}\, 
\approx 1.25 \frac{L_{\rm acc} R_{\star}}{G M_{\star}}\,,
\end{equation}

\noindent{where $M_\star$ and $R_\star$ are the stellar mass and the photospheric radius, respectively, $R_{\rm in}$ is the inner radius of the accretion disk, and 
$G$ is the universal gravitational constant. $R_{\rm in}$ corresponds to the distance at which the disk is truncated, because of the stellar magnetosphere, and 
from which the disk gas is accreted and channeled by the magnetic field lines; therefore, its value is rather uncertain because it depends on how the accretion 
disk is coupled with the star. Following \cite{Gullbringetal1998}, we assume $R_{\rm in}=5\,R_\star$ for all YSOs.}

The median value of the distribution of mass accretion rates is $\sim 2.2 \times 10^{-8}$\,$M_\odot$\,yr$^{-1}$ including upper limits, and $\sim 2.6 \times 10^{-8}$\,$M_\odot$\,yr$^{-1}$ without including upper limits.

Concerning the statistical uncertainties on $\dot{M}_{\rm acc}$, they can be derived as error propagation using Eq.\,\ref{eq:Macc} in logarithmic form, e.g. $\log M_{\rm acc}=\log (1.25/G)+\log L_{\rm acc}+\log R_{\star}-\log M_{\star}$. Uncertainty in $\log L_{\rm acc}$ was derived as mentioned in the previous Section and it is $\sim 0.25$\,dex in average. The other sources of uncertainty for $\dot{M}_{\rm acc}$ are the stellar mass and radius. As for the mass and radius, as discussed in Sect.\,\ref{sec:mass_lum} their determinations are linked to the comparison of the location in the HR diagram with the evolutionary tracks. When we interpolate through the PMS evolutionary tracks to estimate the mass and radius, the uncertainties on effective temperature and stellar luminosity imply an error of $\sim 20\%$ on $R_\star$ and of $\sim 25\%$ on $M_\star$. Combining all the sources of errors, mean statistical uncertainty on $\log \dot{M}_{\rm acc}$ is $\sim 0.38$\,dex. Finaal values of $\dot{M}_{\rm acc}$ are listed in Table\,\ref{tab:stellar_parameters}.

The systematic uncertainty on $\dot{M}_{\rm acc}$ is dominated by the knowledge of the ratio $L_{\rm acc}$/$L_{\rm line}$ relations by \cite{Alcalaetal2017}. This ratio, even if uncertain by a factor of $\sim$1.1-1.3 depending on the line, is the same for all stars, therefore the comparison between different objects is not hampered by this uncertainty, as long as the statistical 
errors are small (see also \citealt{DeMarchietal2010}).

Other sources of systematic errors in the derived $\dot{M}_{\rm acc}$ are due mainly to theoretical evolutionary tracks and isochrones, interstellar extinction, line emission generated by processes different from accretion, and contribution of nebular emission to the photometric colors. Concerning the first source of errors, the main uncertainty on the derived mass and age comes from differences between models computed by different authors or from the use of models with a metallicity that might not properly describe the stellar population under study. As shown by \cite{Biazzoetal2019}, if we for instance had used the PARSEC PMS tracks \citep{Nguyenetal2022} instead of those of the MESA tracks at the same metallicity, we would have obtained similar values of mass and age for the PMS phase, to within 2\% and 6\%, respectively, which translates into a negligible shift of less than $\sim 0.01$\,dex in $\log \dot{M}_{\rm acc}$. Regarding the metallicity, had we used tracks with $Z$ lower by 30\%, the masses of our PMS objects would be systematically lower by about 10\% and the ages younger by a negligible amount for the luminosity and temperature ranges typical of our targets (see \citealt{Biazzoetal2019}), resulting in an effect on $\log \dot{M}_{\rm acc}$ of $\sim 0.04$\,dex. For what concerns the extinction, following \cite{DeMarchietal2010}, underestimating the extinction by, e.g., $\sim 0.5$\,mag would lead to a 30\% overestimate of the $R_\star/M_\star$ ratio, translating into a systematic overastimate of $\sim 0.10$\,dex in $\log \dot{M}_{\rm acc}$. Notably, these systematic shifts are well below our nominal average total statistical uncertainty (see above). Finally, the contribution of processes unrelated to accretion (e.g., nebular emission) is negligible, because their effects were accurately removed during the data reduction steps discussed in Section\,\ref{sec:sample_selection}, with the sole exception of ID33.

\subsection{$\dot{M}_{\rm acc}$: comparison with the literature}
\label{sec:Macccomparison} 

We compared our derived accretion properties with those available in the literature, specifically focusing on the works of \cite{KalariVink2015}, who re-evaluated the targets from \cite{Cusanoetal2011} alongside their own new observations, and the recent work by \cite{Ashrafetal2026}. For the nine targets in common with \cite{KalariVink2015}, who derived accretion rates from H$\alpha$ equivalent widths, because flux calibration was not feasible in their study, we find a good overall agreement. The mean discrepancy in accretion luminosity between their work and ours is $-0.22 \pm 0.62$\,dex in $\log L_{\rm acc}$. This small offset propagates to the mass accretion rates, yielding a mean difference of $-0.08 \pm 0.58$\,dex in $\log M_{\rm acc}$. The relatively large scatter ($\sim 0.6$\,dex) is expected, as it likely reflects both the different methodologies employed and the intrinsic long-term source variability known to be significant in young accreting objects \citep[see Sect.\,\ref{sec:accr_var}; see also][]{Costiganetal2014}. Indeed, their observations were conducted nine years before than ours. Furthermore, we compared our results with those of \cite{Ashrafetal2026}, who derived accretion rates from the H$\alpha$ equivalent width and the continuum flux interpolated at the H$\alpha$ wavelength. While three sources are in common between the two samples, those authors derived accretion rates for only two of them using H$\alpha$ equivalent widths. For these two sources (namely, ID60 and ID83), we find a mean difference of $0.09 \pm 0.19$\,dex in $\log M_{\rm acc}$, indicating a high level of consistency between the two studies.

\begin{figure} 
\begin{center}
\includegraphics[width=0.95\linewidth, trim=0 0 0 0, clip]{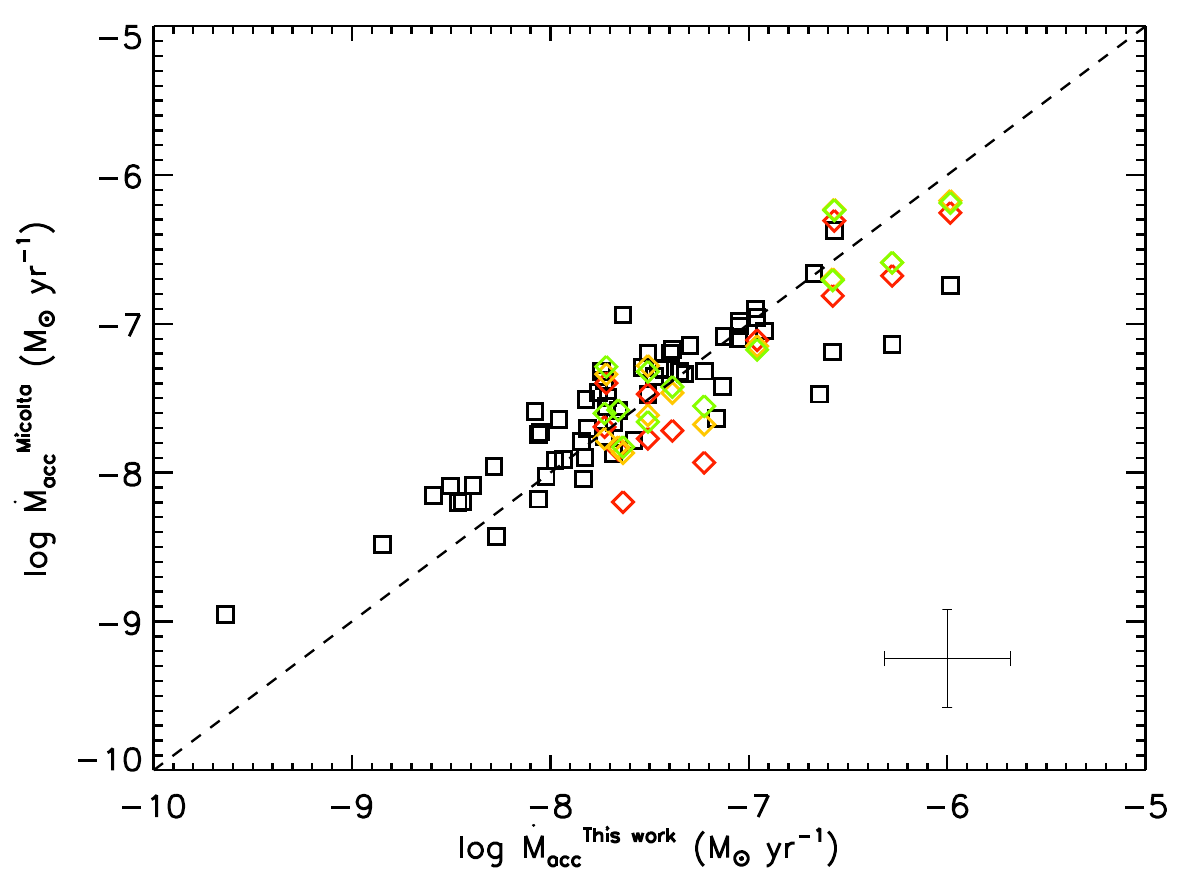}
\caption{Comparison between the mass accretion rates derived in Sect.\,\ref{sec:Maccr} and those obtained from the $\dot M_{\rm acc}$–$L_{\rm line}$ relationships of \cite{Micoltaetal2023}, using H$\alpha$ (squares) and the three \ion{Ca}{2} IRT lines (diamonds). The mean uncertainty is indicated in the lower-right corner of the panel.}
\label{fig:Macc_Micolta}
\end{center}
\end{figure}

Moreover, in Fig.\,\ref{fig:Macc_Micolta} we compare the mass accretion rates derived in this work with those computed through the empirical relationships between $\dot M_{\rm acc}$ and the H$\alpha$ and \ion{Ca}{2} IRT line luminosities obtained by \citet{Micoltaetal2023} for the Chamaeleon\,I star-forming region. We find that our results are compatible with those extracted from these calibrations, with most sources distributed along the 1:1 relation within the uncertainties. This agreement supports the overall consistency of our accretion scale and demonstrates that a generalized empirical framework independent of individual stellar properties can successfully reproduce our physically derived values without introducing severe systematic offsets. These relationships bypass the explicit use of the stellar mass-to-radius ratio, confirming that the bulk accretion properties of our sample align well with standard empirical benchmarks. Furthermore, the comparison also points toward a detection threshold for the \ion{Ca}{2} IRT lines, which are predominantly observed in sources with $\log \dot{M}_{\text{acc}} > -8\,M_{\odot}\,\text{yr}^{-1}$. To quantify the overall dispersion, we calculated the root mean square (rms) of the differences between the two methods ($\log \dot{M}_{\text{acc, our}} - \log \dot{M}_{\text{acc, Micolta}}$), resulting in a value of 0.32 dex for the H$\alpha$ line and $0.24-0.34$\,dex for the calcium lines. This low scatter confirms the robustness of our results across three orders of magnitude in accretion activity. 

Overall, these comparisons with independent studies confirm that our homogeneous characterization of the accretion properties is robust and aligns well with the existing literature despite the inherent challenges of non-simultaneous observations and different adopted methods. A possible key limitation of this comparison, however, is that both methods rely on empirical calibrations derived from nearby, solar-metallicity populations (such as Chamaeleon I and Lupus). By implicitly assuming that line emission coefficients are invariant with metallicity, these frameworks may mask subtle chemical dependencies in the gas cooling rates, a factor that we explore down below in Sect.\,\ref{sec:flux_flux_met}.

\section{Results and discussion}
\label{sec:result_discussion}

\subsection{Metallicity dependence of flux–flux reelations}
\label{sec:flux_flux_met}

\begin{figure*}[!t]
\begin{center}
\includegraphics[width=16cm]{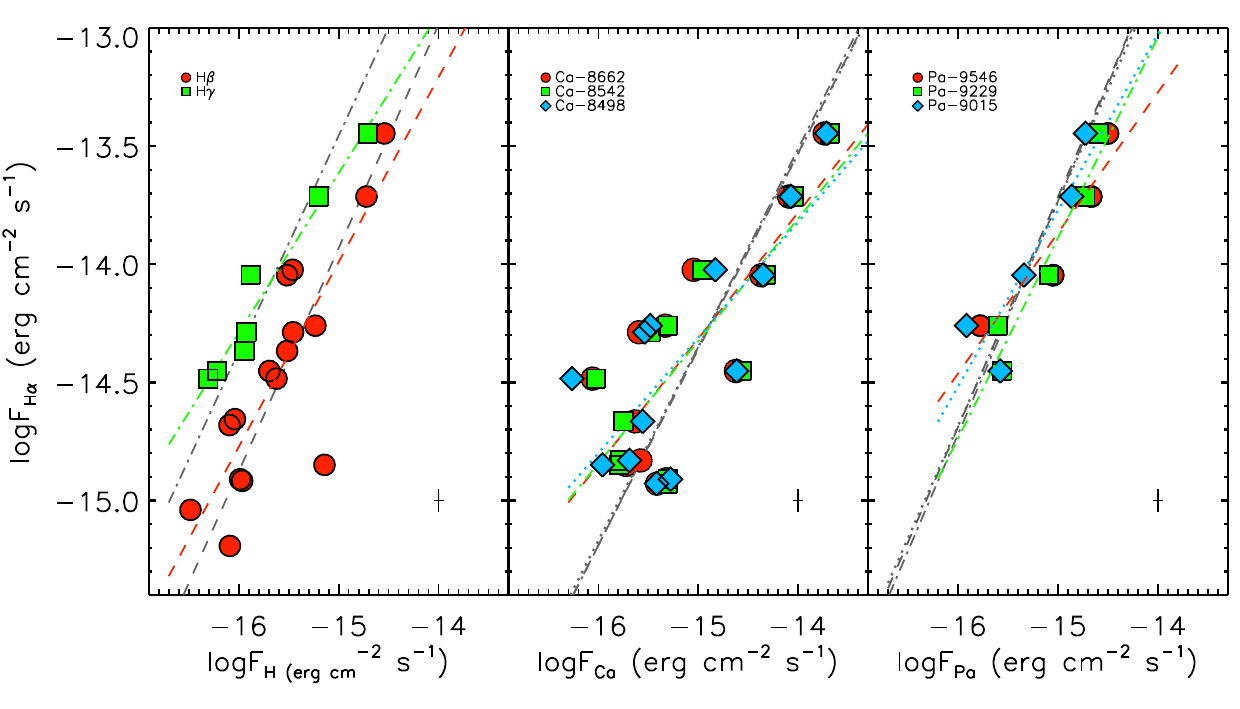}
\caption{Flux-flux relations between H$\alpha$ and Balmer-series, and Ca-IRT, Paschen-series lines. The colored red dashed, green dot-dashed, and blue dotted lines refer to the least-square regressions for the different lines marked in the upper-left corner of each panel. The grey dashed, dot-dashed, and dotted lines refer to the corresponding flux-flux relations found by \cite{Alcalaetal2017} for the hydrogen lines and \cite{Frascaetal2017} for the calcium lines and for the solar-metallicity Lupus region.}
\label{fig:flux_flux} 
\end{center}
\end{figure*}

The H$\alpha$ fluxes are well correlated with the fluxes obtained through the other diagnostics used in this work. Figure~\ref{fig:flux_flux} presents indeed the flux–flux relations between H$\alpha$ and H$\beta$, H$\gamma$, Ca \textsc{ii} infrared triplet lines, Paschen lines) for our sample. The Kendall's rank correlation coefficients $\rho$ range from $\sim$0.5 to $\sim$1.0 with significances $\sigma$ of $10^{-4}-10^{-2}$, depending on the diagnostics and the number of targets per each diagnostics, while least-squares regressions provide the following relations:

\begin{eqnarray*}
\log F_{\rm H\alpha} & = & -2.26(\pm2.16) + 0.78(\pm0.14) \log F_{\rm H\beta} [16]\nonumber  \\
\log F_{\rm H\alpha} & = & -3.48(\pm1.06) + 0.68(\pm0.07) \log F_{\rm H\gamma} [7]\nonumber  \\
\log F_{\rm H\alpha} & = & -6.97(\pm1.75) + 0.49(\pm0.12) \log F_{\rm Ca{\sc II}-8498} [13]\nonumber  \\
\log F_{\rm H\alpha} & = & -6.55(\pm1.73) + 0.52(\pm0.11) \log F_{\rm Ca{\sc II}-8542} [13]\nonumber  \\
\log F_{\rm H\alpha} & = & -6.31(\pm1.88) + 0.53(\pm0.12) \log F_{\rm Ca{\sc II}-8662} [13]\nonumber  \\
\log F_{\rm H\alpha} & = & -2.59(\pm3.05) + 0.75(\pm0.20) \log F_{\rm Pa\zeta} [5]\nonumber  \\
\log F_{\rm H\alpha} & = & -1.15(\pm2.02) + 0.85(\pm0.13) \log F_{\rm Pa\eta} [5]\nonumber  \\
\log F_{\rm H\alpha} & = & -4.95(\pm2.28) + 0.59(\pm0.15) \log F_{\rm Pa\epsilon} [4]\,.
\label{Eq:FHa-FCaIRT}
\end{eqnarray*}
\noindent{where the number of targets per diagnostics is shown among square brackets.} 

In the same figure, we compare our relations with examples of those 
derived in the nearby, solar-metallicity Lupus star-forming regions (\citealt{Alcalaetal2017, Frascaetal2017}). Although the number statistics is rather low, all line fluxes display clear correlations with H$\alpha$, and a trend 
can be seen, with the solar-metallicity samples exhibiting steeper slopes. We explicitly caution the reader that several of our secondary diagnostics (such as H$\gamma$ with 7 sources and the Paschen series with 4–5 sources) are limited by the small number statistics. Consequently, the fitted linear regression slopes for these specific lines carry wide confidence intervals, and these results should be treated as preliminary indications rather than definitive statistical constraints on environmental behavior. In the left panel of Fig.\,\ref{fig:flux_flux} (H$\alpha$ versus H$\beta$ and H$\gamma$), the best-fit slope for our metal-poor sample is $\sim$0.68-0.78, shallower than the value of $\sim$0.95 obtained for Lupus, alghough they remain consistent within their 1-$\sigma$ uncertainty bounds. Because of this statistical overlap, our current dataset cannot definitively reject the null hypothesis of a universal, metallicity-independent flux-flux relation. Therefore, while we discuss potential physical mechanisms below, these interpretations are presented as tentative models to explain the observed mean trends rather than statistically proven environmental shifts.

\begin{figure} 
\begin{center}
\includegraphics[width=8.9cm]{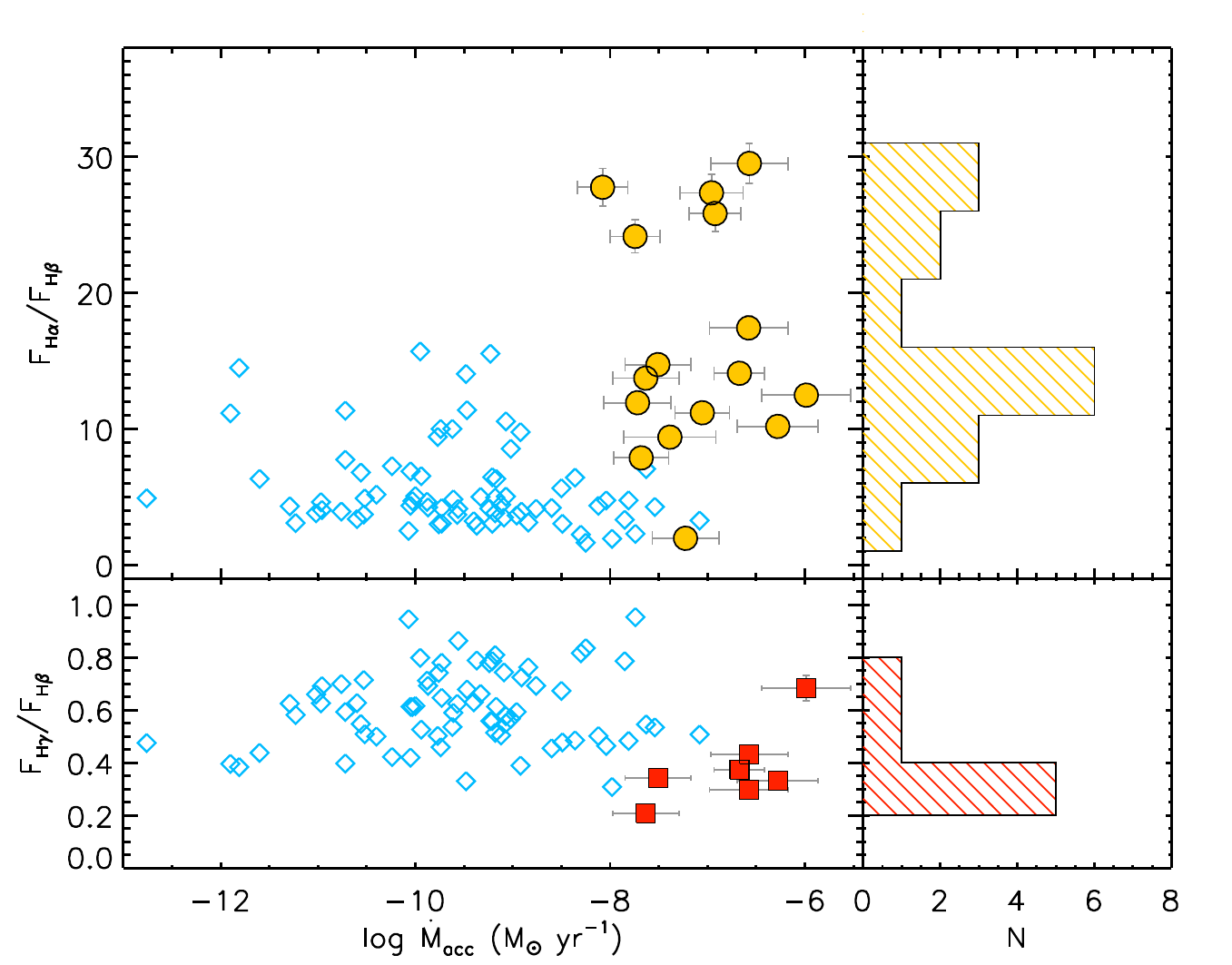}
\caption{H$\alpha$ over H$\beta$ ({\it upper panel}) and H$\gamma$ over H$\beta$ flux ratios as a function of $\log \dot M_{\rm acc}$ for our targets (filled symbols) and those of the Lupus sample (empty diamonds). On the right panels the corresponding histograms.}
\label{fig:Balmer_Macc} 
\end{center}
\end{figure}

If confirmed, this trend would indicate that the relative scaling between Balmer lines differs in low-metallicity environments, with H$\beta$ (and H$\gamma$) responding more sensitively to changes in the emitting conditions than H$\alpha$. This could be consistent with hydrogen emission being dominated by the H$\alpha$ over a broad range of line fluxes, particularly in high-temperature emitting environments where metal-line cooling is less efficient (e.g., \citealt{Muzerolleetal2001}). Infact, as can be seen in Figure~\ref{fig:Balmer_Macc}, the average line flux-ratio $F_{H\alpha}/F_{H\beta}$ in the Sh2-284 sample is about two to three times higher than in the Lupus stars, where the average is about six\footnote{Note that a high $F_{H\alpha}/F_{H\beta}$ ratio in a few objects,  both in metal-poor and solar metallicity objects, might be induced by strong H$\alpha$ emission from outflows. Yet, the average flux ratio is still  higher in metal-poor objects in comparison with that in Lupus YSOs.}, while the average of the $F_{H\gamma}/F_{H\beta}$ ratio
is closer to the Lupus values ($\sim$0.6). This implies higher temperatures and densities for the  emitting regions in the metal-poor sample in comparison with those in solar-metallicity YSOs (see \citealt{Antoniuccietal2017}). Another important behaviour in Figure~\ref{fig:Balmer_Macc} is that the flux ratios in Sh2-284 tend to be constant over a mass accretion rate of more than 2 orders of magnitude within the uncertainties, meaning similar physical conditions in the accreting gas for objects with different accretion rates. The latter is also a characteristic observed in solar metallicity regions (see Figure~\ref{fig:Balmer_Macc} and Sect.\,7.2 in \citealt{Alcalaetal2014} and references therein).

The central panel of Figure~\ref{fig:flux_flux} shows the relationship between H$\alpha$ and the Ca \textsc{ii} infrared triplet. Here, the slopes measured for our sample cluster around $\sim$0.49–0.53, markedly flatter than the slopes of $\sim$0.82–0.84 observed in Lupus. This pronounced difference suggests that in metal-poor environments, the H$\alpha$ emission increases more slowly relative to the Ca \textsc{II} triplet as the accretion rate intensifies. This behaviour could indicate that the reduced efficiency of metal-line cooling may alter the thermal and ionization structure of the accretion flow (\citealt{Fangetal2009}). In the right panel, which compares H$\alpha$ with Paschen-series emission, the slopes derived for the metal-poor sample span the range $\sim$0.59-0.85, whereas the corresponding solar-metallicity relations are again significantly steeper, with slopes of $\sim$0.95–0.99. Since Paschen lines typically remain optically thinner than H$\alpha$, the steeper slopes observed at solar metallicity could indicate a more efficient conversion of accretion power into higher-excitation hydrogen emission, facilitated by enhanced cooling and recombination efficiency in metal-rich environments (\citealt{Edwardsetal2013}).

Taken together, the systematically flatter slopes observed across all three panels for our metal-poor sample might hint at the possibility of a non-universal, metallicity-dependent behavior in flux–flux scaling. However, as stated above, the current statistics prevent us from drawing definitive conclusions. Observations of larger and statistically more significant samples are needed to confirm our results.

\subsection{Accretion luminosity versus stellar parameters}
\label{sec:accr_stpar}

\begin{figure} 
\begin{center}
\includegraphics[width=0.95\linewidth, trim=0 0 0 0, clip]{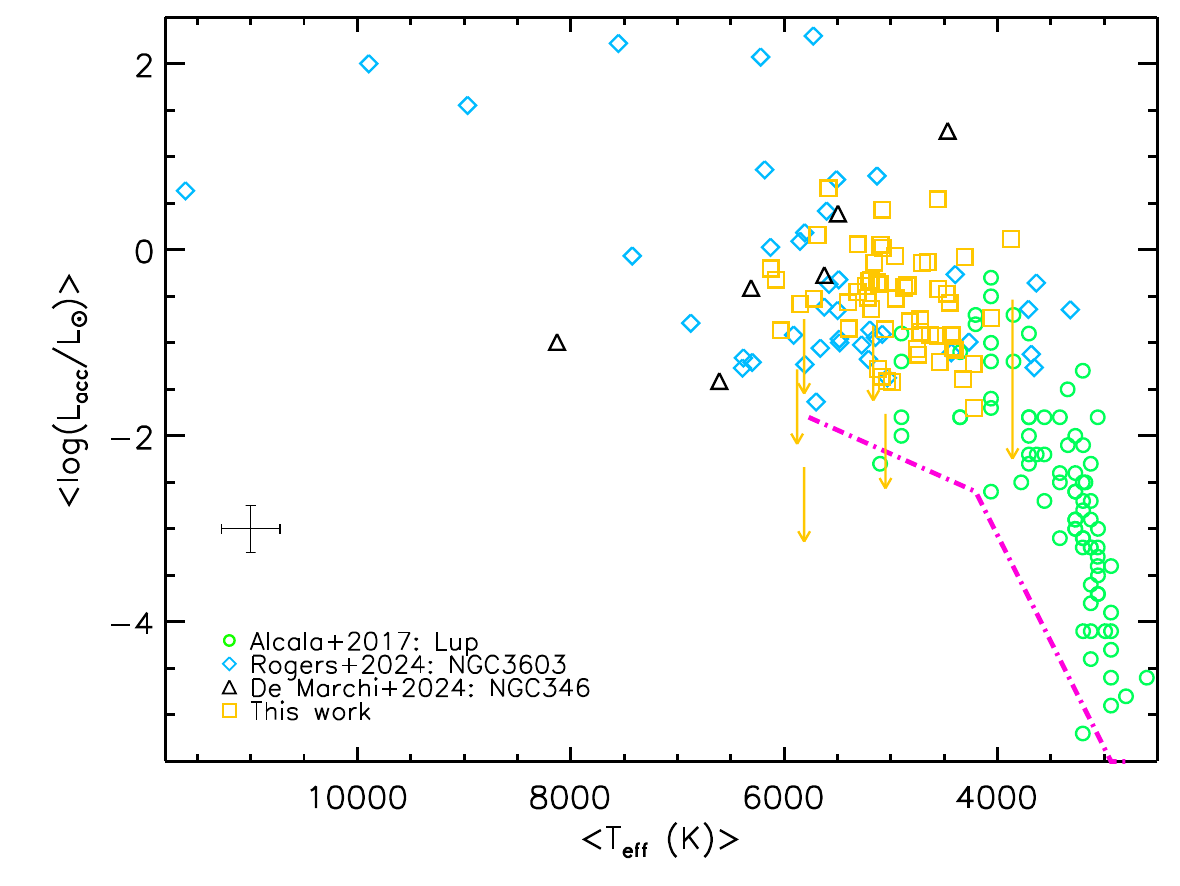}
\caption{Mean accretion luminosity derived from the different emission lines, as described in the text, as a function of the mean effective temperature measurements. Weak accretors are marked with arrows as upper limits. The long arrow on the right refers to the target with bad removal of the background. Symbols refer to different regions analyzed in the literature through spectroscopy, as indicated in the panel. The dot-dashed line in all panels marks the locus below which chromospheric emission is important in comparison with $L_{\rm acc}$ (\citealt{Manaraetal2017}). Mean error is shown in the lower-left part of the panel.}
\label{fig:Lacc_Teff}
\end{center}
\end{figure}

In Fig.\,\ref{fig:Lacc_Teff}, the accretion luminosity is shown as a function of the mean effective temperature given in Table\,\ref{tab:stellar_parameters}. Different symbols refer to different literature estimates obtained through spectroscopic analysis (green circles for Lupus, \citealt{Alcalaetal2017}; cyan diamonds for NGC\,3603, \citealt{Rogersetal2024}; black triangles for NGC\,346, \citealt{DeMarchietal2024}). The dashed line shows the chromospheric level determined by \citet[][see also \citealt{Claesetal2024}]{Manaraetal2017}, and represents the locus obtained for solar-metallicity Class III objects below which the contribution of chromospheric emission starts to be important in comparison with energy losses due to accretion. At a fixed effective temperature, all accreting YSOs show $L_{\rm acc}/L_\odot$ values very similar to the values estimated in the solar nearby Lupus SFR, in the metal-poor targets of NGC\,346 in the SMC and in the solar-metallicity distant Galactic young clusters NGC\,3603, for the same \teff\,range.

\begin{figure} 
\begin{center}
\includegraphics[width=0.95\linewidth, trim=0 0 0 0, clip]{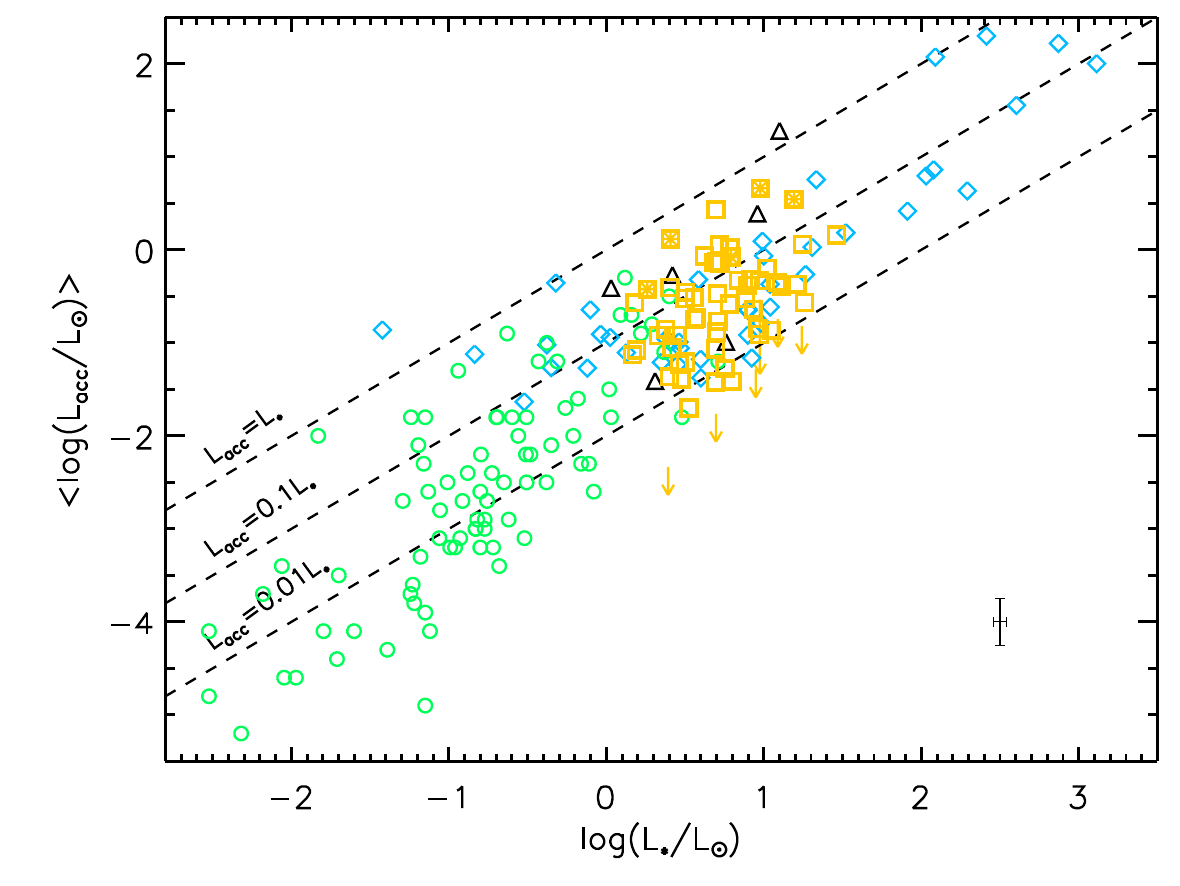}
\caption{Mean accretion luminosity versus mean stellar luminosities. Symbols as in Fig.\,\ref{fig:Lacc_Teff}. Targets with the Paschen lines in emission are marked with crosses. Dashed lines represent the loci of three $L_{\rm acc}$-$L_{\star}$ relations, as labeled. Mean error is shown in the lower-right part of the panel.}
\label{fig:Lacc_Lstar_mean} 
\end{center}
\end{figure}

Figure\,\ref{fig:Lacc_Lstar_mean} displays the accretion luminosity as a function of the stellar luminosity. The sources in Sh2-284 align well with the general trend observed across various SFRs, ranging from the solar neighborhood to the SMC. Within the $L_{\star}/L_{\sun}$ range of 0.17-1.46, our YSOs exhibit accretion luminosities typically spanning from around 0.002\,$L_{\star}$ for the weakest accretors to $\sim$0.5\,$L_{\star}$ for the strongest accretors, with a median $L_{\star}/L_{\sun}$ value of $\sim$0.05 and a median absolute deviation (MAD) of 0.03. This distribution is consistent with the values reported in the other environments overplotted for comparison: namely, a median $L_{\star}/L_{\sun}$ of $\sim$0.02 (MAD=0.01) for Lupus, $\sim$0.06 (MAD=0.04) for NGC\,3606, and $\sim$0.24 (MAD=0.17) for NGC\,346, despite the limited statistics of the latter.

\subsection{Sh2-284: an outer Galaxy region bridging young nearby Galactic and extragalactic environments} 
\label{sec:accr_stmass}

\begin{figure*}[!t]
\begin{center}
\includegraphics[width=16cm, trim=0 0 00 0, clip]{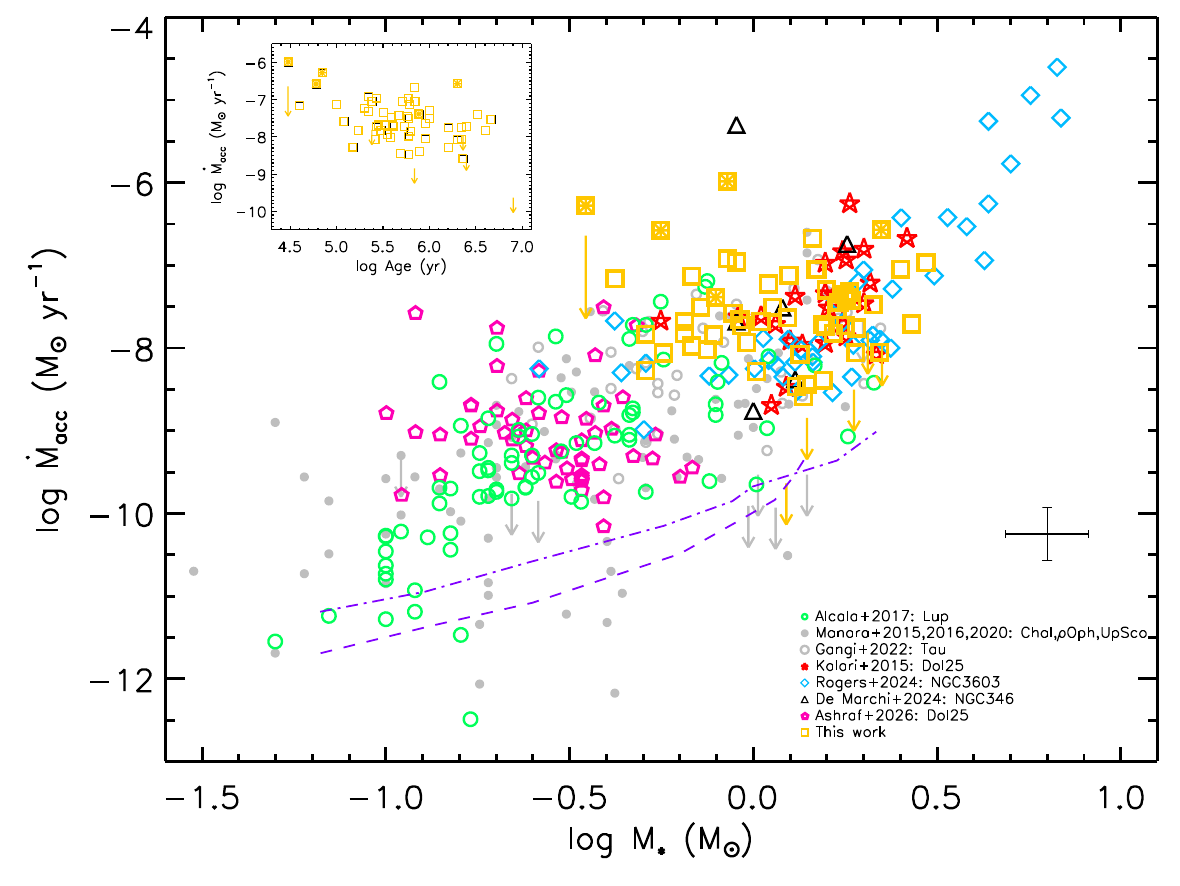}
\caption{Mean mass accretion rate as a function of mean stellar mass for our YSO sample, compared with literature measurements from several well-studied star-forming regions spanning a wide range of metallicities and environments (labels indicated in the panel). Downward arrows mark upper limits. The chromospheric emission thresholds from \citet{Manaraetal2017} are shown as dash-dotted and dashed lines, corresponding to approximate ages of $\sim$3\,Myr and $\sim$10\,Myr, respectively. \textit{Inset:} mass accretion rate versus stellar age for our Sh2-284 sample.}
\label{fig:Macc_Mass_mean} 
\end{center}
\end{figure*}

Figure\,\ref{fig:Macc_Mass_mean} shows the mean mass accretion rate as a function of the mean stellar mass for the YSOs in Sh2-284, together with literature results for SFRs spanning a wide range of metallicities and environments obtained through similar spectroscopic procedure and for targets selected with similar selection criteria as ours. Our data are compared with the Lupus SFR (green circles; \citealt{Alcalaetal2017}), the distant solar-metallicity Galactic cluster NGC\,3603 (blue diamonds; \citealt{Rogersetal2024}), the Dolidze\,25 sample by \citet{KalariVink2015} (red stars), the low-mass MUSE sample in the core of Dolidze\,25 (pink pentagons; \citealt{Ashrafetal2026}, the metal-poor SMC star-forming region NGC\,346 (black triangles; \citealt{DeMarchietal2024}), and four other nearby solar-metallicity SFRs (namely, $\rho$\,Ophiucus, Chamaeleon\,I, Upper Scorpius, and Taurus-Auriga; grey filled dots; \citealt{Manaraetal2015, Manaraetal2016, Manaraetal2020, Gangietal2022}).

Our targets populate the stellar mass range $\sim$0.35–2.95\,$M_\odot$ and exhibit accretion rates between $\sim 2.3\times10^{-10}$ and $\sim 1.0\times10^{-6}\,M_\odot$\,yr$^{-1}$, with a median value of $\sim2.2\times10^{-8}\,M_\odot$\,yr$^{-1}$ (including upper limits). Excluding the four objects characterized by strong emission in all available diagnostics (namely, ID26, ID34, ID54, ID65), which populate the upper envelope of the distribution, the Sh2-284 measurements largely overlap with the broad distributions observed in both nearby and distant Galactic star-forming regions. As suggested by the inset of Fig.\,\ref{fig:Macc_Mass_mean}, higher accretion rates tend to be associated with younger ages, in line with the expected decline of mass accretion during pre-main-sequence evolution (\citealt{Sicilia-Aguilaretal2005, Antoniuccietal2014, Hartmannetal2016}).

At a given stellar mass, the overall scatter in $\dot M_{\rm acc}$ is substantial and spans $\sim$1–2 dex, consistent with what is commonly observed in other nearby regions \citep[e.g.,][and references therein]{Alcalaetal2014, Alcalaetal2017, Manaraetal2023}. This large dispersion, which cannot be explained by accretion variability alone (see Sect.~\ref{sec:accr_var}), together with the intrinsic inhomogeneity of the datasets, arising from differences in observational strategies, accretion diagnostics, and analysis methods adopted in the various studies, limits the ability to identify subtle systematic differences between environments. In this respect, while the Sh2-284 accretion rates are compatible with those measured in other metal-poor systems, such as NGC\,346, the present data do not allow us to establish whether a significant offset with respect to solar-metallicity samples truly exists. 

The accretion rates measured by us for sources in Sh2-284 are broadly comparable to those reported for other metal-poor environments, such as the SMC cluster NGC\,346, despite the different Galactic environments and star formation histories, and are broadly consistent with previous studies of Dolidze\,25. The accretion rates derived by \citet{KalariVink2015} for a smaller sample of YSOs of $0.6$–$2.6\,M_\odot$ overlap with ours within the uncertainties. The low-mass ($0.1$–$0.7\,M_\odot$) sample observed with MUSE by \citet{Ashrafetal2026} extends this behavior toward lower masses. We stress that these datasets and our MODS data sample a different mass range. The difference in mass coverage means that these studies sample complementary stellar parameters, and combining them provides a more complete view across the entire cluster population. Taken together, these results indicate that the physics of gas infall is resilient in this extreme environment and that efficient accretion can be sustained even at low metallicity, while the relative importance of metallicity with respect to other environmental and stellar parameters remains difficult to disentangle. 

To objectively evaluate the apparent alignment in Fig.\,\ref{fig:Macc_Mass_mean}, we performed a localized statistical comparison within the common mass range where our sample overlaps with the literature. When considering results from NGC\,346, NGC\,3603, and Lupus, together with those presented in this work, and restricting the comparison to a homogeneous and common range in stellar mass ($0.9–1.8\,M_\odot$) and age ($<10$\,Myr), a broad trend toward higher accretion rate levels in low-metallicity environments appears to emerge. However, such a trend is not statistically robust. As summarized quantitatively in Table~\ref{tab:Macc_comp}, the median accretion trends across metallicities show substantial statistical overlap when filtering for coeval populations, confirming that any systematic environmental shift remains buried beneath the substantial intrinsic dispersion within each cluster. In particular, this large intrinsic scatter in $\dot{M}_{\rm acc}$ prevents us from drawing firm conclusions on systematic differences between low- and solar-metallicity regions based solely on these samples.

In low-metallicity disks, reduced dust content leads to lower opacity, potentially leading to higher disk temperatures, increased ionization fractions, and more efficient angular momentum transport, which can sustain higher accretion rates for longer timescales \citep[e.g.,][]{ErcolanoClarke2010, Biazzoetal2019, DeMarchietal2024}. These findings seem to challenge some conclusions by \cite{Yasuietal2009, Yasuietal2016}, who suggested that low-metallicity environments in the outer Galaxy lead to extremely short disk lifetimes based on a low fraction of sources with near-infrared excess. However, those studies relied on dust content as a proxy for the disk fraction, whereas our direct measurements of gas infall demonstrate that the accretion process is active, at a mean level of $3 \times 10^{-8}\,M_\odot$/yr, in the vast majority of objects even at the estimated average age of $\sim$1-2\,Myr, i.e. the age at which most disks should disappear at low-metallicity (see \citealt{Yasuietal2021}). The recent census of sub-solar mass YSOs in the $0.2\,Z_\sun$ NGC\,346 environment of the SMC by \cite{Jonesetal2023} and \cite{Habeletal2024} via JWST NIRCam/MIRI observations seems to reinforce this perspective. By detecting a significant population of low-mass YSOs with strong infrared excess, these studies have proven that substantial dust reservoirs exist even at very low metallicity, confirming that the ``raw material" to form rocky planets is present and persists long enough for planetary embryos to grow. Again, \cite{DeMarchietal2024} observed YSOs in the same cluster with NIRSpec@JWST obtaining mass accretion rates at the level of $\sim 10^{-8}$\,$M_\odot$\,yr$^{-1}$ at ages older than $\sim$10\,Myr, thus implying for circumstellar disks longer timescales than previously thought. Very recently, \cite{Yasuietal2026} found rates $\gtsim 10^{-6}$\,$M_\odot$\,yr$^{-1}$ for targets in the distant Galactic metal-poor cluster Digel Cloud 2 analyzed through the F405N JWST narrow-band filter covering the Br$\alpha$ line. When coupled with the sustained gas accretion observed in Sh2-284, these results suggest that planetary formation is a viable and potentially common process in low-metallicity environments similar to those of the early Universe.

\begin{table*}
\centering
\small
\setlength{\tabcolsep}{3pt}
\caption{Mean mass accretion rates, together with dispersions, and median age for clusters with different metallicity and analyzed with spectroscopy. Only targets with $0.9-1.8\,M_\sun$ and younger than $\sim 10$\,Myr were considered.}
\begin{tabular}{lclcccc}
\hline
SFR & Distance & Reference & $Z$ & $<\log \dot{M}_{\rm acc}>$ & $\sigma$ & $Med(Age)$ \\
  & (kpc) &  &   &  (dex) & (dex) & (Myr) \\
\hline
NGC346 & $\sim 61$ & \cite{DeMarchietal2024, DeMarchietal2011} & $\sim 1/8\,Z_{\odot}$ & $-$6.52 & 1.12 & $\sim 1.5$\\
Sh2-284 & $\sim 4.7 $ & This work & $\sim 1/4-1/3\,Z_{\odot}$ & $-$7.76 & 0.64 & $\sim 1$\\
NGC3603 & $\sim 7.2 $ & \cite{Rogersetal2024} & $\sim Z_{\odot}$ & $-$8.07 & 0.36 & $\sim 8$ \\
Lupus & $\sim 0.15-0.20$ & \cite{Alcalaetal2017, Frascaetal2017} & $\sim Z_{\odot}$ & $-$8.73 & 0.72 &  $\sim 1$\\
\hline
\end{tabular}
\label{tab:Macc_comp}
\end{table*}

Overall, the distribution shown in Fig.~\ref{fig:Macc_Mass_mean} found by us and by \cite{Ashrafetal2026}, even obtained through different facilities, suggests that Sh2-284 occupies an intermediate regime between nearby Galactic star-forming regions and more extreme metal-poor and distant environments of other external galaxies. This behavior places the outer Galaxy in an intermediate yet crucial position, bridging nearby Galactic star-forming regions and more extreme metal-poor systems in the Magellanic Clouds, and provides an important empirical benchmark for interpreting JWST observations of star formation under early-Universe-like conditions. This highlights the need for larger and homogeneous samples in wide range of stellar masses to robustly quantify metallicity-dependent trends.

\section{Conclusions}
\label{sec:conclusions}
We have presented a comprehensive spectroscopic investigation of 68 young stellar objects (YSOs) in the metal-poor ($Z \sim 1/3 - 1/2 Z_\odot$) star-forming region Sh2-284, located in the Galactic anticenter. This work represents the first wide-field spectroscopic census of the region, utilizing a multi-diagnostic approach to characterize the connection between stellar properties and magnetospheric accretion in a low-metallicity environment. Investigating a cluster at this large distance and with sub-solar metallicity inherently pushes the boundaries of standard observational and analytical frameworks, introducing constraints on sample completeness and spectral signal-to-noise. However, by leveraging our multi-tracer strategy, we have been able to robustly address these challenges. Our main findings are summarized as follows:
\begin{itemize}
\item We selected the sample using Spitzer data in combination with optical and near-infrared photometric calalogues. Using Gaia DR3 data, we confirmed that the vast majority of our targets are kinematically consistent with the Dolidze\,25 cluster. 
\item Our analysis identified $\sim$95\% of our targets as {\it bona fide} YSOs, validating our selection criteria based on H$\alpha$ excess and infrared colors. Notably, this targeted selection is conceptually similar to the standard strategy widely applied to characterize accreting populations in nearby SFRs using classical long-slit or single-object spectroscopy, which are structurally tailored to isolate actively accreting, optically visible and NIR disk-bearing objects.
\item By adopting a distance of $4.7 \pm 0.5$\,kpc and comparing our targets with theoretical evolutionary tracks, we derived stellar masses in the range $\sim$0.35–2.95 $M_\odot$ and age distribution peaking at a mean value of $\sim$5.8\,yr in $\log(Age)$, although several targets appear older. Given the large physical scale spanned by our survey ($\sim 60$ pc), the observed age spread likely reflects a combination of localized sequential star formation history across the complex, as previously suggested for the region, and the expected observational error propagation at these distances.
\item We provided exploratory and indicative iron abundance estimates for three low-mass targets showing purely photospheric spectra, alongside lithium abundances for 36 targets also accounting for veiling estimates. These measurements constitute a first attempt to chemically characterize low-mass members in this cluster, paving the way for future high-resolution spectroscopic follow-ups.
\item Our analysis reveals a shallower slope for the empirical relations between H$\alpha$ flux and other accretion tracers (including H$\beta$, H$\gamma$, the Ca II IRT, and the Paschen series) in Sh2-284 compared to solar-metallicity environments like Lupus. If confirmed by larger samples, this tentative trend would hint at the non-universality of the flux-flux relations, suggesting that accretion diagnostics could be sensitive to the chemical composition of the star-forming environment. However, because of the current statistical limitations and the substantial data overlap, our dataset cannot definitively reject the null hypothesis of a universal, metallicity-independent scaling; therefore, these specific results should be treated as preliminary indications rather than definitive statistical constraints. These findings could have significant implications for the study of distant or unresolved star-forming regions, where the application of solar-metallicity calibrations could introduce systematic biases in the derived accretion rates. Consequently, our results emphasize the necessity of developing metallicity-dependent frameworks, a crucial step for the accurate interpretation of JWST observations targeting low-metallicity systems in the outer Galaxy and nearby dwarf galaxies. 
\item We derived mass accretion rates ranging from $\sim 2.3 \times 10^{-10}$ to $\sim 1.0 \times 10^{-6} M_\odot$ yr$^{-1}$. Notably, the accretion is resilient also at low metallicities, remaining remarkably high, at a mean level of $\sim2.2\times10^{-8}\,M_\odot$\,yr$^{-1}$, even for targets with estimated ages of $\sim 1–2$ Myr. This demonstrates that gas infall can be efficiently sustained despite the lower dust-to-gas ratio. 
\end{itemize}
Our results place Sh2-284 in a unique position, bridging the gap between nearby solar-metallicity Galactic SFRs and extragalactic metal-poor environments like those in the SMC and LMC. When restricting the comparison to a localized, coeval mass range, the median accretion trends across different environments display a substantial statistical overlap. This indicates that the large intrinsic dispersion of the accretion process itself dominates over environmental shifts across different metallicities. Ultimately, our study demonstrates that gas accretion remains highly resilient in YSOs in the outer Galaxy. The presence of active gas accretion in most of our sample at an average age of $\sim 1-2$\,Myr proves that the physical mechanisms driving accretion are robust, gas reservoir persist long enough to sustain accretion activity even in reduced-metallicity environments, and raw material for planet growth remains available even in the presence of reduced metal content.

The spectroscopic observations presented here provide accretion diagnostics that are complementary to the indicators accessible with JWST and to the forthcoming unbiased ground-based surveys with MOONS at the VLT.
\begin{acknowledgements}
The authors are very grateful to the referee for carefully reading the paper and for his/her useful remarks that allowed us to significantly improve the previous version of the manuscript. This work has been supported by the Grant INAF-2022 TRacing the Accretion Metallicity rElationship (TRAME; PI: K. Biazzo), the Grant INAF-2024 Spectral Key features of Young stellar objects: Wind-Accretion LinKs Explored in the infraRed (SKYWALKER; PI: J. M. Alcal\'a), the European Union - NextGenerationEU, M4C2 1.2 CUP C83C25000450006 within the project Tracing the staR and plAnet formation in different Circumstellar Environments (TRACE; PI: K. Biazzo), and the project Space Observations of Proto-planetary Accretion Disc Evolution (SPADE; PI: B. Nisini) in the framework of the PROgramma di RIcerca Spaziale di base (PRORIS), financed by the Italian Ministry of University and Research (Ministero dell’Università e della Ricerca; MUR). This paper makes use of data products from the following catalogues: $i.$ INT Photometric H$\alpha$ Survey of the Northern Galactic Plane carried out at the Isaac Newton Telescope (INT). The INT is operated on the island of La Palma by the Isaac Newton Group in the Spanish Observatorio del Roque de los Muchachos of the Instituto de Astrofisica de Canarias. All IPHAS data are processed by the Cambridge Astronomical Survey Unit, at the Institute of Astronomy in Cambridge; $ii.$ The Two Micron All Sky Survey, which is a joint project of the University of Massachusetts and the Infrared Processing and Analysis Center/California Institute of Technology, funded by the National Aeronautics and Space Administration and the National Science Foundation; $iii.$ The NASA/IPAC Infrared Science Archive, which is funded by the National Aeronautics and Space Administration and operated by the California Institute of Technology. This work has made use of the VALD and NIST atomic line databases, and also of the SIMBAD database, operated at CDS (Strasbourg, France). This work has made use of data from the European Space Agency (ESA) mission {\it Gaia} (\url{https://www.cosmos.esa.int/gaia}), processed by the {\it Gaia} Data Processing and Analysis Consortium 
(DPAC, \url{https://www.cosmos.esa.int/web/gaia/dpac/consortium}). Funding for the DPAC has been provided by national institutions, in particular the institutions participating in the {\it Gaia} Multilateral Agreement. We acknowledge the work of the Cambridge Astronomical Survey Unit (CASU) and the support from the LBT-Italian Coordination Facility for the execution of observations, data distribution, and reduction. NOIRLab IRAF is distributed by the Community Science and Data Center at NSF NOIRLab, which is managed by the Association of Universities for Research in Astronomy (AURA) under a cooperative agreement with the U.S. National Science Foundation. KB thanks Alessandro Sozzetti and Ciaran Rogers for fruitful discussions about Gaia and NIRSpec data analysis, respectively. 

\end{acknowledgements}

\bibliography{Biblio}
\bibliographystyle{aasjournal}

\appendix
\restartappendixnumbering  

\section{Spectra of the target sample}
Here, we show the spectra of all targets analyzed in the present work.

\begin{figure*}[!t]
\begin{center}
\includegraphics[width=8.5cm]{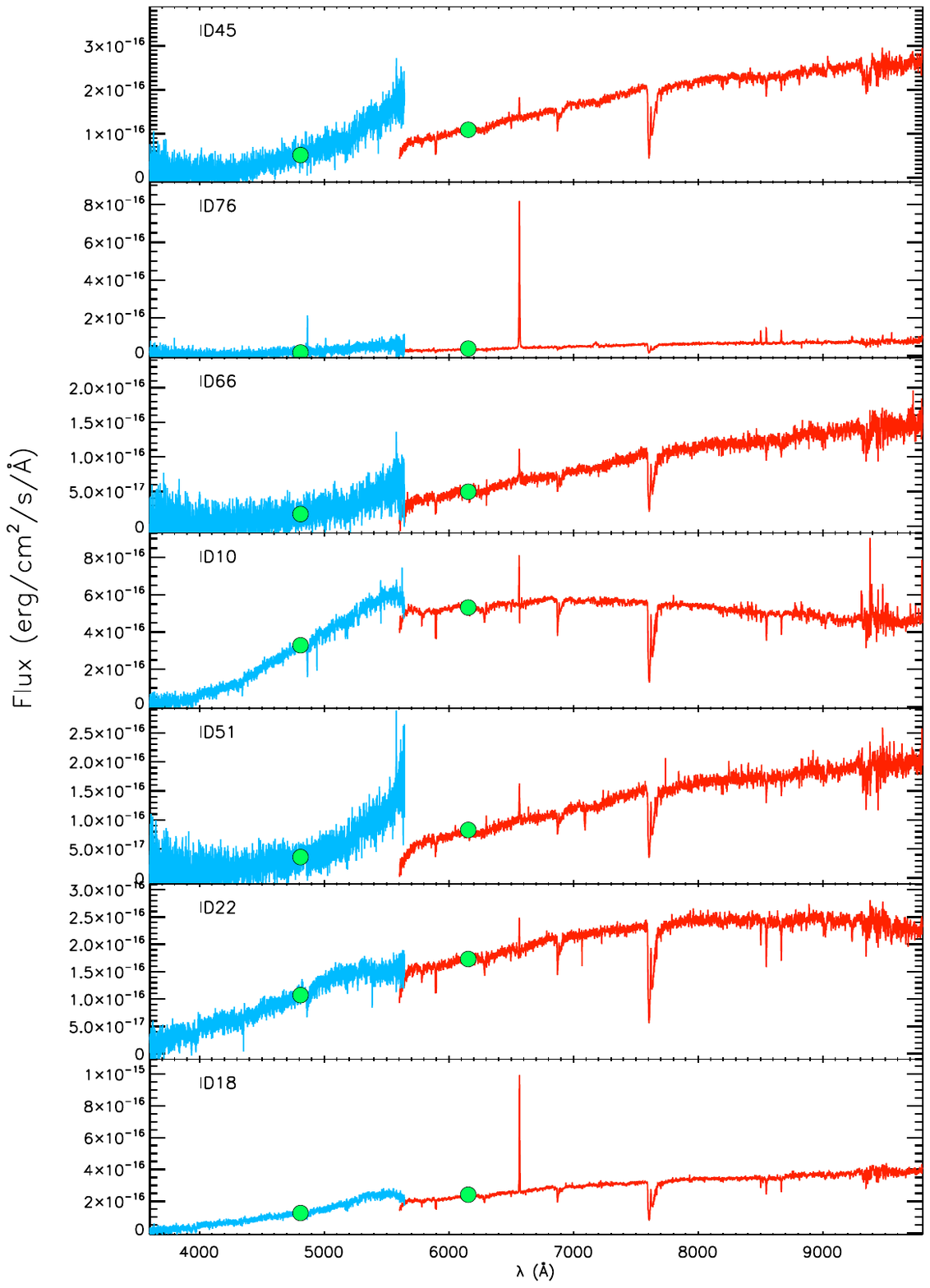}
\includegraphics[width=8.5cm]{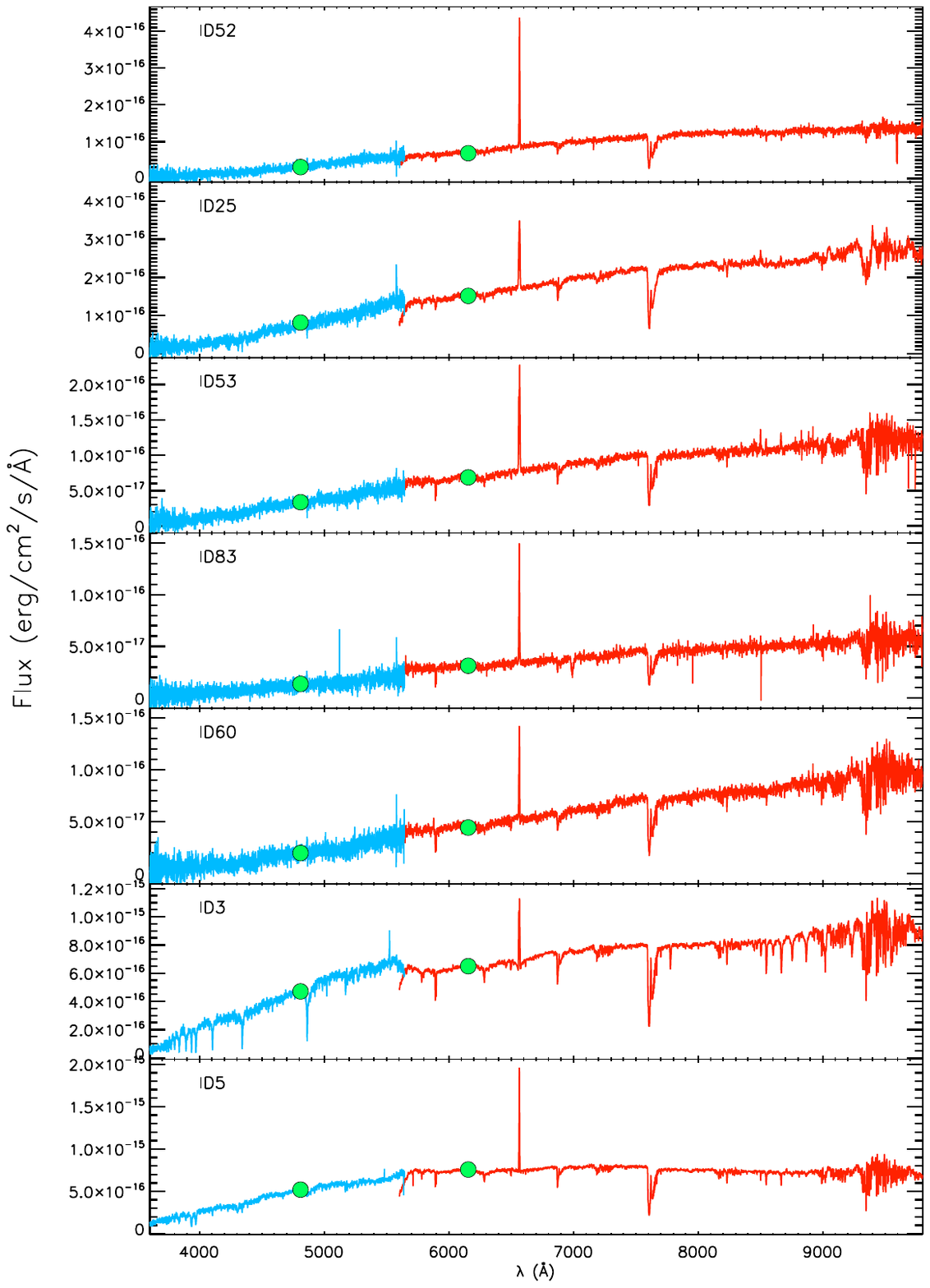}
\includegraphics[width=8.5cm]{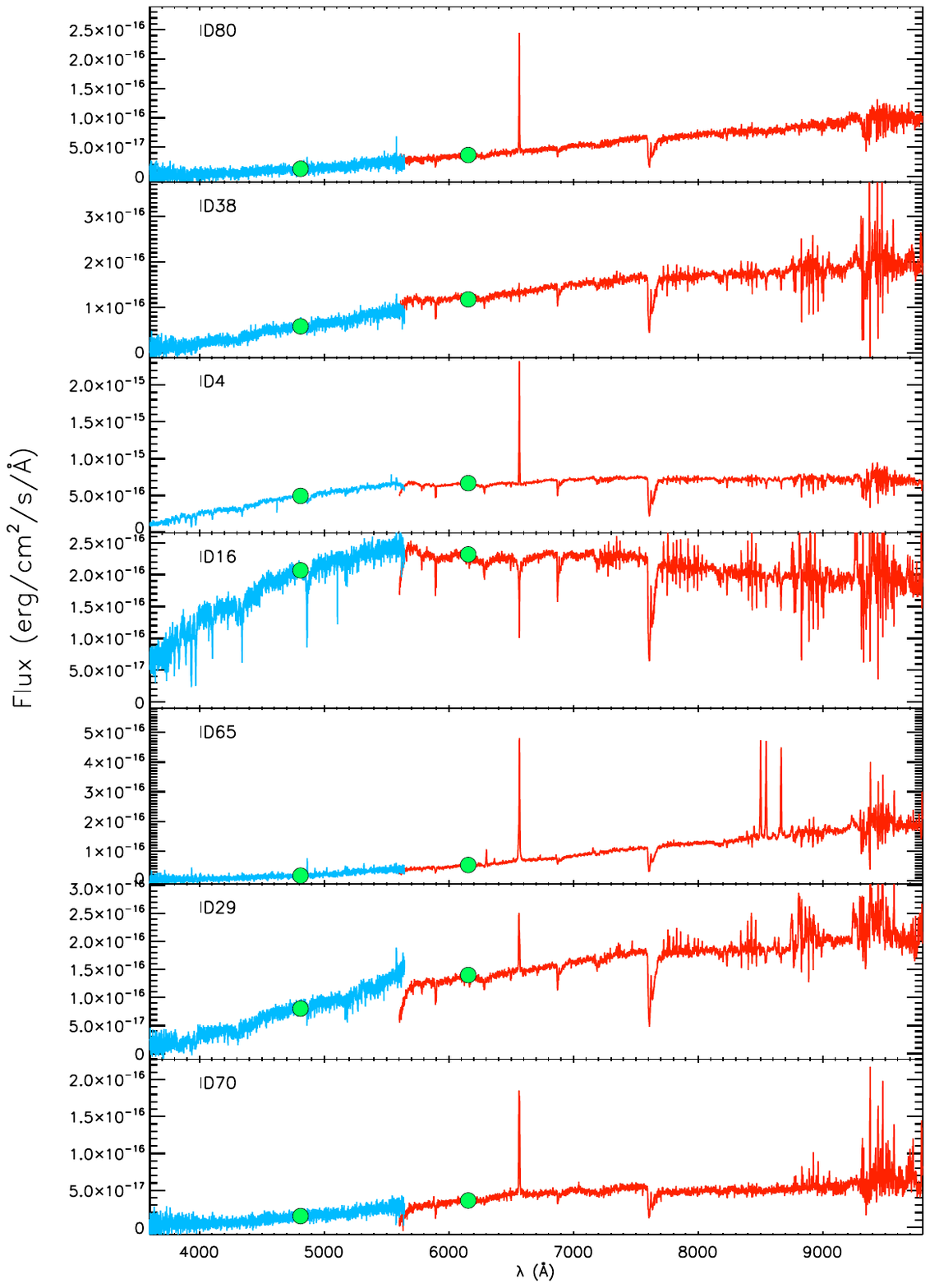}
\includegraphics[width=8.5cm]{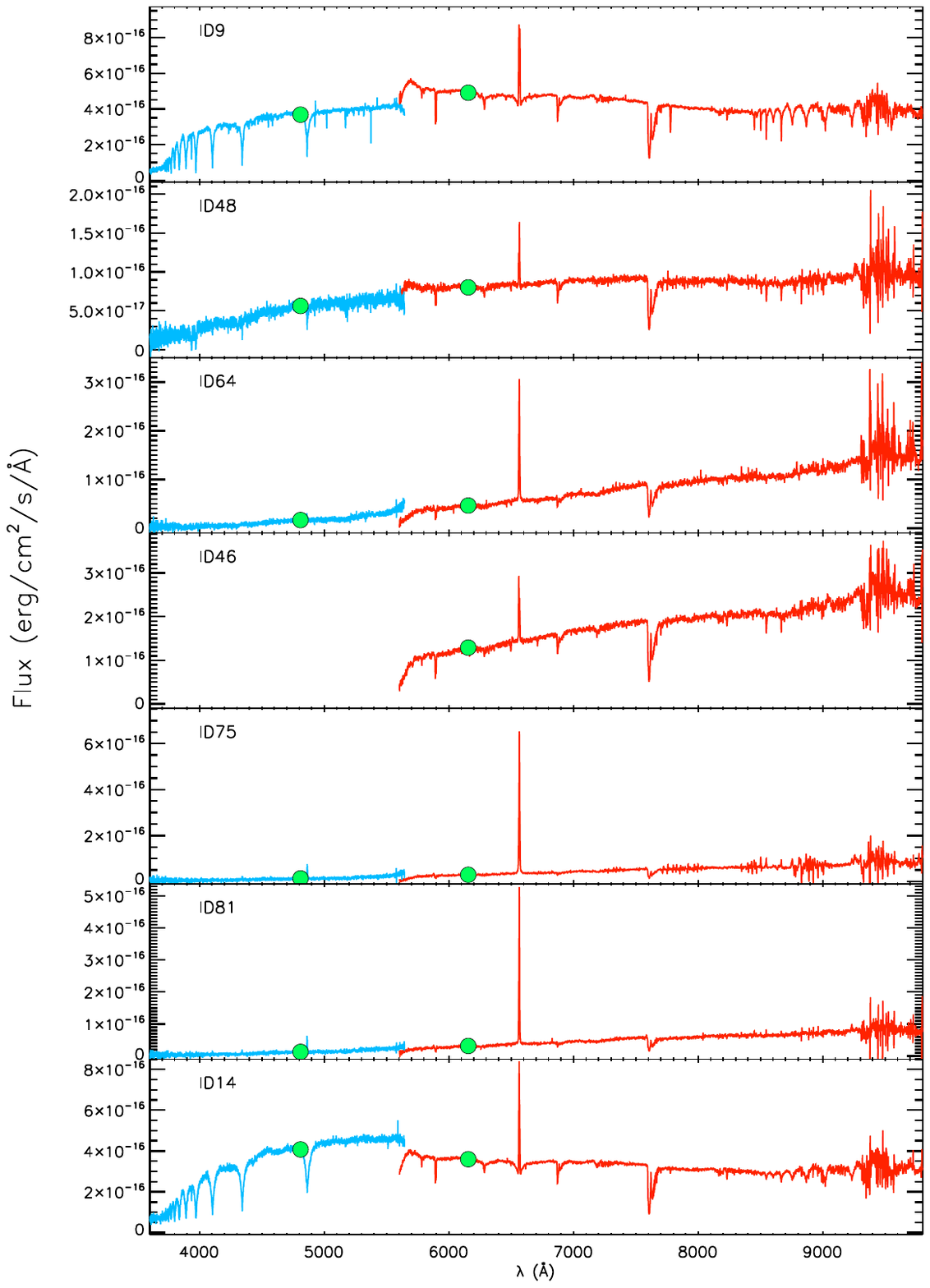}
\caption{Flux-calibrated spectra for 28 targets, displaying both the blue and red spectral regions; target IDs are indicated in in each panel. Accretion properties were derived from diagnostics in the red spectral region, with the exception of ID76, 52, 60, 80, 65, 64, 75, and 81, where additional emission lines were identified and utilized in the blue region.}
\label{fig:spectra1234} 
\end{center}
\end{figure*}

\begin{figure*}[!t]
\begin{center}
\includegraphics[width=8.5cm]{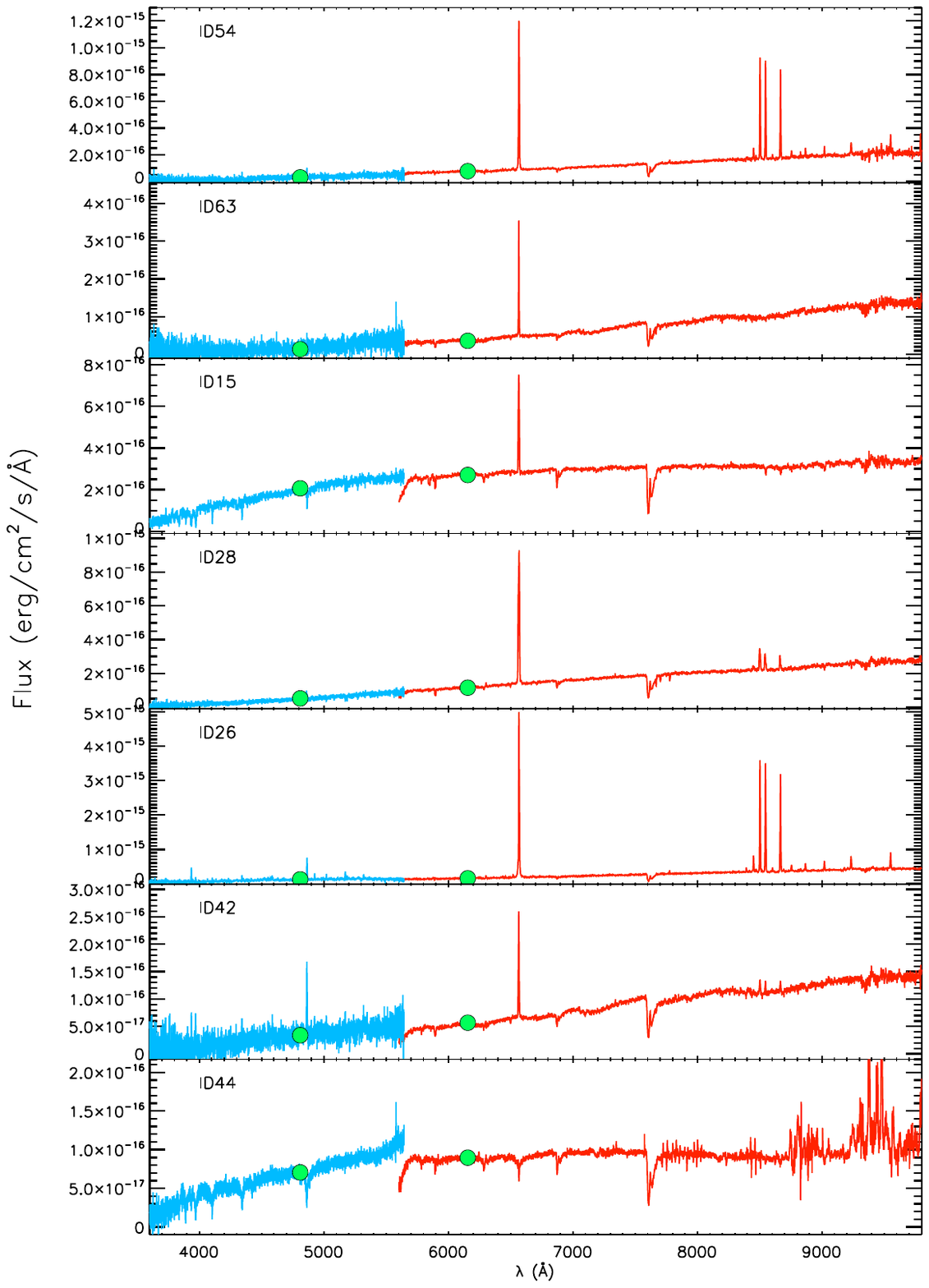}
\includegraphics[width=8.5cm]{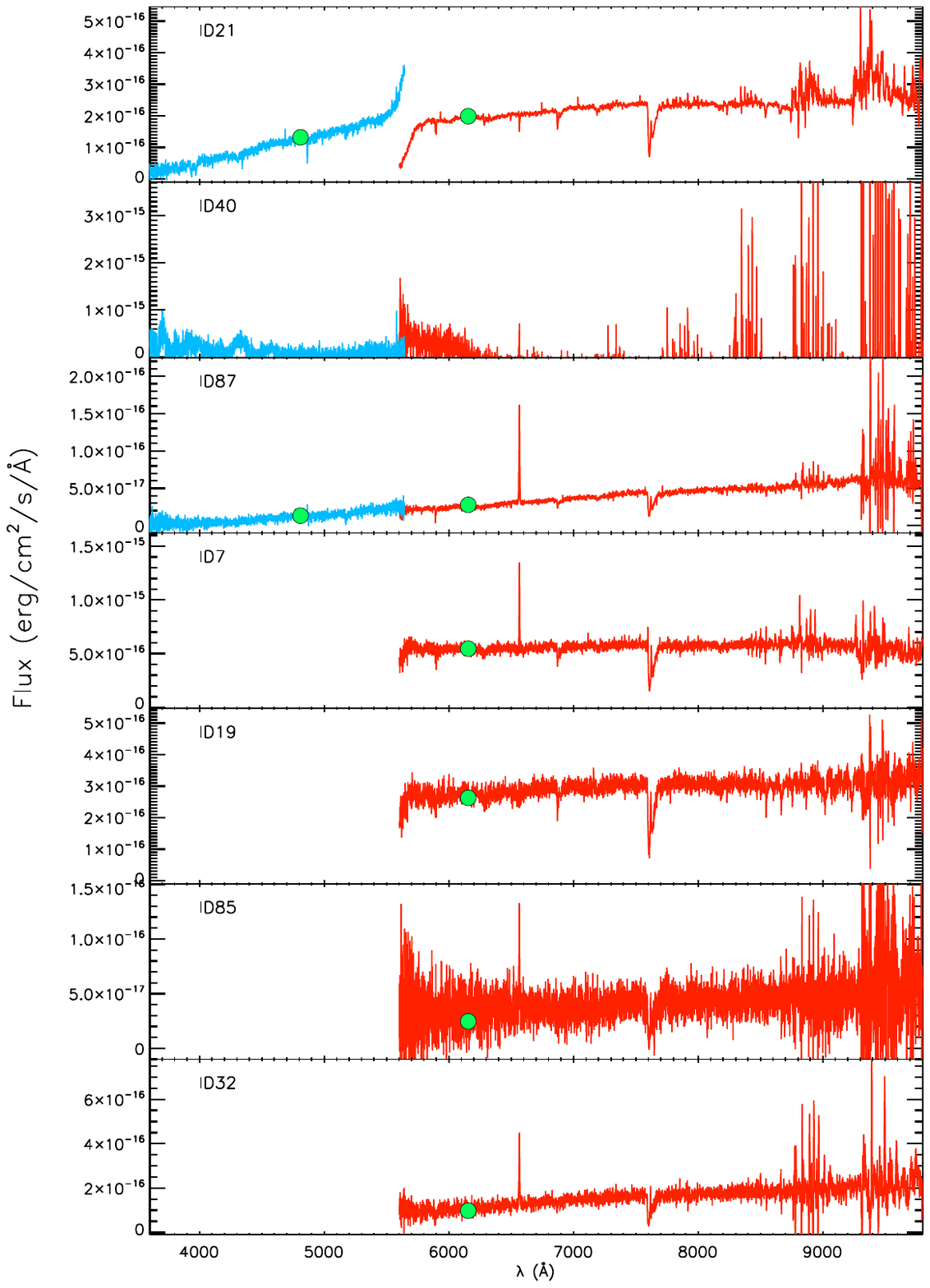}
\includegraphics[width=8.5cm]{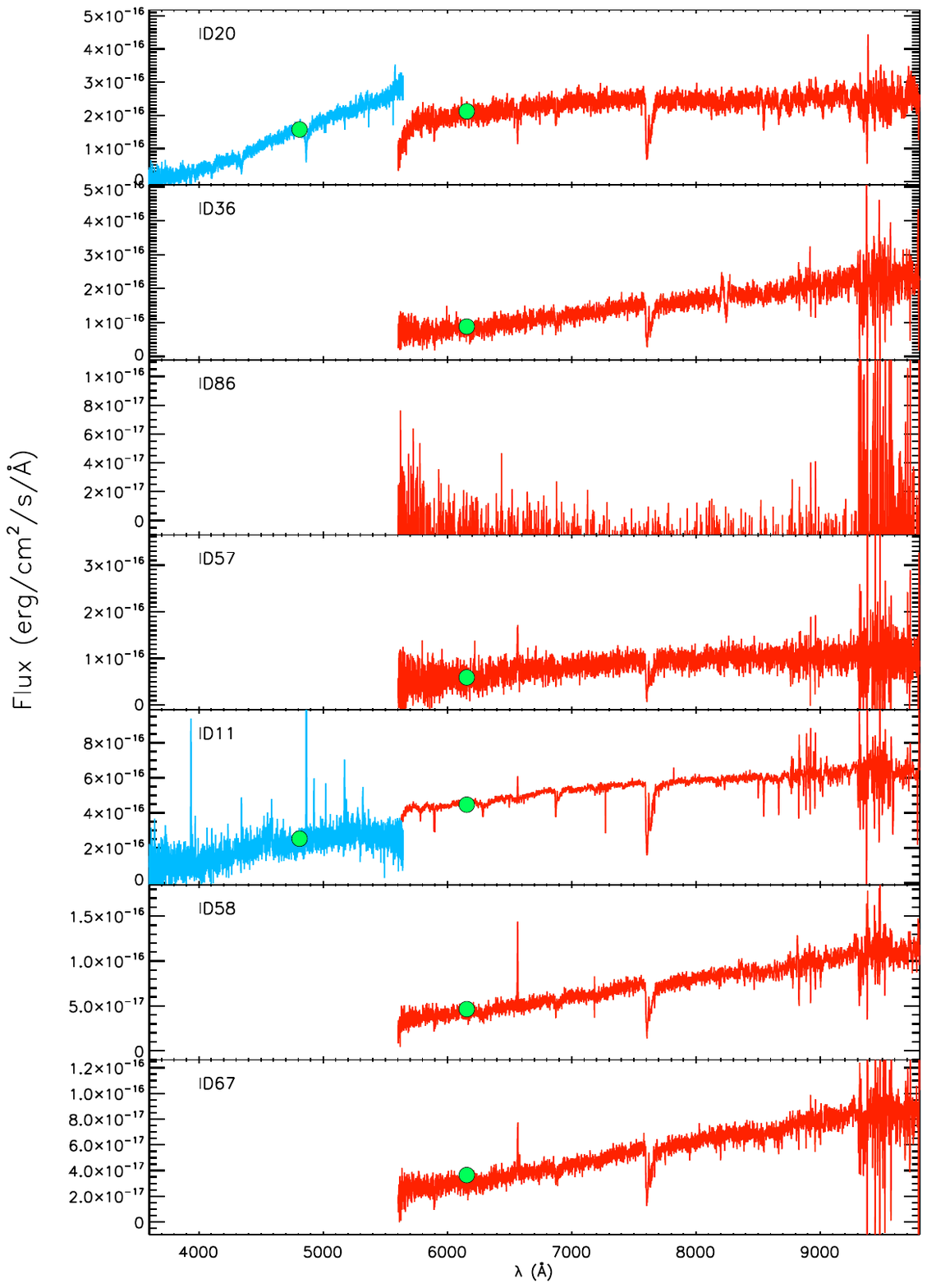}
\includegraphics[width=8.5cm]{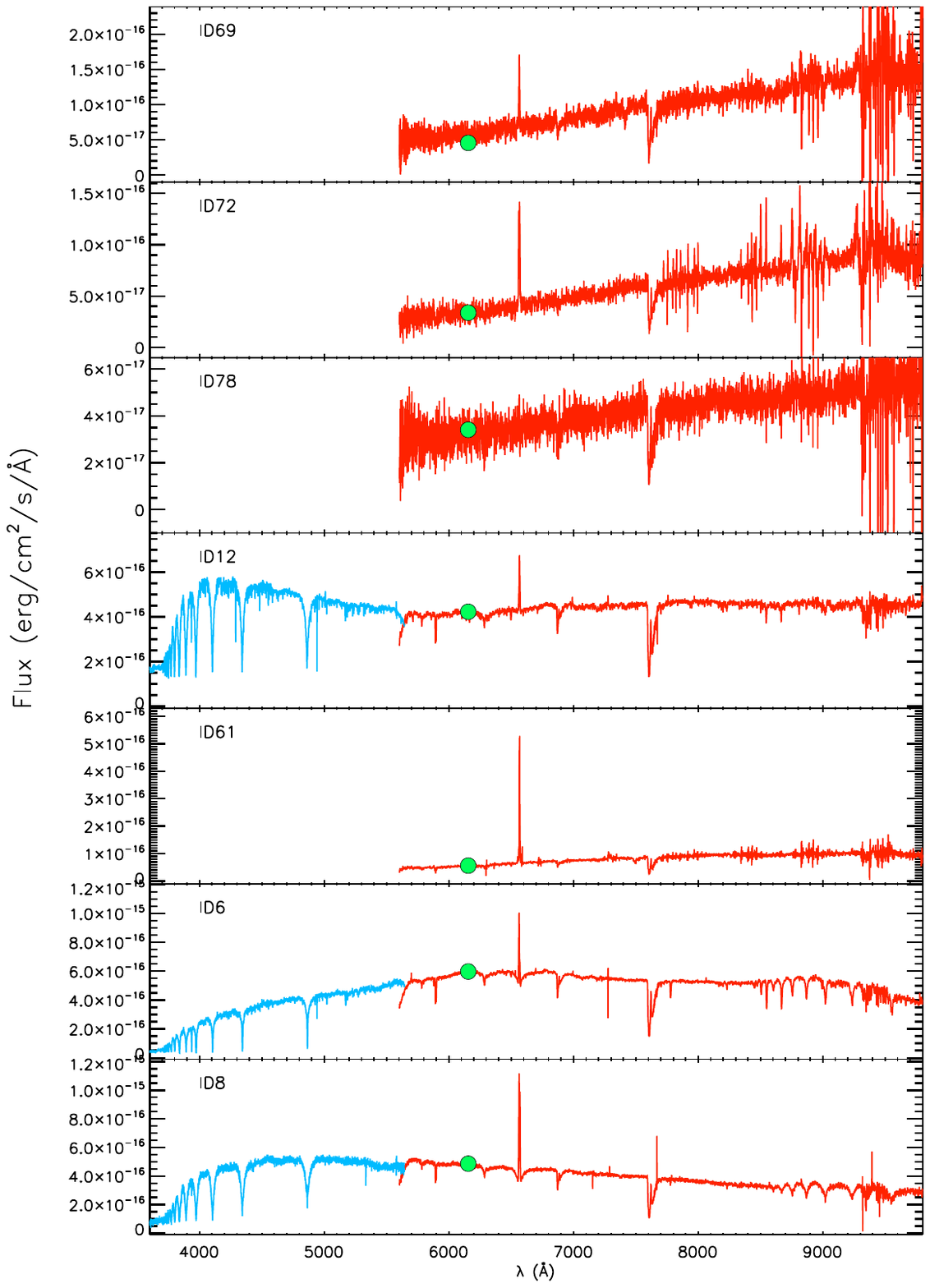}
\caption{Flux-calibrated spectra for 28 targets, displaying both the blue and red spectral regions; target IDs are indicated in in each panel. For sources ID12, 6, and 8, where no simultaneous photometry was available, the blue segments were vertically scaled to match the flux levels of the red segments within their overlapping wavelength range. Accretion properties were derived from diagnostics in the red spectral region, with the exception of ID54, 28, 26, 42, and 87, where additional emission lines were identified and utilized in the blue region. The spectra for ID40 and 86 were excluded from the analysis due to insufficient $S/N$ ratios.}
\label{fig:spectra5678} 
\end{center}
\end{figure*}

\begin{figure*}[!t]
\begin{center}
\includegraphics[width=8.5cm]{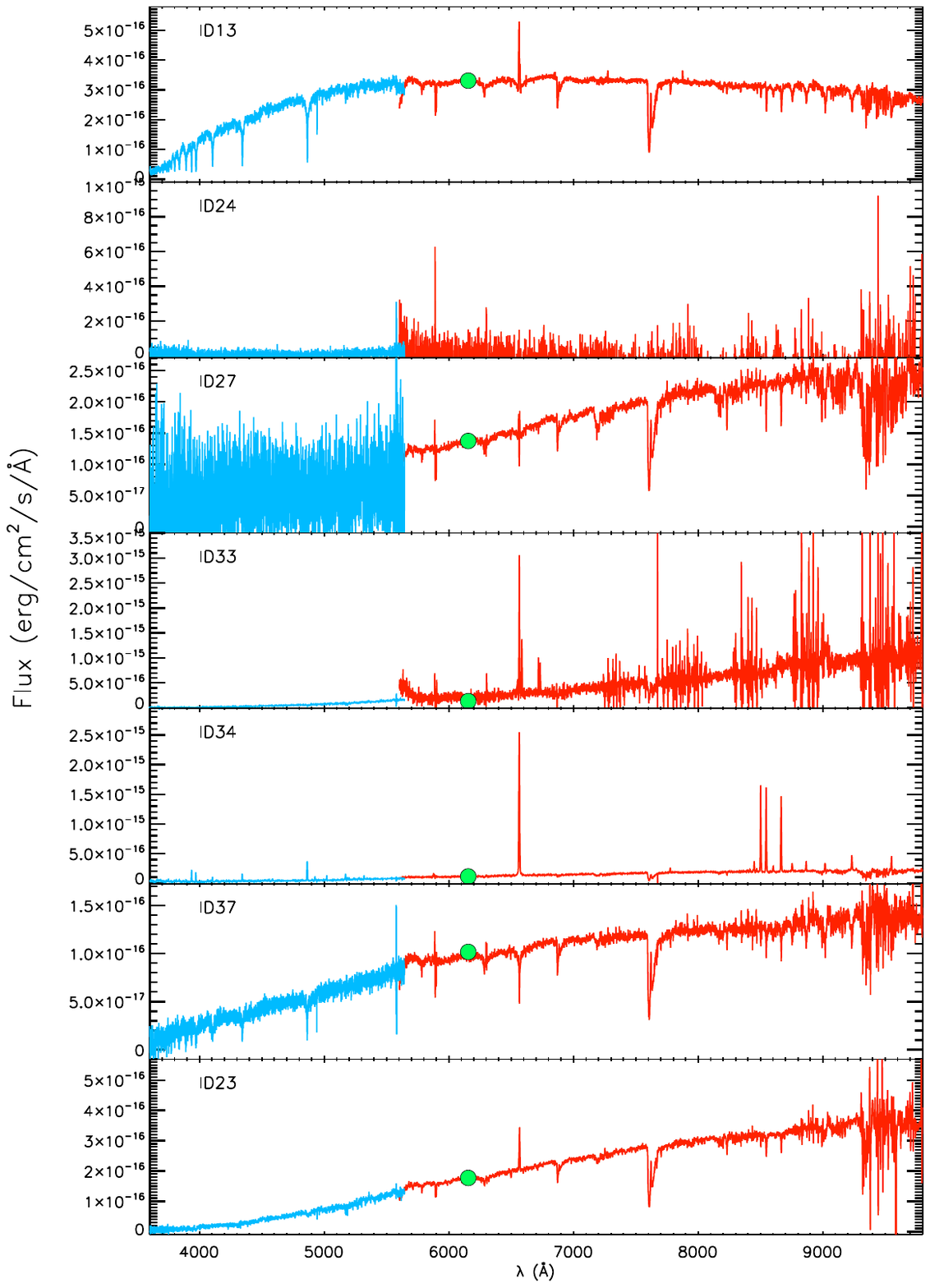}
\includegraphics[width=8.5cm]{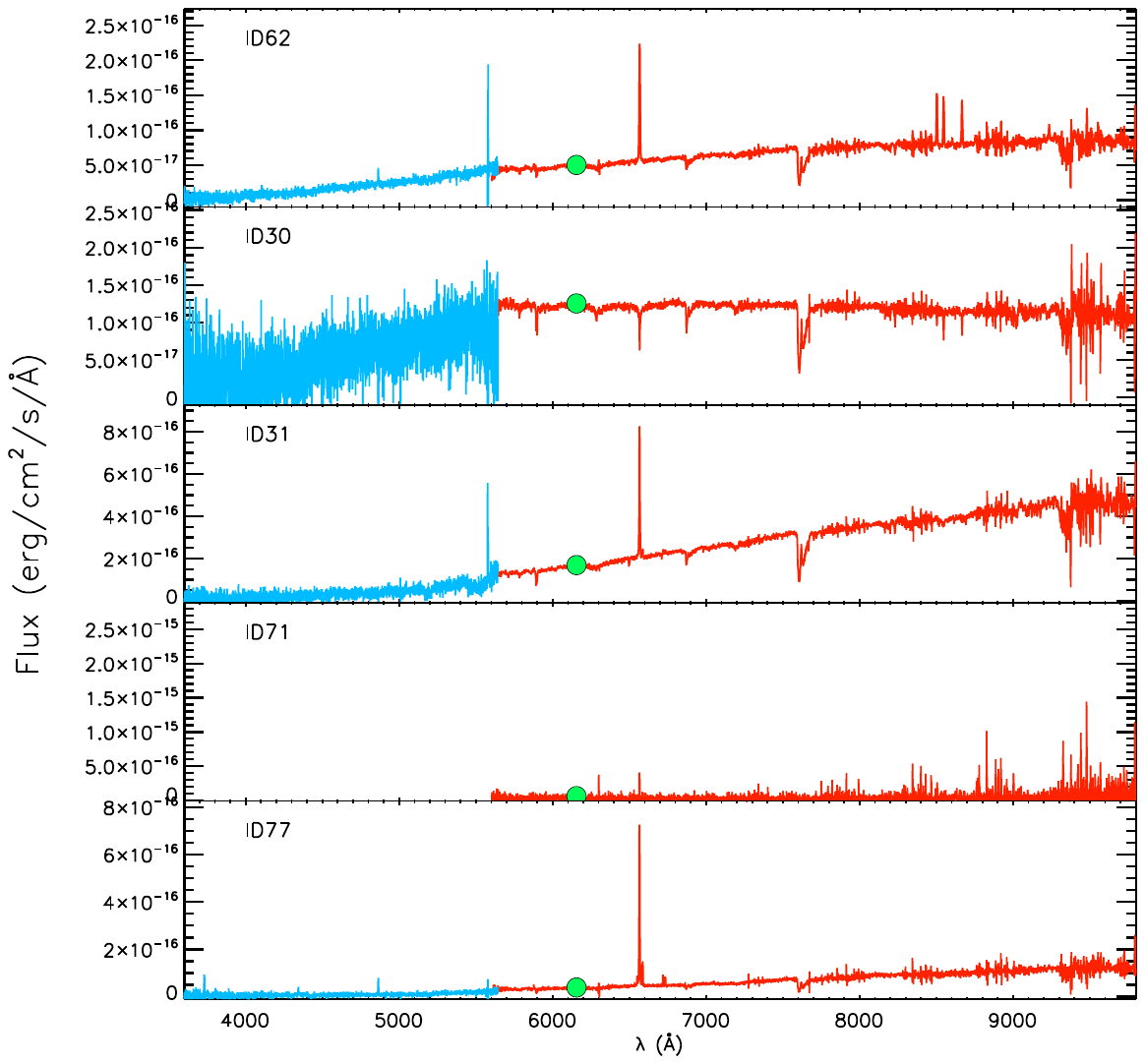}
\caption{Flux-calibrated spectra for 12 targets, displaying both the blue and red spectral regions; target IDs are indicated in in each panel. For sources ID13, 27, 33, 34, 37, 23, 62, 30, 31, and 77, where no simultaneous photometry was available, the blue segments were vertically scaled to match the flux levels of the red segments within their overlapping wavelength range. Accretion properties were derived from diagnostics in the red spectral region, with the exception of ID34, 23, 62, and 77, where additional emission lines were identified and utilized in the blue region. The spectra for ID24 and 71 were excluded from the analysis due to insufficient $S/N$ ratios.}
\label{fig:spectra910} 
\end{center}
\end{figure*}

\section{Accretion variability}
\label{sec:accr_var}

Understanding the impact of variability is crucial for properly interpreting single-epoch accretion diagnostics. Young stellar objects are intrinsically variable across multiple timescales (\citealt{Herbstetal1994}): short-term variations (hours to days) typically stem from rotational modulation by surface spots \citep{Costiganetal2012}, whereas long-term variability (months to years) is driven by accretion rate changes \citep{Biazzoetal2012, Costiganetal2012} or by circumstellar dust \citep{Schisanoetal2009}.

\begin{figure*}[!t]
\begin{center}
\includegraphics[width=13cm]{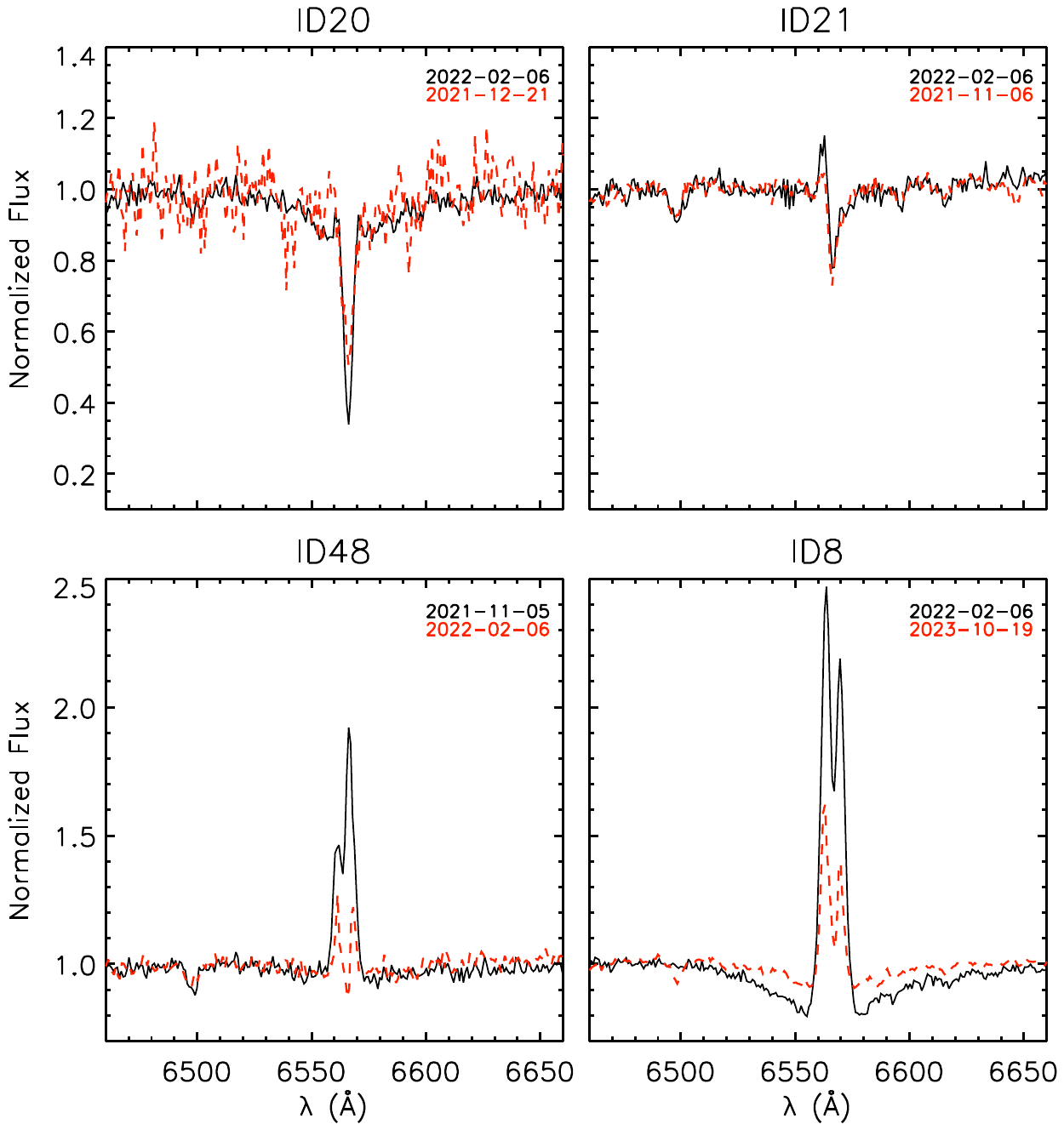}
\vspace{-1cm}
\caption{Observed H$\alpha$ emission line profile variability for the four targets observed at two epochs. All spectra are continuum-normalized. Solid lines show the profiles with the higher intensity, while dashed lines mark those with the lower intensity.}
\label{fig:Halpha_variability} 
\end{center}
\end{figure*}

While most of our analysis relies on single-epoch measurements providing instantaneous estimates of $L_{\rm acc}$ and $\dot{M}_{\rm acc}$, four targets (ID20, ID21, ID48, and ID8) were observed twice over timescales of a few months to nearyly two years (see Fig.~\ref{fig:Halpha_variability}). All four objects were analyzed via spectral subtraction (Sect.~\ref{sec:emission_lines}), which was necessary to detect the net emission features in ID20 and ID21.

Across the two epochs, the net H$\alpha$ equivalent widths vary by factors of $\sim 2–6$, corresponding to changes of $\sim 0.3–0.6$\,dex in $\log \dot{M}_{\rm acc}$. These amplitudes match previous studies \citep{Nguyenetal2009, Costiganetal2014, Fischeretal2023}, but are insufficient to explain the full spread in $\log \dot{M}_{\rm acc}$ observed at a given stellar mass \citep[e.g.,][and references therein; see Sect.~\ref{sec:accr_stmass}]{Nattaetal2006, HerczegHillenbrand2008, Alcalaetal2014, Costiganetal2014, Venutietal2014, Manaraetal2021, Manaraetal2023}. This suggests that variability alone does not dominate the observed accretion rate dispersion, implying that additional stellar or environmental properties must influence the behavior \citep[][and references therein]{Nattaetal2006, HerczegHillenbrand2008, Costiganetal2014, Venutietal2014, Alcalaetal2017}.

Given our small sample of four objects observed over two points in time separated by 2 up to 20 months, this analysis remains as a preliminary evaluation. While the detected amplitudes align with expected YSO behavior, they cannot definitively constrain the role of variability across the entire cluster spread.

\end{document}